\documentclass[prl,amsmath,amssymb,twocolumn, showpacs, superscriptaddress,10pt]{revtex4-1} %retex4-1

\usepackage{amsmath}
\usepackage{hyperref}
\usepackage{graphicx}
\usepackage{amsfonts}
\usepackage{amsthm}
\usepackage{cases}
\usepackage{bm}
\usepackage[utf8]{inputenc}
\usepackage{braket}

\hypersetup{
    colorlinks=true,       % false: boxed links; true: colored links
    linkcolor=black,          % color of internal links (change box color with linkbordercolor)
    citecolor=blue,        % color of links to bibliography
    filecolor=magenta,      % color of file links
    urlcolor=cyan           % color of external links
}

\usepackage[normalem]{ulem}

\usepackage{color}
\definecolor{Blue}{rgb}{0.00, 0.00, 1.00}
\definecolor{Red}{rgb}{1.00, 0.00, 0.00}
\definecolor{Green}{rgb}{0.00, 0.70, 0.00}

\newcommand{\blue}{\color{Blue}}

\hypersetup{
    colorlinks=true,       % false: boxed links; true: colored links
    linkcolor=black,          % color of internal links (change box color with linkbordercolor)
    citecolor=blue,        % color of links to bibliography
    filecolor=magenta,      % color of file links
    urlcolor=cyan           % color of external links
}

\newcommand{\nn}{\nonumber}
\newcommand{\be}{\begin{equation}}
\newcommand{\ee}{\end{equation}}
\newcommand{\bea}{\begin{eqnarray}}
\newcommand{\eea}{\end{eqnarray}}

\newcommand{\dblim}{\lim\limits_{\substack{\ell\to\infty\\t\to\infty}}}

\newcommand{\sr}{{\sf{r}}}
\newcommand{\st}{{\sf{t}}}

\begin{document}

\title{Stationary time correlations for fermions after a quench in the presence of an impurity}  
%{Relaxed wave function for fermionic transport}
\author{Gabriel Gouraud}
\affiliation{Laboratoire de Physique de l'Ecole Normale Sup\'erieure, CNRS, ENS \& Universit\'e PSL, Sorbonne Universit\'e,
Universit\'e de Paris, 75005 Paris, France}
\author{Pierre Le Doussal}
\affiliation{Laboratoire de Physique de l'Ecole Normale Sup\'erieure, CNRS, ENS \& Universit\'e PSL, Sorbonne Universit\'e,
Universit\'e de Paris, 75005 Paris, France}
%\author{Satya N. \surname{Majumdar}}
%\affiliation{LPTMS, CNRS, Univ. Paris-Sud, Universit\'e Paris-Saclay, 91405 Orsay, France}
\author{Gr\'egory \surname{Schehr}}
\affiliation{Sorbonne Universit\'e, Laboratoire de Physique Th\'eorique et Hautes Energies, CNRS UMR 7589, 4 Place Jussieu, Tour 13, 5\`eme \'etage, 75252 Paris 05, France}
\date{\today}

\begin{abstract}
We consider the quench dynamics of non-interacting fermions in one dimension in the presence of a finite-size impurity at the origin.
This impurity is characterized by general momentum-dependent reflection and transmission coefficients which are
changed from $\sr_0(k), \st_0(k)$ to $\sr(k), \st(k)$ at time $t=0$. The initial state is at equilibrium 
with $\st_0(k)=0$ such that the system is cut in two independent halves with $\sr_0^R(k)$, $\sr_0^L(k)$ respectively to the right and to the left of the impurity. 
We obtain the exact large time limit of the multi-time correlations. These correlations become time translationally
invariant, and are non-zero in two different regimes: (i) for $x=O(1)$ where the system reaches a non-equilibrium steady state (NESS) 
(ii) for $x \sim t$, i.e., the ray-regime. For a repulsive impurity these correlations are independent of
$\sr_0^R(k)$, $\sr_0^L(k)$, while in the presence of bound states they oscillate and memory effects persist. We show that 
these nontrivial relaxational properties can be retrieved in a simple manner from the large time 
behaviour of the single particle wave functions. 
\end{abstract}
\maketitle

{\it Introduction}. There is a fast growing interest in noninteracting fermions in the presence of external potentials as solvable models capturing non trivial
quantum correlations. For fermions at equilibrium, either in a trap \cite{castin,CMV2012,eisler_prl,marino_prl,us_review_PRA} or in the presence of an impurity \cite{Fuchs1,Fuchs2,DLMS2021}, spatial correlations have been computed using various analytical methods, including determinantal processes and connections to random matrices \cite{eisler_prl,marino_prl,us_review_PRA,Calabrese_RMT,us_review_JPA}, as well as inhomogeneous bosonisation~\cite{Dubail_stat}. 
Extensions of these methods allow to also obtain the temporal correlations at equilibrium \cite{Dubail_stat,us_eq_dyn}. 

%In the presence of a localized impurity, the Fermi gas exhibits Friedel oscillations \cite{Friedel}, which were recently revisited in~\cite{Fuchs1,Fuchs2,DLMS2021}. 

Noninteracting fermions also provide interesting tractable models to study non-equilibrium quantum dynamics. A seminal example is the Landauer-B\"uttiker theory for transport between two reservoirs \cite{buttiker_prl,Buttiker_review,levitov} where the system is assumed to be in a stationary state from the start. 
A more general setting to study out of equilibrium dynamics and convergence to a stationary state are quantum quenches. 
Noninteracting fermions have been much studied in this context for translationally invariant systems. The resulting equilibrium state reached at large time can be predicted
using the generalised Gibbs ensemble (GGE), which more generally describes integrable interacting systems, see e.g. \cite{FagottiDynamicalCorrelations2012}.  
In these cases, the dynamics can also be computed using the generalised hydrodynamics (GHD) \cite{GHDDoyonYoshimura,review_GHD_fagotti} which predicts correlations in the so-called ray regime $x \sim t$. For noninteracting fermions, the GHD amounts to consider a semi-classical version of the Wigner function. To address the transport properties using quantum quenches a standard protocol is to consider an
initial state %which is 
inhomogeneous in space in one dimension 
\cite{AntalXX1999,Krapivsky2008,DubailViti2016,EislerRacz2013,DeLucaVitiBernardDoyon2013,EislerMaislinger2016,PerfettoGambassi2017,Kormos2017,
Sasamoto2019,GiamarchiRuggieroQuenchLL2021,GHDBertiniDeNardis,Sotiriadis}. 
At late time the system reaches a non-equilibrium stationary state (NESS) characterized by stationary currents, density profiles and counting statistics~\cite{ColluraKarevski2014}. 

An important question is how the transport in the NESS as well as in the ray regime is affected by the presence of inhomogeneities in space.
The simplest case is a local impurity which has been studied, e.g. in \cite{BertiniFagottiDefect2016,Mazza2016,DeLucaDefect2021},
or using conformal field theory \cite{BernardDoyonViti2015,GiamarchiCalabreseQuenchLL2021,BernardDoyonEnergy2012}. Recently,
for noninteracting fermions in one-dimension, several analytical results have been obtained 
%for the current, the correlations, the full counting statistics and the growth of entanglement entropy, 
both for discrete \cite{Prosen2018,deLucaMovingDefect1,Gamayun1,eisler_entropy,calabrese_entropy,bertini_entropy} and continuum \cite{GLS,Gamayun22} models. In these models, the system is initially prepared either with a domain wall or partitioned into two separate halves. 
It then evolves in the presence of an impurity, modeled either by a so-called ``conformal defect'' \cite{bertini_CFT_defect,eisler_entropy,calabrese_entropy}
which is simpler to analyze, or by an external local potential \cite{Prosen2018,deLucaMovingDefect1,Gamayun1,GLS,Gamayun22}.

Despite these works there is no first-principle derivation of the quantum correlations in the NESS for a general defect for noninteracting fermions in the continuum setting. 
%
%There remain many interesting open questions. For instance at present there is no first-principle derivation of the quantum correlations in the NESS for a general defect 
%in the continuum setting. 
% In particular it would be interesting to quantify the memory of the initial conditions, in presence
% of bound states, which are known to lead to persistent oscillations. 
In addition, the stationary dynamics, such as the temporal correlations
in the NESS, and how they differ from equilibrium have not been characterized so far. A standard approach in quantum quenches, e.g., in the context of GHD, relies on a semi-classical version of the Wigner function. In the presence of an impurity it has been improved by taking into account
reflection and transmission probabilities~\cite{Prosen2018,GLS}. 
It was shown to predict the correct large time correlations inside a local region within a single ray $\xi=x/t$. 
However, this approach, although appealing, does not predict the correlations in the NESS ($x=O(1)$) and furthermore, 
even in the ray regime, it does not capture the non-trivial correlations along opposite rays i.e., for $(x,x')\simeq(\xi t, -\xi t)$ \cite{GLS}. 

% {\blue However, this approach, although appealing wasn't designed to capture  nonetheless interesting features. First local quantity, namely correlations in the microscopic regime, when $x\simeq O(1)$, and also more surprisingly, ray regime related quantity such as correlations along opposite rays i.e. when $(x,x')\simeq(\xi t, -\xi t)$ \cite{GDS22}} 

% However, this approach, although appealing does not provide a complete description of the correlations in the NESS and even in the ray regime: 
% for instance it fails to capture the correlations
% between two opposite rays \cite{GDS22}. 

In this paper we address these questions, by considering non interacting fermions in one dimension prepared 
in the presence of an arbitrary localized impenetrable barrier with a corresponding Hamiltonian $\hat H_0$. The initial condition is 
a product state of two half-spaces, each at equilibrium at different temperatures and densities.
This state is evolved at time $t>0$ with Hamiltonian $\hat H$ on the full line in the presence of a penetrable impurity localized
near the origin. The system reaches a NESS at large time, with a stationary current. 

We first show that the eigenfunctions of $\hat H_0$ evolved with $\hat H$ have a large time limit 
which we obtain explicitly in terms of the scattering coefficients of the impurity. The 
interferences between these eigenfunctions allow us to compute all space-time quantum correlations
in the NESS as well as in the ray regime. Furthermore they have a nice physical interpretation in terms of
scattered trajectories. This goes beyond the aforementioned semi-classical approximation of the Wigner function,
and incorporates all quantum interference effects, including those leading to correlations between opposite rays. As a byproduct we find that the 
memory of the initial details is erased in the large time limit, except in the presence of
bound states where oscillations persist at all times.

{\it Model.} We consider $N$ noninteracting fermions in one dimension in the presence of
a finite size impurity modeled by a potential. %$v(x)$. 
The evolution at $t>0$ is governed by the single particle Hamiltonian
\be  \label{def_H}
\hat H=-\frac{1}{2}\partial_x^2+V(x) \;.
\ee 
We work in units where $\hbar=1$ and the mass $m=1$. For simplicity we start with a potential $V(x)$ which is symmetric ($V(x) = V(-x)$), localized in the region $[-\frac{a}{2},\frac{a}{2}]$ and does not possess any bound state (i.e., a repulsive impurity). Such a potential is characterized by a scattering matrix 
\be \label{def_S_intro}
{\cal S}(k) =  \left( {\begin{array}{cc}
    \st(k) & \sr(k) \\
    \sr(k) & \st(k) \\
  \end{array} } \right)\; {\rm with} \;\;
\begin{cases}  
\hspace*{-0.4cm}&|\sr(k)|^2 + |\st(k)|^2 =1 \\
\hspace*{-0.4cm}&{\rm Re}{\left[\sr(k) \st(k)^*\right]}  = 0 
\end{cases}  \;,
\ee
where $\sr(k)$ and $\st(k)$ are momentum dependent reflection and transmission coefficients. 
%We will also use the re-parametrisation
%\bea
%&r+t=e^{-2i\delta_k^+}\\
%&r-t=e^{-2i\delta_k^-}
%\eea
The fermions are confined in a hard box of size $\ell$ with $\ell > a$ (i.e., the wave function vanishes outside $[-\ell/2,\ell/2]$). 

The system is prepared at $t=0$ at Gibbs equilibrium and is described by the single particle Hamiltonian $\hat H_0$
\be  \label{def_H0}
\hat H_0=-\frac{1}{2}\partial_x^2+V_0(x) \;,
\ee 
where $V_0(x)$ is another potential localized in the region $[-\frac{a}{2},\frac{a}{2}]$ and characterized by a scattering matrix
\be \label{def_S}
{\cal S}_0 =  \left({\begin{array}{cc}
    0 & \sr_0^R(k) \\
    \sr_0^L(k) & 0 \\
  \end{array} }\right)\quad, \; |\sr_0^L(k)|^2  = |\sr_0^R(k)|^2 =1 \;.
\ee
The potential $V_0(x)$ is sufficiently divergent at $x=0$ so that the system is cut in two halves with 
$\st^R_0(k)=\st_0^L(k)=0$. The initial $N$-body density matrix $\hat D = \hat {D}_{L} \otimes \hat {D}_{R}  $ is  
the tensor product of left and right density matrices $\hat {D}_{L/R}$, 
each describing equilibrium at temperature $T_{L/R}$ with chemical potentials $\mu_{L/R}$.
In the zero temperature case, this amounts to consider the ground state with a fixed number of fermions $N_{L/R}$
on each side. 

The normalized eigenfunctions ${\phi_k^{R/L}}(x)$ of the initial Hamiltonian $\hat H_0$ vanish
for $x \in \mathbb{R}^{-/+}$ and can be written outside the interval $[-\frac{a}{2},\frac{a}{2}]$
\bea\label{ci}
&&\phi_k^{R}(x) = c^0_\ell \cos(k(|x|-\delta_k^{R}))\theta(x-\frac{a}{2}),\, k\in\Lambda^R\\
&&\phi_k^{L}(x) = c^0_\ell \cos(k(|x|-\delta_k^{L}))\theta(-\frac{a}{2}-x),\, k\in\Lambda^L
\eea
with $c^0_\ell \simeq \sqrt{\frac{4}{\ell}}$ at large $\ell$ and where $\theta(x) = 0$ if $x \leq 0$ and $\theta(x)= 1$ if $x>0$. The phase shifts $\delta_k^{R/L}$ are related to the reflection coefficients as
\bea
\sr_0^{R/L}(k)=e^{-2ik\delta_k^{R/L}} \;,
\eea
while the lattices $\Lambda^R$ and $\Lambda^L$ are defined as follows
\bea \label{quantification_initial}
\Lambda^{R/L}=\left\{k \in \mathbb{R}^+|\phi_k^{R/L}(\pm \frac{\ell}{2})=0\right\} \;.
%&&\Lambda^L=\{k \in \mathbb{R}^+|\phi_k^{L}(-\frac{\ell}{2})=0\} \;.
\eea

{\it Observables.} We are interested in the space-time $m$ point density correlation functions, defined 
for $0 \leq t_1 \leq \dots \leq t_m$ and all distinct space time points $(x_i,t_i)$ as
%{\red P: this cannot be correct, unless all points are distinct,
%since it is already incorrect for $m=2$ and $t_1=t_2$ it misses the delta function}
\be\label{multitimecorr1}
{\cal C}_m(x_1,t_1;...;x_m,t_m)={\rm Tr}( \hat D \hat \rho(x_m,t_m) \ldots \hat \rho(x_1,t_1))
\ee 
where $\hat \rho(x,t)$ is the density operator in the Heisenberg representation. For 
noninteracting fermions they can be expressed as an $m \times m$ determinant involving the so-called space-time extended kernel (see \cite{DLSM2019,DLSM2019arXiv})
\bea\label{multitimecorr2}
{\cal C}_m(x_1,t_1;...;x_m,t_m)=\det_{1\leq i,j\leq m}K(x_i,t_i;x_j,t_j) \;.
\eea
{In particular, the total fermion density reads $\rho(x,t) = K(x,t;x,t)$.} The current correlations can also be obtained from the kernel \cite{SM}. 
The space-time extended kernel can be written as the sum 
\bea\label{initial_kernel}
K(x,t;x',t')=K_R(x,t;x',t')+K_L(x,t;x',t') \;,
\eea
where \cite{footnote1} 
\bea
&&K_{R/L}(x,t;x',t') \label{timedepkernel}  \\
&& =\sum\limits_{k\in\Lambda^{R/L}} (f_{R/L}(k)-\theta(t'-t))  \psi_k^{R/L,*}(x,t)\psi_k^{R/L}(x',t') \;.
\nonumber 
\eea 
Here $f_{R/L}(k)=1/(1+ e^{\beta_{R/L}(\mu_{R/L} - \frac{k^2}{2})})$ are the right and left Fermi factors,
and $\psi_k^{R/L}(x,t)$ is the solution of the Schrödinger equation $i \partial_t \psi_k^{R/L}(x,t) = \hat H \psi_{k}^{R/L}(x,t)$
with the initial condition $\psi_k^{R/L}(x,0)=\phi_k^{R/L}(x)$, where $\phi_k^{R/L}(x)$ are given in Eq. (\ref{ci}).  

{\it Large time limit: the NESS}. We now want to take a double limit: first the thermodynamic limit $\ell \to +\infty$ with fixed left and right mean densities
(i.e., fixed chemical potentials $\mu_{L,R}$) 
and then the large time limit~\cite{Lippman}. We first consider the system over distances $x = O(1)$ (i.e., ``close to the impurity''). In this regime, the system reaches a NESS where one-point quantities (such as the density) becomes stationary, and multi-time correlations become time-translationally invariant.
%where all times go to infinity, $t\to\infty$ ($\substack{\ell\to\infty\\t\to\infty}$), with fixed
%time differences. This will result in a stationary state, as we will see. 
We can compute exactly the large $\ell$ and large time limit of the kernel starting from the exact formula in Eq. (\ref{timedepkernel}). Since it is 
a bit cumbersome (similar to the calculations in~\cite{Prosen2018,GLS}) we instead present here a shortcut by studying directly the asymptotic form of the single-particle 
time evolved initial eigenfunctions, $\psi_k^{R/L}(x,t)$, in the large $\ell$ and large $t$ limit.  
The exact formula for $\psi_k^{R/L}(x,t)$ is an infinite superposition 
which involves the overlaps of $\phi_k^{R/L}(x)$ with all the eigenstates of $\hat H$. The asymptotic form of this superposition is obtained using a contour-integral representation which leads to \cite{SM}
\bea\label{limitfunction}
\psi_k^{R/L}(x,t) = \frac{1}{\sqrt{\ell}} ( e^{-i\frac{k^2}{2}t}\chi_k^{R/L}(x) + \delta \chi_{k,\ell}^{R/L}(x,t) ) 
\eea
where $\delta \chi_{k,\ell}^{R/L}(x,t)$ decays to zero in the limit of large $\ell$ followed by large $t$. 
%and we find that it takes a simple form \cite{SM} 
%{\red $k$ on the left belongs to $\Lambda$ and on the right to $\mathbb{R}$
%\bea\label{limitfunction}
%\psi_k^{R/L}(x,t)\dbleq \frac{1}{\sqrt{\ell}}e^{-i\frac{k^2}{2}t}\chi_k^{R/L}(x)%=\psi_{\infty,k}^{R/L}(x,t)
%\eea
%where on the left hand side $k \in \Lambda_{R/L}$ while on the right hand side $k \in \mathbb{R}^+$ (since the limit $\ell \to \infty$ has been taken). 
In Eq. (\ref{limitfunction}), the leading contributions $\chi_k^{R/L}(x)$ are given by
\bea 
&\chi_k^{R}(x)= \begin{cases} \label{chiR}
( e^{-ikx}+\sr(k)e^{ikx} ) e^{ik\delta_k^{R}} &\mbox{ if } x>\frac{a}{2}\\
\st(k)e^{-ikx} e^{ik\delta_k^{R}}  &\mbox{ if } x<-\frac{a}{2}
\end{cases}\\
&\chi_k^{L}(x)=\begin{cases} \label{chiL}
\st(k)e^{ikx} e^{ik\delta_k^{L}} &\mbox{ if } x>\frac{a}{2}\\
(e^{ikx}+\sr(k)e^{-ikx}) e^{ik\delta_k^{L}} &\mbox{ if } x<-\frac{a}{2}
\end{cases}
\eea
{where we recall that $\sr(k)$ and $\st(k)$ are the reflection and transmission coefficients (\ref{def_S_intro}). 
In this result (\ref{limitfunction}) the time-dependence $\propto e^{-i\frac{k^2}{2}t}$ is simply the one of a free particle of energy $\frac{k^2}{2}$, while the factor $\frac{1}{\sqrt{\ell}}$ ensures the normalization of $\psi_k^{R/L}(x,t)$. 
The form of $\chi_k^{R/L}(x)$ in \eqref{chiR} and \eqref{chiL}
can be qualitatively understood as follows. Away from the impurity, at time $t=0$, 
from \eqref{ci} a particle can have momentum $k$ or $-k$ (everywhere $k>0$). Consider a space time point $(x,t)$ with $x=O(1)>0$ and $t$ large
and first ask how a particle starting {\it from the left} of the impurity
can reach $(x,t)$. As shown in Fig. \ref{Fig1} (top left), there is a single possible initial position such that a particle with initial momentum $k$ reaches $(x,t)$. Since it crosses the barrier, it collects a factor ${\sf t}(k)$. This accounts for the first line in \eqref{chiL}. It contains a single term, with 
phase factor $e^{ik\delta_k^{L}}$, since particles with initial momentum $-k$ escape to $-\infty$ (the phase factor information they carry $e^{-ik\delta_k^{L}}$ is lost). For particle starting {\it from the right}
of the impurity, one similarly interprets the two terms in the first line in \eqref{chiR}. Indeed, one 
sees from Fig. \ref{Fig1} (top right) that there are two possible initial positions such that a particle with
initial momentum $-k$ reaches $(x,t)$ either (i) directly (leading to the factor $e^{-i k x}$)
or after one reflection (which changes $-k$ into $k$ leading to the factor ${\sf r}(k) e^{i k x}$). 
%  Concerning $\chi_k^{R/L}$ from \eqref{chiL} and \eqref{chiR}, they can be qualitatively understood as follows (see also Fig. \ref{flux_balance}). The wave function $\chi_k^L(x)$ describes the scattering of a free particle which is coming infinitely far away from the left (i.e., the term $\propto e^{ik(x+\delta_k^L)}$ where the phase $e^{ik\delta_k^L}$ is inherited from the initial condition (\ref{ci})). %$\cos(k(|x|-\delta_k^{L}))=\frac{1}{2}(e^{ik(-x-\delta_k^{L})}+e^{-ik(-x-\delta_k^{L})})$). 
% %where only the second exponential survive. The first one does not survive as it propagates to the left). 
% This free particle is then either (i) transmitted, hence multiplied by $\st(k)$ or (ii) reflected, hence multiplied by $\sr(k)$. In the latter case, $e^{ikx}$ is changed to $e^{-ikx}$, since the reflected particle propagates to the left as $k$ is positive. Similarly the wave function $\chi_k^R(x)$ represents the scattering of particle coming from the right. Since they are left moving, the incoming free particle is described by $e^{-ikx}e^{ik\delta_k^R}$.
Of course this semi-classical argument is deceptively simple, since in reality momentum is not a quantum number here and the true wave function
is a complicated superposition. However we show here that it becomes exact at large time. Thus, 
although the final result is intuitively simple, the convergence to the large time limit is nontrivial. It 
can be extracted from the exact expression for the subleading part $\delta \chi_{k,\ell}^{R/L}(x,t)$ that we provide
in (\ref{limitfunction})-- see \cite{SM,GLS}. Although we did not perform an exhaustive analysis it is easy to see
that the decay is generically algebraic in time (with possible oscillations).
\begin{figure}
    \centering
    \includegraphics[width=\linewidth]{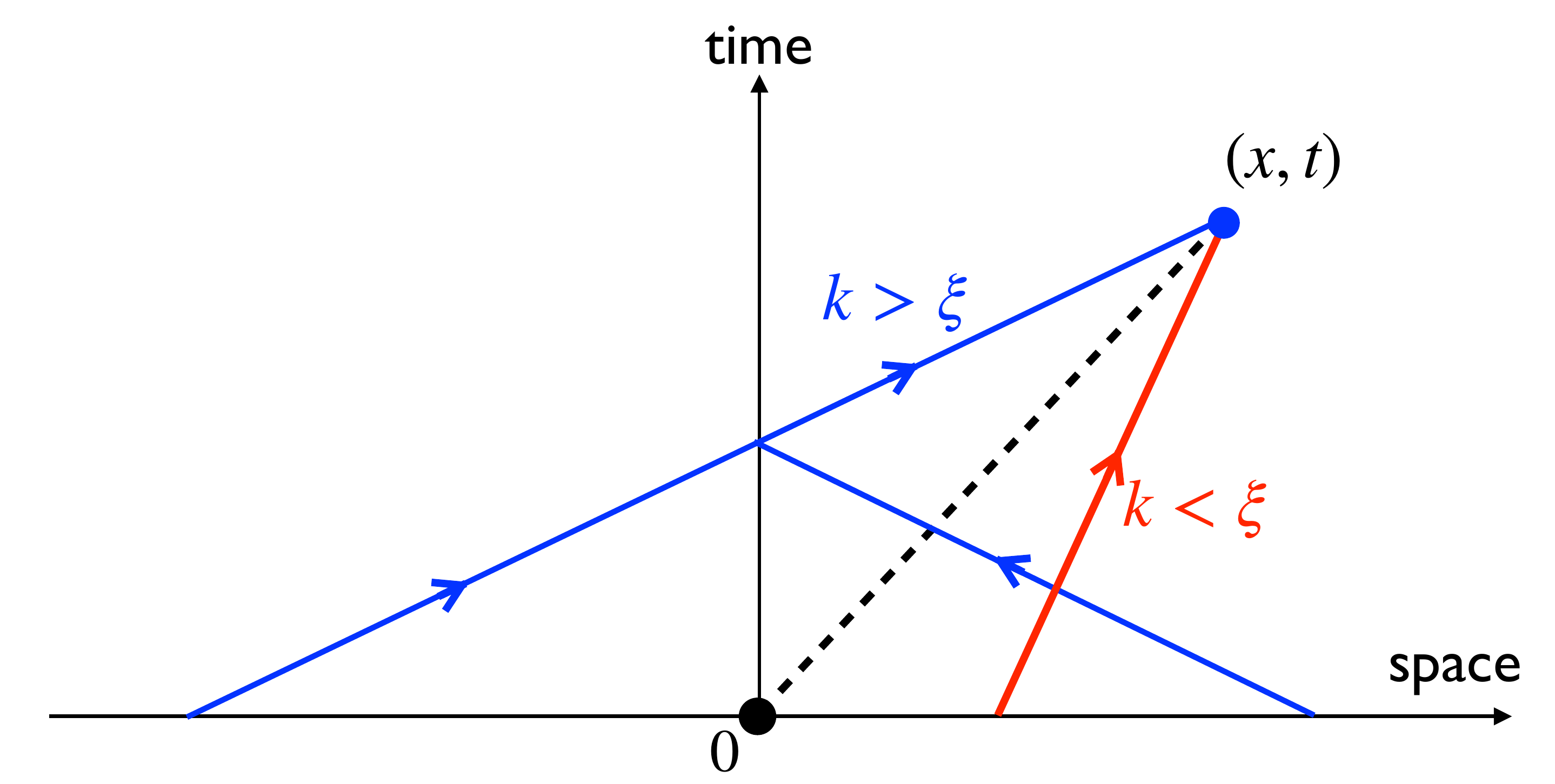}
    \caption{Interpretation of \eqref{chiR}-\eqref{chiL}, see text for details. Top left: a particle starts from the left
    with momentum $k$ and reaches $(x>0,t)$ after being transmitted (solid green), while the $-k$ component (dotted red) get lost at $-\infty$.
    Top right: starting from the right with $-k$ it reaches $(x,t)$ either directly, or after one reflection.
    Bottom left: same as top left for the ray regime, the ray $x/t=\xi$ is black solid line. Bottom right: the trajectories (in red)
    inside the black "hat" shape have $k<|\xi|$: they do not meet the impurity
    and carry the memory of the initial state \eqref{recoverinit}. The other trajectories (in blue) have $k>|\xi|$ and
    are similar to those in the top right panel leading to \eqref{chi}.
    %Sketch of the physical interpretation of the limiting single particle function in Eqs. (\ref{limitfunction})-(\ref{chiR}) in terms of scattering states through the impurity.}
    }
    \label{Fig1}
\end{figure}

%\bea\label{L\sr(k)ernel}
%&&K_R(x,t=0;x',t'=0)=\sum_{k\in\Lambda^R,k\leq k_R}\Psi_k^{R*}(x)\Psi_k^R(x')\\
%&&K_L(x,t=0;x',t'=0)=\sum_{k\in\Lambda^L,k\leq k_L}\Psi_k^{L*}(x)\Psi_k^L(x')
%\eea
%with $k_R$ and $k_L$ denoting the left and right fermi momenta
%with $k_R$ and $k_L$ denoting the left and right fermi momenta
%\bea
%k_R=\frac{2\pi N_R}{\ell},\quad k_L=\frac{2\pi N_L}{\ell}
%\eea
%This regime is obtained after the double limit $\ell\to\infty, N_{R/L}=\frac{\ell\rho_{R/L}}{2}$ followed by $t\to\infty,t'=t+\tau$  with fixed $x\text{, and } x'\not\in [-\frac{a}{2},\frac{a}{2}]$. 
%In this limit the fermion kernel reach a stationary limit.
%\bea
%K(x,t;x',t+\tau)\to K_\infty(x,x',\tau)
%\eea
The next step is to compute the space-time kernel from Eqs. \eqref{initial_kernel} and \eqref{timedepkernel}. It turns out, as we have explicitly checked by an exact independent computation, that in the limit $\ell \to \infty$ and $t \to \infty$, one can simply inject the dominant part of (\ref{limitfunction}) in the formula for the kernel (\ref{initial_kernel}) and \eqref{timedepkernel} and replacing the discrete sums in this formula by integrals (in the limit $\ell \to \infty$). This yields $\dblim K(x,t;x',t-\tau)=K_\infty (x,x',\tau)$ with
\bea \label{Ktau}
&K_\infty (x,x',\tau)=\int_{L,\tau} \frac{dk}{2\pi} e^{i\frac{k^2}{2}\tau}\chi_k^L(x)^*\chi_k^L(x') \nonumber \\ %\int_{k_L,\tau} \frac{dk}{2\pi} \ell\psi_{\infty,k}^L(x,t)^*\psi_{\infty,k}^L(x',t+\tau)\\
%&+\int_{k_R,\tau}\frac{dk}{2\pi}\ell\psi_{\infty,k}^R(x,t)^*\psi_{\infty,k}^R(x',t+\tau)\\
&+\int_{R,\tau} \frac{dk}{2\pi} e^{i\frac{k^2}{2}\tau}\chi_k^R(x)^*\chi_k^R(x') \;.
\eea
Here and below we use the shorthand notation
\bea
&\int_{R/L,\tau}\frac{dk}{2\pi}=\int_0^{\infty}\frac{dk}{2\pi}\left(f_{R/L}(k)-\theta(- \tau)\right) \;,\\
&\int_{L,R}\frac{dk}{2\pi}=\int_0^{\infty}\frac{dk}{2\pi}(f_L(k)-f_R(k))
%&\theta(\tau)=\begin{cases}
%0 \mbox{ if }\tau\leq0\\
%1 \mbox{ if }\tau>0
%\end{cases}
\eea
and we use the convention $\theta(0) = 0$. Injecting the explicit form of the function $\chi_k^{R/L}(x)$ from (\ref{chiR}) and (\ref{chiL}) in (\ref{Ktau}) we find
\bea\label{kernel_extended_NESS}
&K_\infty(x>\frac{a}{2},x'>\frac{a}{2},\tau)=\int_{L,R}\frac{dk}{2\pi}e^{i(\frac{k^2}{2}\tau-  k (x-x'))}|\st(k)|^2\nn\\
%&\int_{k_R,\tau}\frac{dk}{2\pi}e^{-i\frac{k^2}{2}\tau}\left((\sr(k)e^{ik(x+x')}+e^{ik(x-x')})+cc\right)\\
&+\int_{R,\tau}\frac{dk}{\pi}e^{i\frac{k^2}{2}\tau}\left(\cos(k(x-x'))+{\rm Re}[\sr(k)e^{ik(x+x')}]\right)\nn\\
&K_\infty(x>\frac{a}{2},x'<\frac{-a}{2},\tau)= \int_{R,\tau}\frac{dk}{2\pi}e^{i\frac{k^2}{2}\tau}\st(k)e^{ik(x-x')} \nn \\ 
&+ \int_{L,\tau}\frac{dk}{2\pi}e^{i\frac{k^2}{2}\tau}\st^*(k)e^{-ik(x-x')}\nn\\
%\int_{R,\tau}\frac{dk}{\pi}e^{i\frac{k^2}{2}\tau}{\rm Re}[\st(k)e^{ik(x-x')}]\nn\\
&+\int_{L,R}\frac{dk}{2\pi}e^{i\frac{k^2}{2}\tau}\frac{\sr(k)\st(k)^*-\sr(k)^*\st(k)}{2}e^{-ik(x+x')} \;,
\eea
together with the other regions obtained using the symmetry
$K_\infty(x,x',\tau)|_{L,R}=K_\infty(-x,-x',\tau)|_{R,L}$. Note that the initial phase shifts $e^{i k \delta^{R/L}}$
cancel in the kernel.
%\begin{figure}
%    \centering
%    \includegraphics[width=\linewidth]{Intricated-Wigner1.pdf}
%    \caption{Representation of $K^L$ at large time. The different sectors are due to free particle coming form the left, particle transmitted through the defect and particle reflected by the defect. {\blue maybe add a defect size interval?}}
%    \label{fig:NESS_kernel}
%\end{figure}
As expected $K_\infty(x,x',\tau)$ vanishes at large $|\tau|$, with algebraic decay see \cite{SM}. In particular, from (\ref{kernel_extended_NESS}), one obtains the density $\rho_\infty(x) = K_\infty(x,x,0)$ in the NESS, which reads, for $|x|>a/2$
\bea\label{density_NESS}
&\rho_\infty(x)=\int_{L,R}\frac{dk}{2\pi}|\st(k)|^2+\int_{0}^{\infty}\frac{dk}{\pi}f_R(k)\left(1+{\rm Re}[\sr(k)e^{i2kx}]\right)\;. \nn 
\eea
Similarly, from (\ref{kernel_extended_NESS}) one also obtains the current in the NESS $J_\infty = \frac{1}{2i}(\partial_{x'} - \partial_x) K_{\infty}(x,x',0)|_{x'=x}$, which yields
\bea\label{current_NESS}
J_\infty =\int_{L,R}\frac{dk}{2\pi}k|\st(k)|^2 \;.
\eea
For $f_L(k)=f_R(k)$ the first and fourth line in \eqref{kernel_extended_NESS} vanish and one can check that one recovers the thermal equilibrium (in the absence of bound states).

{\it Large time limit: the ray regime}. We also computed the asymptotic kernel at large time when distances are scaled with time $x=O(t)$, i.e.,
setting $x = \xi t + y$ with $\xi, y=O(1)$. Again we first obtain the asymptotic form of the wave function \cite{SM}
\be\label{limitfunctionxi}
\psi_k^{R/L}(x=\xi t + y ,t) \underset{\substack{\ell\to\infty\\t\to\infty}}{\simeq} \frac{1}{\sqrt{\ell}} e^{-i\frac{k^2}{2}t}\chi_{\xi,k}^{R/L}(x=\xi t + y) \;,
\ee
where we have defined
% \bea
% &\chi_{\xi,k}^L(x)=\begin{cases}
% \theta(k>\xi)\st(k) e^{ikx}e^{ik\delta_k^{L}} \hspace{3mm} \mbox{ if } \xi>0\\
% \theta(k<-\xi)e^{-ikx}e^{-ik\delta_k^{L}} \hspace{2mm} \mbox{ if } \xi<0\\
% +(e^{ikx}+\theta(k>-\xi)\sr(k) e^{-ikx})  e^{ik\delta_k^{L}}
% \end{cases}\\
% &\chi_{\xi,k}^R(x)=\begin{cases}
% (e^{-ikx}+\theta(k>\xi)\sr(k) e^{ikx})e^{ik\delta_k^{R}}\\
% +\theta(k<\xi)e^{ikx}e^{-ik\delta_k^{R}} \hspace{8mm} \mbox{ if } \xi>0\\
% \theta(k>-\xi)\st(k) e^{-ikx} e^{ik\delta_k^{R}} \hspace{2mm} \mbox{ if } \xi<0
% \end{cases}
% \eea
\bea \label{chiRray} 
&\chi_{\xi,k}^R(x)=\begin{cases}
\theta(k>-\xi)\st(k) e^{-ikx} e^{ik\delta_k^{R}} \hspace{2mm} \mbox{ if } \xi<0\\
\theta(k<\xi)e^{ikx}e^{-ik\delta_k^{R}} \hspace{11mm} \mbox{ if } \xi>0\\
+(e^{-ikx}+\theta(k>\xi)\sr(k) e^{ikx})e^{ik\delta_k^{R}}
\end{cases}\\
&\chi_{\xi,k}^L(x)=\begin{cases} \label{chiLray}
\theta(k>\xi)\st(k) e^{ikx}e^{ik\delta_k^{L}} \hspace{3mm} \mbox{ if } \xi>0\\
\theta(k<-\xi)e^{-ikx}e^{-ik\delta_k^{L}} \hspace{2mm} \mbox{ if } \xi<0\\
+(e^{ikx}+\theta(k>-\xi)\sr(k) e^{-ikx})  e^{ik\delta_k^{L}} \;.
\end{cases} 
\eea
Since $x=\xi t + y$ %the asymptotic wavefunction 
\eqref{limitfunctionxi} exhibits fast oscillations
in time $\propto e^{ - i \frac{k^2}{2} t \pm i \xi k t}$. The forms \eqref{chiRray} and
\eqref{chiLray} can be understood by an extension to the ray regime of the argument given in the NESS, see
Fig. \ref{Fig1} (bottom). It can also be summarized by considering
the "light cone" with slopes $\pm k$ originating from the impurity, see Fig. \ref{lightconefunction}.
Outside of it, i.e., for $|\xi| > k$,  $\psi^{R/L}(x,t)$ recovers the initial condition up to a time propagation phase $e^{-i\frac{k^2}{2}t}$, i.e.,
\be \label{recoverinit} 
\frac{1}{\sqrt{\ell}}\chi_{\xi,k}^{R/L}(x)=\phi_k^{R/L}(x) \quad \text{for} \quad |\xi|>k \;.
\ee
Inside the cone, i.e., for $|\xi| < k$, $\psi^{R/L}(x,t)$ is given by the extrapolation to the ray regime
(with $x=\xi t+y$) of the form obtained above in the NESS (for $x=O(1)$), in Eqs. \eqref{chiRray} and \eqref{chiLray},
i.e.,
\be \label{chi} 
\chi_{\xi,k}^{R/L}(x)=\chi_{k}^{R/L}(x) \quad, \quad \text{for} \quad |\xi|<k \;.
\ee

\begin{figure}
    \centering
    \includegraphics[width=0.8\linewidth]{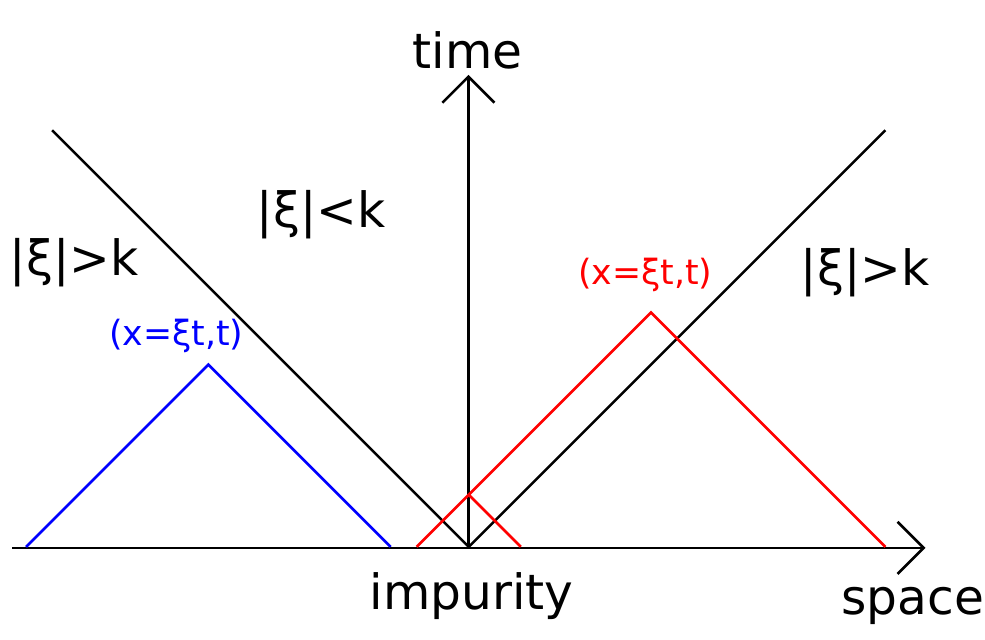}
    \caption{For a given initial momentum $k$, the space-time is divided in two regions (at large time). If $|\xi|>k$ the initial condition is unaffected by the impurity. In blue: the two trajectories arriving at $(x,t)$ with phase factors $e^{\pm ik(x-\delta_k^{L})}$, explaining \eqref{recoverinit}. However for $|\xi|<k$ the possible trajectories (red) are affected by the impurity, giving back the bottom part of Fig. \ref{Fig1}.}
    \label{lightconefunction}
\end{figure}
We now compute the kernel in the ray regime from Eqs. \eqref{initial_kernel} and \eqref{timedepkernel}. Again, it turns out \cite{SM} that in the limit $\ell \to \infty$ followed by $t \to \infty$ with $x,x'=O(t)$, one can simply inject the asymptotic forms (\ref{limitfunctionxi}) in the formula for the kernel (\ref{initial_kernel}) as was done in \eqref{timedepkernel}, and replace the discrete sums by integrals 
%(in the limit $\ell \to \infty$). 
We thus use the same procedure that brought us from \eqref{timedepkernel} to \eqref{Ktau}, but here using $\chi_{k,\xi}^{R/L}$ instead of $\chi_k^{R/L}$.
% \bea\label{Kxi}
% &  K(x,t;x',t+\tau)\underset{\substack{\ell\to\infty\\t\to\infty\\x/t\to\xi\\ x'/t\to\xi'}}{\simeq}\int_{L,\tau} \frac{dk}{2\pi} e^{-i\frac{k^2}{2}\tau}\chi_{\xi,k}^L(x)^*\chi_{\xi',k}^L(x')\nn\\
% &+\int_{R,\tau} \frac{dk}{2\pi} e^{-i\frac{k^2}{2}\tau}\chi_{\xi,k}^R(x)^*\chi_{\xi',k}^R(x')
% \eea
When scaling $x=\xi t+y$, and $x'=\xi' t+y'$, because of oscillatory behaviors of $\chi_{k,\xi}^{R/L}$, the kernel has a non zero limit only if $\xi=\pm\xi'$.
For these two cases, we obtain \cite{footnotetau}
\be
  \underset{\substack{\ell\to\infty\\t\to\infty}}{\lim} K(\xi t+y,t;\xi' t+y',t-\tau)=\begin{cases}
0\mbox{ if } \xi\neq\pm\xi'\\
K_\xi^+(y,y',\tau) , \xi'=\xi\\
K_\xi^-(y,y',\tau) , \xi'=-\xi
\end{cases} 
\ee
with the following explicit expressions 
\bea 
&K_{\xi>0}^+(y,y',\tau) = \int_{R,\tau}\frac{dk}{\pi} e^{i\frac{k^2}{2}\tau}  \cos(k(y-y'))\nn\\
&+\int_{L,R} \frac{dk}{2\pi} e^{i\frac{k^2}{2}\tau} |\st(k)|^2 e^{-ik(y-y')} \theta(k-|\xi|) \label{Kxi1} \\
&K_{\xi>0}^-(y,y',\tau)=\int_{L,R} \frac{dk}{2\pi} e^{i\frac{k^2}{2}\tau}\sr(k) \st(k)^* e^{-ik(y+y')} \theta(k-|\xi|)\nn
\eea
and similar formula for $\xi<0$ obtained by symmetry, see \eqref{symray} in \cite{SM}. 
{\blue }One can check that the ray regime matches the large distance behavior of the NESS
in the following sense
\be \label{K_ray}
\lim\limits_{\substack{y,y'\to\infty\\y\mp y'=O(1)}}K_{\infty}(y,y',\tau)=\lim\limits_{\xi\to0^+}K_\xi^\pm(y,y',\tau) \;.
\ee
}
%&\lim\limits_{\substack{x\to\infty\\x'\to-\infty\\x+x'=O(1)}}K_{\infty}(x,x',\tau)=\lim\limits_{\xi\to0^+}K^-(x,x',\tau)
%\eea
%
Interestingly, while the above formula for $K^+$ can also be obtained from a semi-classical argument based on the Wigner function \cite{Prosen2018, GLS},
this is not the case for $K^-$. Instead, in the present framework, the correlations between $(-x,t)$ and $(x,t)$ 
arise naturally from trajectories which start from the same point (e.g. from the right with momentum $-k$) and are either reflected or transmitted.
In the kernel, the interferences between these two trajectories result from the product of the first and the third line in \eqref{chiRray} when
computing $\chi^R_{-\xi,k}(x,t)^* \chi^R_{\xi,k}(x,t)$ [see \eqref{Ktau}].

%The symmetry $(\xi,\xi',y,y',L,R) \to (-\xi,-\xi',-y,-y',R,L)$ allows to obtain the other values.

%Note: taking $x'=\xi'(t+\tau)+y'$ amounts to taking  $x'=\xi't+\tilde{y'}$ with $\tilde{y'}=\xi'\tau+y'$
%\begin{figure}
%    \centering
%    \includegraphics[width=\linewidth]{Intricated-Wigner3.pdf}
%    \caption{Representation of $K^L$ at large time when $x$ is scaled with $t$.}
%    \label{lightconekernel}
%\end{figure}
At large time the mean density along rays thus converges $\rho(\frac{\xi}{t},t) \to \tilde \rho(\xi)$ given by
\bea
\tilde \rho(\xi>0)= \rho_{R} + \int_{L,R} \frac{dk}{2\pi}  |\st(k)|^2  \theta(k-|\xi|) \;,
\eea
and the same for $\xi<0$ exchanging $R$ and $L$. Here $\rho_{R/L}=\int_{0}^\infty\frac{dk}{\pi} f_{R/L}(k)$ are the initial mean densities. 
The function $\tilde \rho(\xi)$ exhibits a jump discontinuity at $\xi=0$, $\tilde{\rho}(0^+)-\tilde{\rho}(0^-)=\int_{L,R} \frac{dk}{\pi}  |\st(k)|^2$.
Similarly the current along rays converges to $J(\frac{\xi}{t},t) \to \tilde J(\xi)$ with 
\be
\tilde J(\xi)=\int_{L,R}\frac{dk}{2\pi}k |\st(k)|^2\theta(k-|\xi|) \;.
\ee

{\it Bound states}. Let us now discuss the NESS regime when $V(x)$ admits a sequence of bound states $\phi_{\kappa}(x)$, $\kappa \in \Lambda_b$
of energies $- \frac{\kappa^2}{2}$. In that case the asymptotic large time kernel 
$K=K_s+K_b$ is the sum of two pieces: (i) one due to scattering states, $K_s$, identical
to the one obtained above (ii) one due to bound states, $K_b=K_b^R+K_b^L$, where 
\bea 
&& K_b^{R/L}(x,t;x',t') \\
&& = \sum_{ \kappa',\kappa'' \in \Lambda_b } \phi_{\kappa'}(x)  \phi_{\kappa''}(x') e^{-i(\frac{\kappa'^2}{2}t-\frac{\kappa''^2}{2}t')} C_{\kappa',\kappa''}^{R/L} \;,
\nonumber 
\eea 
with $C_{\kappa',\kappa''}^{R/L} = \langle \phi_{\kappa'} | (1 + e^{ \beta_{R/L} (\hat H_0^{R/L} - \mu_{R/L} )})^{-1}  | \phi_{\kappa''} \rangle$ \cite{SM}.
%\be 
%C_{\kappa',\kappa''}^{R/L} = \langle \phi_{\kappa'} | \frac{1}{1 + e^{ \beta_{R/L} (\hat H_0^{R/L} - \mu_{R/L}) } } | \phi_{\kappa''} \rangle
%\ee 
When there are at least two bound states the total NESS kernel exhibits permanent oscillations in time. These oscillations also
occur in the density and current, see \cite{SM}. Note that some information about the
form of the initial wave functions remains relevant (and whether $V_0$ has or not bound states itself), while
in the absence of bound states of $V(x)$, only the scattering coefficients remains relevant at large time. Finally, bound states do not contribute to the kernel in the ray regime since their wave-functions decrease exponentially at large $|x|$. For related results in the presence of bound states
%on oscillations and memory
see \cite{Rossi2021,Prosen2018,Gamayun1}. 
 
\begin{figure}[t]
\includegraphics[scale=0.6]{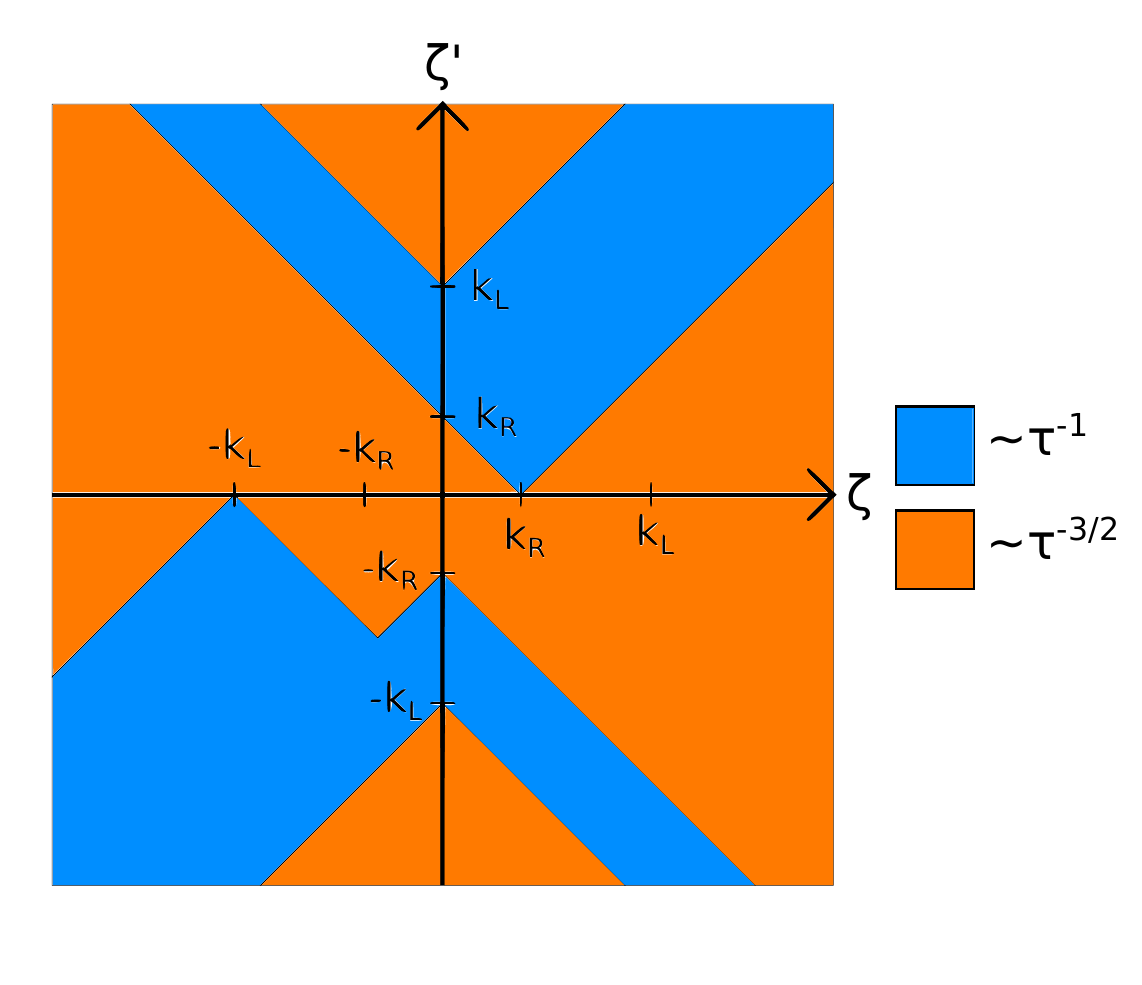}
\caption{The density-density correlation in the NESS (\ref{density_correl}) exhibits different power law regimes for large $\tau=t'-t$ and $x=\zeta \tau ,\quad x'=\zeta' \tau$ (the figure is plotted for $k_R<k_L$).}\label{phasediagram}
\end{figure}

We now study the connected correlation function of the density
at two distinct space-time points in the NESS. 
It is obtained from the kernel in the NESS using \eqref{multitimecorr1} and \eqref{multitimecorr2} as
\be \label{density_correl}
\lim_{t \to +\infty} \langle \hat \rho(x',t + \tau) \hat \rho(x,t) \rangle^c =
 - K_\infty(x,x';-\tau) K_\infty(x',x,\tau) \;.
\ee
In the large $\tau$ limit the behavior of this correlation depends on the ratios $\zeta=\frac{x}{\tau}$
and $\zeta'=\frac{x'}{\tau}$. As shown in Fig. \ref{phasediagram} there are several sectors in the $(\zeta,\zeta')$
plane, where the decay of the correlation at large time is of the form $\sim \tau^{-\alpha} C(\zeta,\zeta')$ where
$\alpha$ depends on the sector, up to oscillating and diffusive factors respectively of the form
$e^{\pm i k_{R/L} (x \pm x')}$ and $e^{- \frac{(x-x')^2}{\tau}}$ (see \cite{SM} for details). As compared to the equilibrium case
in the absence of a defect for which $\alpha = 3/2$ for all $(\zeta, \zeta')$ \cite{us_eq_dyn}, the stationary temporal correlations in the NESS in the presence of an impurity exhibits a richer behavior, in particular regions with a slower decay. In addition, we have shown that the
response function reaches a stationary limit $\lim_{t \to +\infty} \frac{\delta \langle \hat \rho(x,t) \rangle}{\delta f(x',t-\tau)}|_{f=0}$
which we have expressed in terms of the kernel $K_\infty(x,x',\tau)$ \cite{SM}.

% we have also computed the large tim 
% {\red Interestingly, the kernel allows to recover the response function $\chi(x,t;x',t')$ to an infinitesimal perturbation of the evolution Hamiltonian of the form
% $\delta H(t) =\int dx \hat \rho(x,t) f(x,t)$, defined as $\delta \langle \hat \rho(x,t) \rangle= \int dx' dt' f(x',t') \chi(x,t;x',t')$. In the large time limit with fixed $\tau$ it reaches a stationary limit $\chi_\infty(x,x',\tau)$ which depends on the large time kernel $K_\infty(x,x',\tau)$ see \cite{SM}.}

Finally we have computed the Wigner function $W(x,p)$ in the NESS. It recovers the semi-classical expression for $|x| \to +\infty$.
However for $x\simeq O(1)$ it exhibits quantum properties such as a $\delta(p)$ peak, which is
a signature of the finite limit of the correlations along opposite rays, $K_\infty(x,-x)$ for $|x| \to +\infty$ which relate to the ray regime $K_\xi^-(y,y',\tau)$ see \eqref{K_ray}.

In conclusion, we have obtained {in Eqs. \eqref{limitfunction} and \eqref{limitfunctionxi} the limiting forms of the time-evolved initial single particle} wave-functions at large time after a quantum quench in the presence of an arbitrary impurity. Here we have presented the case of a symmetric impurity $V(x)=V(-x)$,
the extension to the general case being presented in \cite{SM}. 
{These} limiting wave-functions allow to obtain elegant expressions for the space-time correlation kernel {of non interacting fermions} in terms of the scattering coefficients, both in the
NESS and in the ray regime. They incorporate the quantum interference effects and allow to overcome the shortcomings of the semi-classical
approach. These correlation kernels allow in principle calculation of the full counting statistics
for arbitrary intervals (see \cite{Gamayun1,Gamayun22} in the semi-infinite case). 
It would be interesting to realize such partitioning protocol in cold Fermi gas transport experiments \cite{krinner}.
In bosonic cold atoms the relaxation after a quantum quench has been observed
\cite{boson_splitting, boson_splitting2}, including in the noninteracting limit~\cite{quantum_walk,Olshanii}.
In fact, our results for the wave-functions at large time immediately 
extend to non-interacting bosons. This opens the way to the study of
the correlations in the bosonic NESS in the presence of an impurity, with possible
extensions to the interacting case~\cite{Sotiriadis}. {Finally, it is possible to
extend our $1d$ calculation to other geometries, such as tubes, $2d$ sheets.}

\medskip

\let\oldaddcontentsline\addcontentsline% Store \addcontentsline
\renewcommand{\addcontentsline}[3]{}% Make \addcontentsline a no-op
\begin{acknowledgements}
\noindent{\it Acknowledgements:} We thank J. Dalibard, A. De Luca, F. H. L. Essler, O. Gamayun for useful discussions. This research was supported by ANR grant ANR-17-CE30-0027-01 RaMaTraF.    
\end{acknowledgements}
\let\addcontentsline\oldaddcontentsline% Restore \addcontentsline

% {\red check what happens for (i) bound states (ii) bound state in the initial condition 
% (iii) cylindre ou plaque.. }

\let\oldaddcontentsline\addcontentsline% Store \addcontentsline
\renewcommand{\addcontentsline}[3]{}% Make \addcontentsline a no-op

\let\addcontentsline\oldaddcontentsline% Restore \addcontentsline

\bibliographystyle{iopart-num}

\newpage

\begin{widetext}

\setcounter{secnumdepth}{2}

\begin{large}
\begin{center}

Supplementary Material for\\  {\it Stationary time correlations for fermions after a quench in the presence of an impurity}

\end{center}
\end{large}

\bigskip

\tableofcontents

\section{Scattering matrix and eigenstates}

\subsection{Eigenstates of the evolution Hamiltonian $\hat H$}
We consider $N$ noninteracting fermions in one dimension evolving through a defect, described by the single particle Hamiltonian
\be  \label{def_H}
\hat H=-\frac{1}{2}\partial_x^2+V(x) \;.
\ee 
Here, $V(x)$ is a potential without bound state, and localized in the region $[-\frac{a}{2},\frac{a}{2}]$. 
For convenience we add two infinite walls at $x=\pm\ell/2$, 
with $\ell > a$, so that the wave function vanishes outside $[-\ell/2,\ell/2]$. 

We denote $\phi_k(x)$ an eigenfunction of $\hat H$ with eigenenergy $\frac{k^2}{2}$. Outside the defect, i.e., for $|x|>a/2$, it is a plane wave with momentum $\pm k$
\bea
\label{def_phi}
\phi_k(x)=(A_k e^{ikx}+B_k e^{-ikx})\theta(x<-\frac{a}{2})+(C_k e^{ikx}+D_k e^{-ikx})\theta(x>\frac{a}{2}) \;. 
\eea
We will not need below its detailed behavior inside the defect.
The amplitudes of each terms are related through
\bea\label{Srelation}
\left( {\begin{array}{cc}
    C_k  \\
    B_k  \\
  \end{array} } \right)={\cal S}_k\left( {\begin{array}{cc}
    A_k  \\
    D_k  \\
  \end{array} } \right) \;.
\eea
Here ${\cal S}_k$ is the scattering matrix which characterizes the potential 
\be \label{def_S}
{\cal S}_k=  \left( {\begin{array}{cc}
    \st_L(k) & \sr_R(k) \\
    \sr_L(k) & \st_R(k) \\
  \end{array} } \right)
\ee
where $\sr_{R/L}(k)$ and $\st_{R/L}(k)$ are the reflection and transmission coefficients for waves coming either from the right or from the left.
Note that the scattering matrix ${\cal S}_k$ and the scattering coefficients are defined for any real $k$ and
depend only on $V(x)$ for $x \in [-a,a]$: they are independent of the boundary conditions chosen at $x=\pm \ell/2$,
which only restricts the possible values of $k$ and fixes some amplitude relations (see below).
The scattering coefficients satisfy some general relations (independently of the boundary conditions) which we now discuss.
Let us denote $W(f,g)(x)=f(x) g'(x) - f'(x) g(x)$. When $f,g$ are two eigenfunctions of $\hat H$ with the same energy $W(f,g)(x)$ is called the Wronskian and is independent of $x$. Since the potential is real,
if $\phi_k(x)$ is eigenfunction, then $\phi^*_k(x)$ is also an eigenfunction with the same value of the energy. 
Hence the probability current $J(x) = \frac{1}{2 i} W(\phi_k^*,\phi_k)(x)$ is independent of $x$, i.e.,  
one has current conservation. Here this implies that $|A_k|^2-|B_k|^2=|C_k|^2 - |D_k|^2$, i.e.,  that ${\cal S}_k$ is unitary, ${\cal S}^\dagger_k {\cal S}_k=I$,
which leads to the constraints
\be \label{unitarity} 
|\sr_{R}(k)|^2 + |\st_{R}(k)|^2 =1 \quad , \quad |\sr_{L}(k)|^2 + |\st_{L}(k)|^2 =1 \quad , \quad \sr_R(k) \st_L(k)^* + \sr_L(k)^* \st_R(k) = 0 \;.
\ee
Since one has also ${\cal S}_k {\cal S}^\dagger_k=I$ it also implies $|\sr_L(k)|=|\sr_R(k)|$, $|\st_L(k)|=|\st_R(k)|$
and $\sr_L(k)^* \st_L(k) + \sr_R(k) \st_R(k)^* = 0$. Now, since $\phi^*_k(x)$ is also an eigenfunction,
it means that \eqref{Srelation} also holds upon exchanging $(A,B) \to (B^*,A^*)$ and $(C,D) \to (D^*,C^*)$. This 
implies that ${\cal S}_k^* = \Omega {\cal S}_k^{-1} \Omega$ where $\Omega=\left( {\begin{array}{cc}
    0 & 1 \\
    1 & 0 \\
  \end{array} } \right)$. Using unitarity it also implies ${\cal S}_k^* = \Omega {\cal S}_k^\dagger \Omega$
  and finally ${\cal S}_k = \Omega {\cal S}_k^T \Omega$. This in turn implies
  \be 
 \st_L(k) =  \st_R(k) := \st (k)
  \ee 
which is thus a consequence of time reversal invariance. A natural parameterization of the scattering amplitudes is thus
\be  \label{param} 
\sr_L(k) = e^{i \varphi_L(k)} \cos \theta(k) \quad , \quad \sr_R(k) = e^{i \varphi_R(k)} \cos \theta(k) 
\quad , \quad \st(k) = i e^{\frac{i}{2} (\varphi_R(k)+ \varphi_L(k))} \sin \theta(k)
\ee 

Although it is sufficient to consider $k \in \mathbb{R}^+$ it is instructive to consider the change $k \to -k$. 
One obtains that ${\cal S}_{-k} = \Omega {\cal S}^{-1}_k \Omega = \Omega {\cal S}^\dagger_k \Omega$.
This implies that the scattering coefficients obey $\st(-k)=\st(k)^*$, $\sr_R(-k)=\sr_R(k)^*$ and $\sr_L(-k)=\sr_L(k)^*$,
i.e., the functions $\varphi_{L,R}(k)$ in \eqref{param} are odd in $k$ and $\theta(k)$ is even.

% {\red  pierre tell me about this other way
% \bea
% &i\partial_t \psi_k = \hat H \psi\\
% &\psi_k(x,t)=\phi_k(x)e^{-iE_k t}\\
% \eea
% if
% \be
% \phi_k(x)=A_1 e^{ikx} + A_2 e^{-ikx}
% \ee
% then
% \be
% \psi_k(x,t)=A_1 e^{ik(x-c t)} + A_2 e^{-ik(x+c t)}
% \ee
% with $c=\frac{E_k}{k}$ the part $A_1$ propagates to the right and the part $A_2$ propagates to the left, form this we deduce \eqref{Srelation} (35 et non pas 189) and \eqref{def_S} which is just a link between incoming and outcoming flux.\\
% Now we take the complexe conjugate

% \bea
% &-i\partial_t \psi_k^* = \hat H \psi^*\\
% &\psi_k^*(x,t)=\phi_k^*(x)e^{iE_k t}\\
% \eea
% Because $\hat H^*=\hat H$\\
% if
% \be
% \phi_k^*(x)=A_1^* e^{-ikx} + A_2^* e^{ikx}
% \ee
% then
% \be
% \psi_k^*(x,t)=A_1^* e^{-ik(x-c t)} + A_2^* e^{ik(x+c t)}
% \ee
% We see that $A_1^*$ propagates to the right and the part $A_2^*$ propagates to the left. We conclude that 
% \bea
% \left( {\begin{array}{cc}
%     C_k^*  \\
%     B_k^*  \\
%   \end{array} } \right)={\cal S}_k\left( {\begin{array}{cc}
%     A_k^*  \\
%     D_k^*  \\
%   \end{array} } \right) \;.
% \eea
% so $S=S^*$??

% Or maybe I'm doing something wrong and (because $S$ is related to propagation) we rather have this
% \bea
% \left( {\begin{array}{cc}
%     C_k^*  \\
%     B_k^*  \\
%   \end{array} } \right)={\cal S}_k^*\left( {\begin{array}{cc}
%     A_k^*  \\
%     D_k^*  \\
%   \end{array} } \right) \;.
% \eea
% which has no contradiction
% }

From now on we restrict ourselves to the case of a symmetric potential $V(x)=V(-x)$. In that case an orthonormal eigenbasis of
$\hat H$ can be chosen in terms of eigenfunctions denoted $\phi_{\sigma,k}(x)$ 
which are even (with $\sigma=+1$) or odd (with $\sigma=-1$) in $x$, which
in \eqref{def_phi} correspond to 
\be \label{Sparity} 
C_k=\sigma B_k \quad , \quad 
D_k=\sigma A_k
\ee
One can check using \eqref{Srelation} and \eqref{def_S} that it implies the additional symmetry of the scattering coefficients
$\sr_{R}(k)=\sr_{L}(k)=\sr(k)$. Since we showed that one also has $\st_{R}(k)=\st_{L}(k)=\st(k)$, this 
recovers the expression \eqref{def_S_intro} for the scattering matrix in the main text. Inserting \eqref{Sparity}
in \eqref{def_phi} and using \eqref{Srelation} and \eqref{def_S} we see that the even and odd eigenfunctions 
must have amplitude ratios
\be  \label{ampratio} 
\frac{C_k}{D_k} = \frac{B_k}{A_k} = \sr(k) +\sigma \st(k)
\ee 
which is a complex number of modulus unity, thanks to the relations \eqref{unitarity}, i.e.,
$|\sr(k)|^2 + |\st(k)|^2 =1$ and $\sr(k) \st(k)^* + \sr(k)^* \st(k) = 0$. Note that if
the problem is considered on the whole axis $x \in \mathbb{R}$ the odd and even eigenfunctions have the same energy.

Here, since the boundary conditions at $x=\pm \ell/2$ are also even in $x$, the eigenfunctions are indeed even or odd,
but for finite $\ell$ they have different energies. Since furthermore we will use the
hard wall condition at $x= \pm \ell/2$, the wave-vector $k$ belongs necessarily to one of the two lattices $\Lambda^+$ or 
$\Lambda^-$ for even and odd wave-functions respectively, defined by 
\bea \label{quantification}
\Lambda^\sigma=\{k \in \mathbb{R}^+|\phi_{\sigma,k}( \frac{\ell}{2})=0\} \;.
\eea
Because of \eqref{ampratio} the odd and even eigenfunctions of $\hat H$ can be chosen real 
and parameterized for $|x|>a/2$ as 
\bea\label{phi}
&\phi_{-,k}(x)={\rm sgn}(x)c_{\ell,-,k}\cos(k(|x|-\delta^{-}_k))\quad k\in\Lambda^-\\
&\phi_{+,k}(x)=c_{\ell,+,k}\cos(k(|x|-\delta^{+}_k))\quad k\in\Lambda^+\nn
\eea
This amounts to choose $C_k=D_k^* \propto e^{- i k \delta^{\sigma}_k}$ where the 
phase shifts $k \delta^{\pm}_k$ are related to the scattering coefficients as 
\be \label{rtodelta} 
\sr(k)+\st(k)=e^{-2ik\delta_k^+} \quad , \quad \sr(k)-\st(k)=e^{-2ik\delta_k^-} \;.
\ee
These eigenfunctions are normalized on $[-\frac{\ell}{2},\frac{\ell}{2}]$. The 
normalization prefactor $c_{\ell,\sigma,k}$ becomes $k$-independent and $\sigma$-independent at
large $\ell$ with 
\be 
c_{\ell,\sigma,k} \simeq \sqrt{\frac{2}{\ell}}
\ee
Indeed denoting $\phi_{\pm,k}(x) = c_{\ell,\pm,k} \tilde \phi_{\pm,k}(x)$ the norm is 
\be  \label{41} 
c_{\ell,\pm,k}^2 \left( \int dx |\tilde \phi_{\pm,k}(x)|^2 \theta(|x| < \frac{a}{2}) + \int_{-\ell/2}^{\ell/2} 
dx  \cos^2(k(|x|-\delta^{\pm}_k)) \theta(|x| > \frac{a}{2}) \right) = 1
\ee 
The second integral behaves as $\simeq \frac{\ell}{2} + O(a,1)$ at large $\ell$. It is natural to
assume that the solution $|\tilde \phi_{\pm,k}(x)|^2$ of the Schr\"odinger equation which behaves as a cosine outside the impurity
region is $O(1)$ inside the region, hence the first integral is $O(a)$. 

% Because of the
% hard wall condition at $x= \pm \ell/2$ the wavevector $k$ belongs necessarily to one of the two lattices $\Lambda^+$ or 
% $\Lambda^-$ for even and odd wavefunctions respectively, defined by 
% \bea \label{quantification}
% \Lambda^\sigma=\{k \in \mathbb{R}^+|\phi_{\sigma,k}( \frac{\ell}{2})=0\} \;.
% \eea
% In that case the wave functions can be chosen real with $D_k=C_k^* \propto e^{i k \delta^{\sigma}_k}$.
% The phase shifts $k \delta^{\pm}_k$ are related to the scattering coefficients as follows.
% Injecting \eqref{phi} in \eqref{Srelation} and \eqref{def_phi}
% and using the relations \eqref{Sparity},
% one obtains the two relations for $k \delta^{\pm}_k$ 
% \bea\label{deltarelation}
% &&e^{2ik\delta^{\sigma}_k}=\frac{D_k}{C_k}=\frac{1}{\sr(k) + \sigma \st(k)} = \sr(k)^*+\sigma \st(k)^*
% \eea
% since $\sr(k)+\sigma \st(k)$ is of modulus one, thanks to the relations \eqref{unitarity}, i.e.
% $|\sr(k)|^2 + |\st(k)|^2 =1$ and $\sr(k) \st(k)^* + \sr(k)^* \st(k) = 0$. More explicitly one has 
% \be \label{rtodelta} 
% \sr(k)+\st(k)=e^{-2ik\delta_k^+} \quad , \quad \sr(k)-\st(k)=e^{-2ik\delta_k^-} \;.
% \ee

%\bea
%&r=e^{-ik(\delta_k^++\delta_k^-)}\cos(k(\delta_k^--\delta_k^+))\\
%&t=ie^{-ik(\delta_k^++\delta_k^-)}\sin(k(\delta_k^--\delta_k^+))
%\eea
Note that in the absence of impurity, $V(x)=0$, one has $\delta_k^+=0$ and $\delta_k^-=\frac{\pi}{2}$,
i.e., $\sr(k)=0$ and $\st(k)=1$.

\subsection{Eigenstates of $\hat H_0$}

The analysis of the eigenstates and the scattering matrix associated with the initial
potential $V_0(x)$ is quite similar to the previous section. There are some differences however
as we now summarize. First $V_0(x)$ is divergent around $x=0$ so that the system factorizes into independent two half space
systems, hence one has $\st_0^R(k)=\st_0^L(k)=0$. Note that we do not assume that $V_0(x)$ is even in $x$.

Another difference is that we use the basis where the eigenfunctions are neither even nor odd but
vanish on one side with $\phi_k^{R}(x)=0$ for $x<0$ and $\phi_k^{L}(x)=0$ for $x>0$. Outside
the interval $[-\frac{a}{2},\frac{a}{2}]$ they read
\bea \label{phiRL} 
&\phi_k^{R}(x)=c_{\ell,R,k}^0 \cos(k(|x|-\delta_k^{R}))\theta(x-\frac{a}{2}),\quad k\in\Lambda^R\\
&\phi_k^{L}(x)=c_{\ell,L,k}^0 \cos(k(|x|-\delta_k^{L}))\theta(-\frac{a}{2}-x),\quad k\in\Lambda^L
\eea
where the normalizing constants $c_{\ell,R/L,k}^0$ become independent of $k$ and $L/R$ in the large 
$\ell$ limit, with 
\be 
c^0_{\ell,R/L,k} {\simeq}\sqrt{\frac{4}{\ell}}
\ee
by a similar argument as in the previous section, see Eq. (\ref{41}), remembering that the wave-functions $\phi_k^{R/L}(x)$ vanish on $\mathbb{R}^{\mp}$. The wave-vectors $k$ belong
to the lattices $\Lambda^{R/L}$ respectively, which are defined in \eqref{quantification_initial} in the text. 
The relation between the phase shifts and the reflection coefficients can be obtained as 
\bea \label{phaseshiftH00} 
\sr_0^{R/L}(k)=e^{-2ik\delta_k^{R/L}} \;.
\eea

\subsection{Some examples of impurity potentials and their scattering coefficients}

It is useful to give a few examples of potentials for which the scattering coefficients are known %from \cite{Griffiths} {see  \cite{doubledelta}}
\\

{\bf The delta barrier} : in the case $V(x)=g\delta(x)$ one has
\be
\st(k)=\frac{k}{k+ig}
%=\frac{1}{1+i\beta} 
\quad , \quad \sr(k)=\frac{-ig}{k+ig}
%=\frac{-i\beta}{1+i\beta} 
%\quad \beta=\frac{g}{k}
\ee
\\

{\bf The double delta barrier}: in the case $V(x)=g(\delta(x-\frac{a}{2})+\delta(x+\frac{a}{2}))$ \cite{doubledelta} one has
%\be
%{\st}(k)=\frac{4k^2}{g^2(e^{2ika}-1)+4k^2+4ikg}\\
%\quad , \quad 
%{\sr}(k)=-ig\frac{e^{ika}(2k-ig)+e^{-ika}(2k+ig)}{g^2(e^{2ika}-1)+4k^2+4ikg}
%\ee
\be
{\st}(k)=\frac{k^2}{g^2(e^{2ika}-1)+k^2+2ikg}\\
\quad , \quad 
{\sr}(k)=-ig\frac{e^{ika}(k-ig)+e^{-ika}(k+ig)}{g^2(e^{2ika}-1)+k^2+2ikg}
\ee\\

{\bf The square barrier potential}: in the case $V(x)=V_0 \theta(|x|<a/2)$, i.e., 
a square barrier of length $a$ and height  $V_0$  (which can be negative if the potential is an attractive well) with $k_1=\sqrt{k^2-2 V_0)}$, one has \cite{Griffiths}
%\bea
%&\st(k)=\frac{4 k k_1 e^{-ia(k-k_1)}}{(k+k_1)^2-e^{iak_1}(k-k_1)^2} 
\\%= \frac{e^{iak_1}}{\cos(ak_1)-i\frac{k^2+k_1^2}{2kk_1}\sin(ak_1)}\\
%&\sr(k)=\frac{ (k^2-k_1^2) \sin(ak_1) }{2ik k_1 \cos(a k_1)+(k^2+k_1^2)\sin(a k_1)}
%= \frac{e^{iak_1}}{\cos(ak_1)-i\frac{k^2+k_1^2}{2kk_1}\sin(ak_1)}i\frac{k_1^2-k^2}{2kk_1}\sin(ak_1)
%\eea
\bea
&\st(k)=\frac{ e^{-iak}}{\cos(k_1 a)-i\frac{k_1^2+k^2}{2k k_1}\sin( k_1 a)} ,\quad \quad \sr(k)=\frac{ ie^{-ika}(k_1^2-k^2) \sin(ak_1) }{2k k_1 \cos(a k_1)-i(k^2+k_1^2)\sin(a k_1)}
\eea
Note that $\sr(0)=-1$ unless $\sin(ak_1)=0$.\\
%{\bf The delta derivative barrier}: for the case $V(x)=g\delta'(x)$
%Hand made, I get weird result tho
%\bea
%&t^L=1-g,\quad r^R=g\\
%&r^L=-g,\quad t^R=1+g\\
%\eea
% Wikipedia gives 
% \bea
% &\st(k)=\frac{ e^{-iak}}{\cos(k_1 a)-i\frac{k_1^2+k^2}{2k k_1}\sin( k_1 a)} ,\quad \quad 
% \sr(k)=\frac{ (k^2-k_1^2) \sin(ak_1) }{2ik k_1 \cos(a k_1+(k^2+k_1^2)\sin(a k_1)}
% \eea

 {\bf The delta derivative barrier}: From \cite{Lange} for the more general potential with no symmetry, $V(x)=g_1 \delta(x) + g_2 \delta'(x) $, one has
 %$V(x)=g_1 \delta(x) + g_2 \delta'(x) + g_3 \partial_x(\delta(x) \partial_x)$
 %\bea
 %&& \st^{L}(k) = \frac{1-D + 2 i Im[g_2]}{1+D + i (\frac{g_1}{k}-kg_3)},\quad \sr^{R}(k) = \frac{- 2 Re[g_2] - i (\frac{g_1}{k}+ k g_3)}{1+D+i(\frac{g_1}{k}-kg_3)}\\
 %&& \sr^{L}(k) = \frac{- 2 Re[g_2] - i (\frac{g_1}{k}+ k g_3)}{1+D+i(\frac{g_1}{k}-kg_3)},\quad  \st^{R}(k) = \frac{1-D - 2 i Im[g_2]}{1+D + i (\frac{g_1}{k}-kg_3)} \\
 %&& D=g_1g_3+|g_2|^2
 %\eea

%   \bea
%  && \st^{L}(k) = \frac{1-|g_2|^2 + 2 i Im[g_2]}{1+|g_2|^2 + i \frac{g_1}{k}},\quad \sr^{R}(k) = \frac{- 2 Re[g_2] - i \frac{g_1}{k}}{1+|g_2|^2+i\frac{g_1}{k}}\\
%  && \sr^{L}(k) = \frac{- 2 Re[g_2] - i \frac{g_1}{k}}{1+|g_2|^2+i\frac{g_1}{k}},\quad  \st^{R}(k) = \frac{1-|g_2|^2 - 2 i Im[g_2]}{1+|g_2|^2 + i \frac{g_1}{k}} 
%  \eea
  \bea
 \st(k) = \frac{1-g_2^2 }{1+g_2^2 + i \frac{g_1}{k}},\quad \sr^{R}(k) = \frac{ g_2 - i \frac{g_1}{k}}{1+g_2^2+i\frac{g_1}{k}}
 \quad , \quad \sr^{L}(k) = \frac{- g_2 - i \frac{g_1}{k}}{1+g_2^2+i\frac{g_1}{k}} 
 \eea
  If $g_1=0$, this is an example of a defect such that the scattering coefficients are independent on $k$. Defects with that property have been studied 
in the context of conformal defects. Additionally we have $\sr(0)\neq -1$.\\

 {\bf Impurities in series}. It is useful to consider the transfer matrix ${\cal M}$, which reads
\be 
\left( {\begin{array}{cc}
    C  \\
    D  \\
  \end{array} } \right)={\cal M} \left( {\begin{array}{cc}
    A  \\
    B  \\
  \end{array} } \right) \quad , \quad {\cal M} = 
  \left( {\begin{array}{cc}
    t_L - \frac{r_L r_R}{t_R} & \frac{r_R}{t_R} \\
    - \frac{r_L}{t_R}   & \frac{1}{t_R}  \\
  \end{array} } \right)
\ee
Consider now two distinct symmetric impurities, the first one at $x=0$, the second at $x=a$. The total transfer matrix is
\be 
{\cal M} = \left( {\begin{array}{cc}
    t_2 - \frac{r_2^2}{t_2} & \frac{r_2}{t_2} \\
    - \frac{r_2}{t_2}   & \frac{1}{t_2}  \\
  \end{array} } \right) 
 \left( {\begin{array}{cc}
    e^{ i k a}  & 0 \\
    0   & e^{- i k a}  \\
  \end{array} } \right) 
  \left( {\begin{array}{cc}
    t_1 - \frac{r_1^2}{t_1} & \frac{r_1}{t_1} \\
    - \frac{r_1}{t_1}   & \frac{1}{t_1}  \\
  \end{array} } \right) 
\ee 
which leads to the scattering coefficients of the combined system
\be 
t_R= t_L =  \frac{t_1 t_2   e^{i a k}}{1-r_1 r_2 e^{2 i a   k}}
\quad , \quad r_R = \frac{r_2- r_1 e^{2 i a k} \left(r_2^2-t_2^2\right)}{1-r_1 r_2
   e^{2 i a k}} \quad , \quad 
   r_L=
   \frac{r_1-r_2 e^{2 i a k}
   \left(r_1^2-t_1^2\right)}{1-r_1 r_2
   e^{2 i a k}}
\ee 

The most general case is 
\bea 
t_R = \frac{e^{i a k}
   t_{\text{R1}}
   t_{\text{R2}}}{1-e^{2 i a k}
   r_{\text{L2}}
   r_{\text{R1}}} \quad , \quad 
   r_R = 
   r_{\text{R2}}+\frac{e^{2 i a k}
   t_{\text{L2}} r_{\text{R1}}
   t_{\text{R2}}}{1-e^{2 i a k}
   r_{\text{L2}}
   r_{\text{R1}}}
   \quad , \quad r_L = 
   r_{\text{L1}}+\frac{e^{2 i a k}
   t_{\text{L1}} r_{\text{L2}}
   t_{\text{R1}}}{1-e^{2 i a k}
   r_{\text{L2}}
   r_{\text{R1}}}
   \quad , \quad t_L = \frac{e^{i
   a k} t_{\text{L1}}
   t_{\text{L2}}}{1-e^{2 i a k}
   r_{\text{L2}}
   r_{\text{R1}}} \nonumber 
   \\
\eea 
Hence if $t_{\rm R1}=t_{\rm L1}$ and $t_{\rm R2}=t_{\rm L2}$ from time invariance, one has also 
$t_{R}=t_{L}$ as expected. 

{\bf Remark}. If one translates both impurities by $\delta$ the scattering matrix is multiplied by $e^{2 i k \delta}$.
This allows to check the case of two delta impurities given above.

\section{Dynamics of initial eigenstates: regime $x=O(1)$}

In this section we solve the time evolution of the individual initial eigenfunctions $\psi_k^{R/L}(x,t)$ under 
the evolution Hamiltonian $\hat H$. By definition at $t=0$ they are the eigenstates of $\hat H_0$
so that
\bea
\psi_k^{R/L}(x,0)=\phi^{R/L}_k(x)
\eea
where $k$ belongs to the initial lattices $\Lambda^{R/L}$. 
Their
time evolution is expressed as a superposition over the eigenstates of $\hat H$  with amplitudes given by the overlaps
\be \label{exact_psik}
\psi_k^{R/L}(x,t) =\sum\limits_{\sigma=\pm 1,k'\in\Lambda^\sigma}\phi_{\sigma,k'}(x)e^{-i\frac{k'^2}{2}t}\braket{ \phi_{\sigma, k'}|\psi_k^{R/L}(t=0)}
\ee

In the large $\ell$ limit both lattices for $k \in \Lambda^{R/L}$ and $k' \in \Lambda^{\pm}$ converge to $\mathbb{R^+}$. 
The main result that we will show in the next two subsections is that in the double limit $\ell \to +\infty$ followed by $t \to +\infty$
the superposition \eqref{exact_psik} concentrates on $k' \to  k \in \mathbb{R}^+$ so that for any $x=O(1)$ fixed
\bea \label{Psifinalres} 
&\psi_k^{R/L}(x,t) \underset{\substack{\ell\to\infty\\t\to\infty}}{\simeq} e^{-i\frac{k^2}{2}t} \left( \phi_{+,k}(x)e^{ik(\delta^{R/L}_k-\delta_k^{+})}\pm \phi_{-,k}(x)e^{ik(\delta^{R/L}_k-\delta_k^{-})} \right)
\eea
where the $\pm$ refers to $R/L$. Inserting the explicit form \eqref{phi} of the functions $\phi_{\pm,k}(x)$ for $|x|>a$ with now $k \in \mathbb{R}^+$
leads to the result \eqref{limitfunction} stated in the text (see also below). The main mechanism for this reduction 
is that the overlaps $\braket{ \phi_{\sigma, k'}|\psi_k^{R/L}(t=0)}$ develop a pole at $k'=k$
whose residue leads to \eqref{Psifinalres}. The details of the derivation of the convergence however are quite involved.
We proceed in two stages: in the next subsection we obtain a formula for $\ell \to +\infty$ and fixed $t$ and in the following one we
study $t \to +\infty$. Finally in the last subsection we obtain the kernel in the NESS by inserting \eqref{Psifinalres} in the general formula

\subsection{Large $\ell$ limit} 

We now obtain a formula for $\psi_k^{R/L}(x,t)$ in the limit of
large $\ell$ and for $|x|>\frac{a}{2}$. We first give the result and provide the details of the derivation below. 

{\bf Main result in the thermodynamic limit}. We obtain the leading behavior 
\bea\label{largeellpsi}
 \psi_k^{R/L}(x,t) \underset{\ell\to\infty}{\simeq} \frac{1}{\sqrt{\ell}}(e^{-i\frac{k^2}{2}t}\chi_{k}^{R/L}(x)+\delta \chi_{k}^{R/L}(x,t)),\quad \mbox{ for }\quad |x|>\frac{a}{2}
\eea
where we denote 
\bea \label{chifinal} 
&\chi_{k}^{R}(x)=e^{ik\delta_k^{R}}(\theta(x)(e^{-ikx}+\sr(k)e^{ikx})
+\theta(-x)\st(k)e^{-ikx}) \\
&\chi_{k}^{L}(x)=e^{ik\delta_k^{L}}(\theta(-x)(e^{ikx}+\sr(k)e^{-ikx})
+\theta(x)\st(k) e^{ikx}) \nonumber 
\eea
as well as 
\\
\bea\label{deltapsi}
&\delta \chi_k^{R/L}(x,t)=-\int_{\Gamma_-}\frac{dk'}{\pi}2 e^{-i\frac{k'^2}{2}t}\frac{F^{R/L}_+(k,k',x) \pm {\rm sgn}(x) F^{R/L}_-(k,k',x)}{k-k'}\\
&+\sqrt{2}\,\int_0^\infty\frac{dk'}{2\pi}e^{-i\frac{k'^2}{2}t}\left(\cos(k'(|x|-\delta_{k'}^{+}))M^{R/L}(\phi_k^{R/L},\phi_{+,k'},a)\right. \nonumber\\
&\left.+{\rm sgn}(x)\cos(k'(|x|-\delta_{k'}^{-}))M^{R/L}(\phi_k^{R/L},\phi_{-,k'},a)\right) \nonumber
\eea
We have defined the auxiliary functions
\bea \label{def_Fpm}
&F^{R/L}_\pm(k,k',x)=\cos(k'(|x|-\delta_{k'}^{\pm}))\frac{k'\sin(k'(\frac{a}{2}-\delta_{k'}^{\pm}))\cos(k(\frac{a}{2}-\delta_{k}^{R/L}))-k\sin(k(\frac{a}{2}-\delta_k^{R/L}))\cos(k'(\frac{a}{2}-\delta_{k'}^\pm))}{k+k'}
\eea
as well as the contour $\Gamma_-$ in the complex plane for $k'$ such that
\bea \label{def_gammam}
\int_{\Gamma_-}dk'=\int_{+\infty-i \epsilon}^{-i \epsilon} dk'+\int_{-i\epsilon}^0 dk'
\eea
where $\epsilon$ is some strictly positive number. We have also defined the partial overlaps
\bea\label{partialoverlap}
\frac{M^{R/L}(\phi_k^{R/L},\phi_{\pm,k'},a)}{\ell}=\int_{R/L} dy \phi_k^{R/L}(y) \phi_{\pm,k'}(y) \theta(|y|<\frac{a}{2})
\eea
where we used the notations $\int_{R} dy=\int_{0}^{\ell/2} dy$ and 
$\int_{L} dy=\int_{-\ell/2}^{0} dy$. The factor $1/\ell$ is introduced for convenience since the
wave-functions normalization constants are $O(1/\sqrt{\ell})$.
\\

{\bf Derivation}. We start from \eqref{exact_psik}. We first need to calculate the overlaps.
%\bea
%&\braket{+, k'|\psi_k^{R/L}(0)}=\int_{R/L}dy\phi_k^{R/L}(y)\phi_{+,k'}(y)%\int_{R/L}dy\sqrt{\frac{4}{\ell}}\cos(k(|y|-\delta_k^{R/L}))\sqrt{\frac{2}{\ell}}\cos(k'(|y|-\delta_{k'}^{+}))\\
%=\frac{2^{3/2}}{\ell}\frac{k\sin(k\delta_k^{R/L})\cos(k'\delta_{k'}^+)-k'\sin(k'\delta_{k'}^{+})\cos(k\delta_{k}^{R/L})}{k^2-k'^2}\\
%&\braket{-, k'|\psi_k^{R/L}(0)}=\int_{R/L}dy\phi_k^{R/L}(y)\phi_{-,k'}(y)%=\int_{R/L}dy\sqrt{\frac{4}{\ell}}\cos(k(|y|-\delta_k^{R/L}))\sqrt{\frac{2}{\ell}}{\rm s}(y)\cos(k'(|y|-\delta_{k'}^{-}))\\
%=\pm\frac{2^{3/2}}{\ell}\frac{k\sin(k\delta_k^{R/L})\cos(k'\delta_{k'}^-)-k'\sin(k'\delta_{k'}^{-})\cos(k\delta_{k}^{R/L})}{k^2-k'^2}
%\eea
We perform the following decomposition
\bea \label{decomposition}
&\braket{\phi_{\pm, k'}|\psi_k^{R}(t=0)}
=\int_{0}^{a/2} dy \phi_k^{R}(y) \phi_{\pm,k'}(y) + \int_{a/2}^{\ell/2} dy \phi_k^{R}(y) \phi_{\pm,k'}(y)\\
&\braket{\phi_{\pm, k'}|\psi_k^{L}(t=0)}
=\int_{-a/2}^0 dy \phi_k^{L}(y) \phi_{\pm,k'}(y) + \int_{-\ell/2}^{-a/2} dy \phi_k^{L}(y) \phi_{\pm,k'}(y) \nonumber 
\eea
In the second parts of these overlaps (i.e., the integrals over $[-\ell/2,-a/2]$ and $[a/2,\ell/2]$)
the eigenfunctions are known, hence we compute those parts exactly. 
The resulting expressions are then easily extended to the complex plane for $k'$.
We see that they contain a "pole" at $k'=k$, of the form $\frac{1}{\ell(k-k')}$ (the factor $1/\ell$ comes
from the normalization constants). This implies that, when $k'$ in the summation
in \eqref{exact_psik} is as close from $k$ as it can be, which is $k-k'\sim \frac{1}{\ell}$ (recall that
$k$ and $k'$ belong to two distinct lattices), then the second parts of the overlaps in \eqref{decomposition} are of order 1.
The first parts (i.e., the integrals over $[-a/2,0]$ and $[0,a/2]$) correspond to the partial overlaps defined in \eqref{partialoverlap}. Since the eigenfunctions of $\hat H$ are not explicitly known in these intervals, we cannot give an explicit expression. However we will assume that these partial overlaps $M^{R/L}$ are non singular functions when $k' \to k$ in the large $\ell$ limit. This is natural for wave-functions which are uniformly bounded on the finite interval $[-a/2,a/2]$. 
This implies that the first integrals in Eq. (\ref{decomposition}) are of order $\ell^{-1}$. 
%This makes sense for two free states that their overlap over $[-a/2,0]$ and $[0,a/2]$ are only of order $\ell^{-1}$.
%\bea
%&\braket{\phi_{+, k'}|\psi_k^{R}(t=0)}
%=\int_{0}^{\ell/2}dy\phi_k^{R}(y)\phi_{+,k'}(y)\\
%=\int_{0}^{a/2} dy \phi_k^{R}(y) \phi_{+,k'}(y) + \int_{a/2}^{\ell/2} dy \phi_k^{R}(y) \phi_{+,k'}(y)\\
%&=\frac{f^{R}(\phi^R,k,\phi_+,k',a)}{\ell}+\frac{2^{3/2}}{\ell}\frac{k'\sin((\frac{a}{2}-k'\delta_{k'}^{+}))\cos(k(\frac{a}{2}-\delta_{k}^{R}))-k\sin(k(\frac{a}{2}-\delta_k^{R}))\cos(k'(\frac{a}{2}-\delta_{k'}^+))}{k^2-k'^2}
%\eea
%c_{\ell,R/L,k}^0 c_{\ell,+,k'}

To compute easily the integrals in the second parts of \eqref{decomposition} we note that
$\phi_{\pm,k'}''(x)= -  (k')^2 \phi_{\pm,k'}(x)$ and $\phi^{R/L,\prime \prime}_k(x)= -  k^2 \phi^{R/L}_k(x)$
for $|x|>a/2$ and hence $(\phi_{\pm,k'}' \phi_{k} - \phi_{\pm,k'} \phi'_{k})' =  (k^2-(k')^2) \phi_{\pm,k'} \phi_{k}$ 
with $\phi_k=\phi_k^{R/L}$,
which implies
\be\label{overlap0}
 \int_{a/2}^{\ell/2} \phi_{\pm,k'}(x) \phi^R_{k}(x) \, dx  = \frac{1}{(k')^2-k^2} ( \phi_{\pm,k'}'(\frac{a}{2}) \phi^R_{k}(\frac{a}{2}) - \phi_{\pm, k'}(\frac{a}{2}) \phi^{R,\prime}_{k}(\frac{a}{2}))
\ee 
and similarly for $\phi^L_{k}$. We have used that all the wave-functions vanish at $x=\pm \ell/2$. This leads to 
\bea
&\braket{\phi_{+,k'}|\psi_k^{R/L}(t=0)} = \frac{M^{R/L}(\phi^{R/L},k,\phi_+,k',a)}{\ell} + \frac{2^{3/2}}{\ell}\frac{k'\sin(k'(\frac{a}{2}-\delta_{k'}^{+}))\cos(k(\frac{a}{2}-\delta_{k}^{R/L}))-k\sin(k(\frac{a}{2}-\delta_k^{R/L}))\cos(k'(\frac{a}{2}-\delta_{k'}^+))}{k^2-k'^2} \label{overlap1}\\
&\braket{\phi_{-,k'}|\psi_k^{R/L}(t=0)}=\frac{M^{R/L}(\phi^{R/L},k,\phi_-,k',a)}{\ell}\pm \frac{2^{3/2}}{\ell}\frac{k'\sin(k'(\frac{a}{2}-\delta_{k'}^{-}))\cos(k(\frac{a}{2}-\delta_{k}^{R/L}))-k\sin(k(\frac{a}{2}-\delta_k^{R/L}))\cos(k'(\frac{a}{2}-\delta_{k'}^-))}{k^2-k'^2} \label{overlap2}
\eea
Where $\pm$ stands for $R/L$  and $c_{\ell,R/L,k}^0 c_{\ell,\pm,k'}$ has been replaced by its large $\ell$ limit $\frac{2^{3/2}}{\ell}$ which does not affect the final result.

%Note that the residue of the r.h.s for $k=k'$ does not depend on $a$. Next we can argue that the unknown part
%\be 
%\int_0^{a/2} \tilde \phi_k \tilde \phi_{k} 
%\ee 
%where $\tilde \phi_{k'}$ and $\tilde \Psi_{k}$ are the true eigenstates of energies $k'^2$ and $k^2$ respectively, is of order $1/\ell$. This implies that they cannot %contribute to the residue. 
By inserting these expressions (\ref{overlap1}) and (\ref{overlap2}) in Eq. (\ref{exact_psik}), we obtain the exact expression of the wave function at large but finite $\ell$ and finite $t$. This yields
\bea \label{dec2} 
&\psi_k^{R/L}(x,t)=\psi_{k,1,+}^{R/L}(x,t)\pm {\rm sgn}(x)\psi_{k,1,-}^{R/L}(x,t)+\psi_{k,2}^{R/L}(x,t)
\eea
\bea \label{psi1} 
&\psi_{k,1,\pm}^{R/L}(x,t)=\sqrt{\frac{2}{\ell}}\sum\limits_{k'\in \Lambda^\pm}\frac{2^{3/2}}{\ell} \frac{F^{R/L}_\pm(k,k',x)}{k-k'}e^{-i\frac{k'^2}{2}t}\\
%\frac{2^{3/2}}{\ell}\frac{k\sin(k\delta_k^{R/L})\cos(k'\delta_{k'}^+)-k'\sin(k'\delta_{k'}^{+})\cos(k\delta_{k}^{R/L})}{k^2-k'^2}\\
%\frac{2^{3/2}}{\ell}\frac{k\sin(k\delta_k^{R/L})\cos(k'\delta_{k'}^-)-k'\sin(k'\delta_{k'}^{-})\cos(k\delta_{k}^{R/L})}{k^2-k'^2}
&\psi_{k,2}^{R/L}(x,t)=\sum\limits_{k'\in \Lambda^+} \sqrt{\frac{2}{\ell}} \cos(k'(|x|-\delta_{k'}^{+})) e^{-i\frac{k'^2}{2}t} \frac{M^{R/L}(\phi_k^{R/L},\phi_{+,k'},a)}{\ell}\\
&+ {\rm sgn}(x) \sum\limits_{k'\in \Lambda^-} \sqrt{\frac{2}{\ell}} \cos(k'(|x|-\delta_{k'}^{-})) e^{-i\frac{k'^2}{2}t} \frac{M^{R/L}(\phi^{R/L}_k,\phi_{-,k'},a)}{\ell}
\eea
where the functions $F^{R/L}_{\pm}$ and $M^{R/L}$ are defined respectively in Eqs. (\ref{def_Fpm}) and (\ref{partialoverlap}).
This expression is still exact for any $\ell>a$ and $t$, apart from the fact that the normalization constants 
and $c_{\ell,R/L,k}^0$ and $c_{\ell,\pm,k'}$ have been replaced by their large $\ell$ behaviors. 

We now consider the limit $\ell \to +\infty$ of each term in \eqref{dec2}.

{\bf Large $\ell$ limit of $\psi_{k,2}^{R/L}$} Because $M^{R/L}$ is regular, the large $\ell$ limit of $\psi_{k,2}^{R/L}$ is a Riemann sum and the result is straightforward
\bea
\psi_{k,2}^{R/L} \underset{\ell\to\infty}{\simeq}& \sqrt{\frac{2}{\ell}}\int_0^\infty\frac{dk'}{2\pi}e^{-i\frac{k'^2}{2}t}\left(\cos(k'(|x|-\delta_{k'}^{+}))M^{R/L}(\phi^{R/L}_k,\phi_{+,k'},a)\right.\\
&\left.+{\rm sign}(x)\cos(k'(|x|-\delta_{k'}^{-}))M^{R/L}(\phi^{R/L}_k,\phi_{-,k'},a)\right) \label{psi2_largel} 
\eea

This expression involves $M^{R/L}$ which depends on the details of the two potentials $V$ and $V_0$. 
As we will see below, it decays to zero at large time, so that in the total wave-function $\psi_k(x,t)$ only the
scattering coefficients associated to $\mathcal{S}$ and $\mathcal{S}_0$ will matter in the double limit.
\\

{\bf Large $\ell$ limit of $\psi_{k,1,\pm}^{R/L}$.} Because the exact expression \eqref{psi1} for $\psi_{k,1,\pm}^{R/L}$ has a pole at $k'=k$, one cannot "blindly" replace (in the limit $\ell \to \infty$) the discrete sum over $k' \in \Lambda_{\pm}$ by an integral.  
Instead we use a contour integral method very similar to the one introduced in \cite{Prosen2018, GLS} which allows us to take this limit. 

%\bea
%R_k=\sum\limits_{k'\in \Lambda^{\pm}}\frac{2}{\ell}\frac{F(k,k')}{k-k'}e^{-i\frac{k'^2}{2}t}
%\eea\\
%with $F(k,k')$ a regular function. $k$ is an element of $\Lambda^{R/L}$. $\Lambda^{R/L}$ and $\Lambda^\pm$ are characterised by phases $\delta_k^{R/L}$ and $\delta_{k'}^\pm$ through quantization conditions \eqref{quantification} 
%\bea
%&\cos(k (\frac{\ell}{2} - \delta_k^{R/L})) = 0 \\
%&\cos(k' (\frac{\ell}{2} - \delta_{k'}^\pm)) = 0 
%\eea\\
The quantification condition \eqref{quantification} for $k'$, %and \eqref{quantification_initial} 
which defines the lattices $\Lambda^\pm$, can be written
\bea \label{quantification2}
&e^{2 i k' (\frac{\ell}{2} - \delta_{k'}^\pm) } + 1 = 0
\eea
%In our case, this will be applied with $F(k,k')=\frac{\ell c_{\ell,R/L,k}^0 c_{\ell,\pm,k'}}{\sqrt{2}} F_\pm(k,k',x)$, $\delta_{k}=\delta_{k}^{R/L}$ and $\delta_{k'}'=\delta_{k'}^\pm$
Now we will write the sum over $k'$ as a contour integral in the complex plane for $k'$. With this goal in mind we introduce the function
\bea
g_{\delta^\pm,\ell}(k')=\frac{-1}{e^{-ik'(\ell-2\delta_{k'}^\pm)}+1}
\eea
The Eq. \ref{quantification2} implies that $g_{\delta^\pm,\ell}(k')$ has poles on each element $k'$ of the lattice $\Lambda^\pm$ with residue 
\be 
-  \frac{1}{ \partial_{k'} e^{-ik'(\ell-2\delta_{k'}^\pm)} } =
  \frac{1}{ (i \ell - 2 \partial_{k'} (k'\delta_{k'}^\pm)) e^{-ik'(\ell-2\delta_{k'}^\pm)} }  
= - \frac{1}{i \ell - 2 i \partial_{k'} (k' \delta_{k'}^\pm)} =- \frac{1}{i \ell} (1 + O(1/\ell))  
\ee\\
This allow us to write the discrete sum in the expression \eqref{psi1} of $\psi_{k,1,\pm}^{R/L}$ 
as an integral over a contour $\Gamma_0$ equal to the union of small clockwise circular contours around each element of $\Lambda^\pm$ as shown in Fig. \ref{ContourNess}.
\be\label{sumtoint}
\frac{1}{\ell}  \underset{k'\in\Lambda^{\pm}}{\sum} \dots =\int_{\Gamma_0}\frac{dk'}{2\pi} g_{\delta^\pm,\ell}(k') \dots
\ee
leading to 
\begin{figure}
    \centering
    \includegraphics[width=\linewidth]{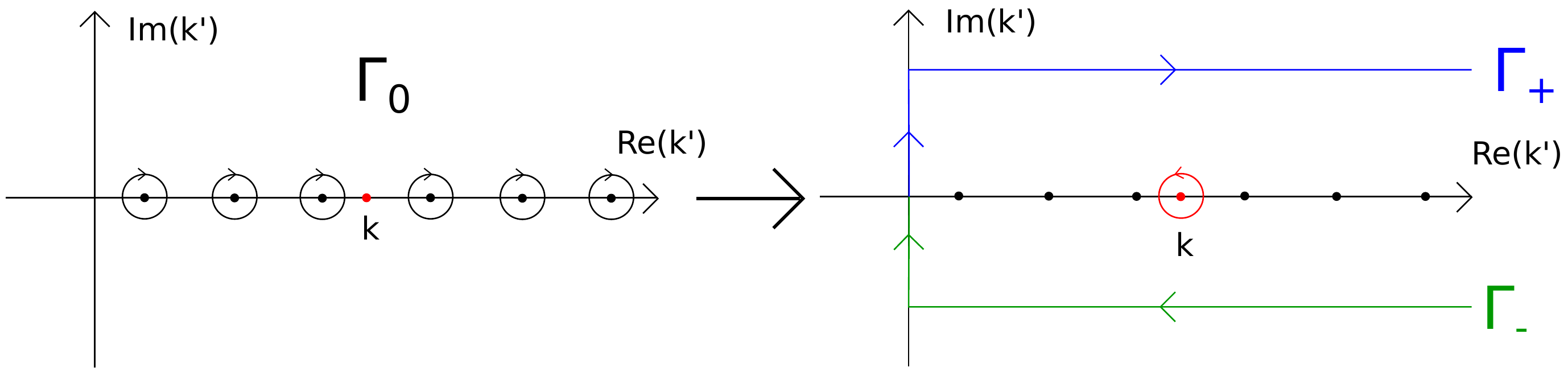}
    \caption{Illustration of the contour $\Gamma_0$ in the complex $k'$ plane, and its deformation into the union of three different contours, i.e., $\Gamma_+$,  $\Gamma_-$ and the small counter-clockwise red circle around $k'=k$.}
    \label{ContourNess}
\end{figure}
\bea
\psi_{k,1,\pm}^{R/L}(x,t) \simeq \frac{1}{\sqrt{\ell}}\int_{\Gamma_0} \frac{dk'}{\pi}2 g_{\delta^\pm,\ell}(k')\frac{F^{R/L}_\pm(k,k',x)}{k-k'}e^{-i\frac{k'^2}{2}t}
\eea
Assuming that the integrand has a singularity only at $k'=k$ (within a strip around the positive real axis) we can now
deform this contour into the union of a larger contour $\Gamma=\Gamma_- \cup \Gamma_+$ that encloses all the previous one, and of a tiny contour around $k$. 
Computing the residue associated to this pole at $k'=k$ we obtain 
\bea\label{complexmethod}
\psi_{k,1,\pm}^{R/L}(x,t)& \simeq - \frac{1}{\sqrt{\ell}}4i   g_{\delta^\pm,\ell}(k)F^{R/L}_\pm(k,k,x)e^{-i\frac{k^2}{2}t} + \frac{1}{\sqrt{\ell}} \int_{\Gamma_- \cup \Gamma_+}  \frac{dk'}{\pi} 2 g_{\delta^\pm,\ell}(k')\frac{F^{R/L}_\pm(k,k',x)}{k-k'}e^{-i\frac{k'^2}{2}t}
\eea
Using the quantification condition for $k \in \Lambda^{R/L}$ which reads $e^{2 i k (\frac{\ell}{2} - \delta_k^{R/L}) } + 1 = 0$
we can eliminate $\ell$ and obtain 
\be\label{prefactor1}
- 2i g_{\delta^\pm,\ell}(k) =
 \frac{e^{ik(\delta_k^{R/L}-\delta_k^\pm)}}{\sin(k(\delta_k^{R/L}-\delta_k^\pm))} =  (\cot(k(\delta_k^{R/L}-\delta_k^\pm))+i)
\ee
So that
\bea
\psi_{k,1,\pm}^{R/L}(x,t)=  \frac{2}{\sqrt{\ell}}\frac{e^{ik(\delta_k^{R/L}-\delta_k^{\pm})}}{\sin(k(\delta_k^{R/L}-\delta_k^\pm))}F^{R/L}_\pm(k,k,x)e^{-i\frac{k^2}{2}t}+\frac{1}{\sqrt{\ell}}\int_{\Gamma_- \cup \Gamma_+}\frac{dk'}{\pi}2g_{\delta^\pm,\ell}(k')\frac{F^{R/L}_\pm(k,k',x)}{k-k'}e^{-i\frac{k'^2}{2}t}
\eea
Under this form, we can now take the large $\ell \to +\infty$ limit. The limiting value of the function $g_{\delta^\pm,\ell}(k')$ for $k' \in \Gamma_- \cup \Gamma_+$ is (for ${\rm Im}(k') \neq 0$
\bea\label{glimit}
g_{\delta^\pm,\ell}(k') \underset{\ell\to\infty}{\to} \begin{cases}
0 \mbox{ if } k'\in\Gamma_+ \\
-1 \mbox{ if } k'\in\Gamma_-
\end{cases}
\eea
So that we are left with the residue part and the $\Gamma_-$ integral part only
\bea
\psi_{k,1,\pm}^{R/L}(x,t) \underset{\ell\to\infty}{\simeq}  \frac{2}{\sqrt{\ell}}\frac{e^{ik(\delta^{R/L}_k-\delta_k^{\pm})}}{\sin(k(\delta_k^{R/L}-\delta_k^{\pm}))}F^{R/L}_\pm(k,k,x)e^{-i\frac{k^2}{2}t}-\frac{2}{\sqrt{\ell}}\int_{\Gamma_-}\frac{dk'}{\pi}\frac{F^{R/L}_\pm(k,k',x)}{k-k'}e^{-i\frac{k'^2}{2}t} \label{psi1_largel}
\eea

We now put together in the decomposition \eqref{dec2} 
our results for the limit $\ell \to + \infty$ for $\Psi_{k,2}^{R/L}$ in \eqref{psi2_largel} and
for $\Psi_{k,1,\pm}^{R/L}$ in \eqref{psi1_largel}. We use that for $k'=k$ Eq. \eqref{def_Fpm} reduces to
\be \label{def_Fpm2}
 F^{R/L}_\pm(k,k,x)= 
%\frac{1}{2} \cos(k(|x|-\delta_{k}^{\pm})) \left( %\sin(k(\frac{a}{2}-\delta_{k}^{\pm}))\cos(k(\frac{a}{2}-\delta_{k}^{R/L}))-\sin(k(\frac{a}{2}-\delta_k^{R/L}))\cos(k(\frac{a}{2}-\delta_{k}^\pm)) \right) \\
  \frac{1}{2} \cos(k(|x|-\delta_{k}^{\pm})) \sin( k (\delta_{k}^{R/L} -\delta_{k}^{\pm} )) 
\ee
This leads to the result announced
above in Eqs. \eqref{largeellpsi} and \eqref{deltapsi}. In particular we obtain 
\bea
&\chi_{k}^{R/L}(x)=   \cos(k(|x|-\delta_k^{+})e^{ik(\delta^{R/L}_k-\delta_k^{+})}\pm {\rm sgn}(x)\cos(k(|x|-\delta_k^{-}))e^{ik(\delta^{R/L}_k-\delta_k^{-})}  \\
&= e^{ik\delta_k^{R/L}}(\frac{1\pm {\rm sgn}(x)}{2}e^{-ik|x|}+\frac{e^{-2ik\delta_k^+}\pm {\rm sgn}(x)e^{-2ik\delta_k^-}}{2}e^{ik|x|})  \\
&= e^{ik\delta_k^{R/L}}(\frac{1\pm {\rm sgn}(x)}{2}(e^{-ik|x|}+r(k)e^{ik|x|})+\frac{1\mp {\rm sgn}(x)}{2}t(k) e^{ik|x|}) 
\eea
leading to the final expression of $\chi_{k}^{R/L}$ in \eqref{chifinal}. 

\subsection{Large time limit}

We now show that in the large time limit, the time evolved initial eigenfunctions take the simple form for $|x|>a/2$ given in the text in (\ref{limitfunction})
\be\label{psilargetime}
\psi_{k}^{R/L}(x,t)\underset{\substack{\ell\to\infty\\t\to\infty}}{\simeq}\frac{1}{\sqrt{\ell}}e^{-i\frac{k^2}{2}t}\chi_k^{R/L}(x)
\ee
where $\chi_{k}^{R/L}(x)$ are given in \eqref{chifinal}. This is achieved by arguing that the remainder \eqref{deltapsi} vanishes in the limit
\be\label{psilargetime2}
\underset{t\to\infty}{\lim}\delta \chi_k^{R/L}(x,t)=0
\ee

Consider the first integral in \eqref{deltapsi} over the contour $\Gamma_-$. Let us parameterize $k'=k_1'+ik_2'$ in terms of
its real and imaginary parts. On the horizontal part of the contour $k_2=-\epsilon<0$ the term in the integrand 
$e^{-i\frac{k'^2}{2} t}=e^{i\frac{k_2'^2-k_1'^2}{2}t}e^{k_1'k_2't} = e^{i\frac{k_2'^2-k_1'^2}{2}t}e^{- k_1' \epsilon t}$ decays exponentially in time.
Upon integration over $k_1' \in [0,+\infty[$ the result decays to zero at least algebraically in time. On the vertical part of $\Gamma_-$, $k_1'=0$ and the term $e^{-i\frac{k'^2}{2} t}=e^{i\frac{k_2'^2}{2}t}$ oscillates rapidly for large $t$. The
integral is dominated by the vicinity of $k'=0$ and one can check that for fixed $k>0$ the integrand is regular at $k'=0$ (and generically of order $O(1)$).  
This integral thus also decays algebraically in time. Concerning the second integral in \eqref{deltapsi}, although we do not
have any analytical expression, a similar argument based on the oscillating factor $e^{-i\frac{k'^2}{2} t}$ suggests that
it also decays to zero algebraically at large time. 
\\

{\bf Remark:} The above arguments were made for $|x| >a/2$, i.e., outside the defect, where the wave-functions and their analytic continuations are controlled.
However given the form \eqref{exact_psik}, and the fact that the asymptotic form arises mainly from some properties of the overlaps, it is natural to conclude 
that the asymptotic formula \eqref{psilargetime} generalizes to any $x=O(1)$  (including inside the impurity) but with an extended formula for $\chi_k^{R/L}$
%{\red the following is false one should replace$\phi$ by $\sqrt{\frac{\ell}{2}}\phi$}
\bea \label{remarksym} 
& \frac{1}{\sqrt{\ell}} \chi_{k}^{R/L}(x)=\frac{1}{\sqrt{2}} \left( \phi_{+,k}(x)e^{ik(\delta^{R/L}_k-\delta_k^{+})}
+ \sigma_{R/L} \phi_{-,k}(x)e^{ik(\delta^{R/L}_k-\delta_k^{-})} \right) 
\eea
where $\sigma_R=1$ and $\sigma_L=-1$,
as announced above in \eqref{Psifinalres}. Here $k \in \mathbb{R}^+$ and $\phi_{\pm,k}(x)$ denote the two eigenstates of $\hat H$ (even and odd)
which become degenerate in the infinite size limit with eigenenergy $\frac{k^2}{2}$.

% {\red the asymmetric impurity gives
% \be
% \chi_{k}^{R/L}(x)=\sqrt{\frac{\ell}{4}}2(\phi_{+,k}(x)e^{ik(\delta^{R/L}_k-\delta_k^{+})} P_k^\mp \mp \phi_{-,k}(x)e^{ik(\delta^{R/L}_k-\delta_k^{-})}P_k^\pm )
% \ee
% Where $\pm$ stands for $\pm=\begin{cases}
% + \quad R\\
% - \quad L
% \end{cases}$ and $\pm$ the opposite
% }

\subsection{Computation of the kernel in thermodynamic and large time limit}\label{sectionKernelNess}

By definition of the time dependent kernel is given by
\bea \label{Kt} 
&K_{R/L}(x,t;x',t')=\sum\limits_{k\in\Lambda^{R/L}}\left( f_{R/L}(k)-\theta(t'-t)\right)(\psi_k^{R/L}(x,t))^*\psi_k^{R/L}(x',t')
%&=\sum\limits_{k\in\Lambda^{R/L}}\frac{1}{\ell}\left( f_{R/L}(k)-\theta(\tau)\right)(e^{-i\frac{k^2}{2}t}\chi_{k}^{R/L}(x)+\delta \psi_{k,\ell}^{R/L}(x,t))^*(e^{-i\frac{k^2}{2}t'}\chi_{k}^{R/L}(x')+\delta \psi_{k,\ell}^{R/L}(x',t'))\\
%&= K_{\infty,R/L}(x,t;x',t') + \delta K_{R/L}(x,t;x',t')
\eea
where $f_{R/L}(k)=\frac{1}{e^{\beta_{R/L}(\mu_{R/L}-\frac{k^2}{2})}+1}$ is the Fermi factor and we use the convention $\theta(0)=0$.
Since we have already obtained the large $\ell$ limit of the individual wave-functions $\psi_k^{R/L}(x,t)$,
see \eqref{largeellpsi}, it is natural to inject their expressions into the formula \eqref{Kt}. Indeed, as $\ell \to \infty$, 
it turns out that one can safely replace $\frac{1}{\ell} \sum_{k\in\Lambda^{R/L}} $ 
by $\int_0^{+\infty} \frac{dk}{2 \pi}$, i.e., there is no singularity in this summation. This leads to
\bea\label{largelK}
\underset{\ell\to\infty}{\lim}K_{R/L}(x,t;x',t')&=\int_0^{\infty}\frac{dk}{2\pi}\left( f_{R/L}(k)-\theta(t'-t)\right)(e^{-i\frac{k^2}{2}t}\chi_{k}^{R/L}(x)+\delta \chi_{k}^{R/L}(x,t))^*\\
&\times(e^{-i\frac{k^2}{2}t'}\chi_{k}^{R/L}(x')+\delta \chi_{k}^{R/L}(x',t')) \nonumber \\
&= K_{\infty,R/L}(x,t;x',t') + \delta K_{R/L}(x,t;x',t')
\eea
where we have split the limiting kernel in two parts, defined as
\bea
&K_{\infty,R/L}(x,t;x',t-\tau) =  \int_{{R/L},\tau} \frac{dk}{2\pi} e^{i\frac{k^2}{2}\tau}\chi_k^{R/L}(x)^*\chi_k^{R/L}(x')\\
&\delta K_{R/L}(x,t;x',t-\tau)=\int_{R/L,\tau} \frac{dk}{2\pi}\left((\delta \chi_{k}^{R/L}(x,t))^*\delta \chi_{k}^{R/L}(x',t-\tau)\right.\\
&\left. + \; (e^{-i\frac{k^2}{2}t}\chi_{k}^{R/L}(x))^*\delta \chi_{k}^{R/L}(x',t-\tau)+(\delta \chi_{k}^{R/L}(x,t))^*e^{-i\frac{k^2}{2}(t-\tau)}\chi_{k}^{R/L}(x')\right) \;,
\eea
and where we used the shorthand notations 
\bea
&\int_{R/L,\tau}\frac{dk}{2\pi}=\int_0^{\infty}\frac{dk}{2\pi}\left(f_{R/L}(k)-\theta(-\tau)\right) \;,\\
&\int_{L,R}\frac{dk}{2\pi}=\int\frac{dk}{2\pi}(f_L(k)-f_R(k))
\eea
Note that we have set $t'=t - \tau$. 

Let us consider now the limit $t, t' \to +\infty$ with fixed $\tau$. Since we have shown in the previous section that $\delta \chi_{k}^{R/L}(x,t)$ 
decays to zero in the large time limit
it is natural to conclude that 
\bea
\underset{\substack{\ell\to\infty\\t\to\infty}}{\lim}\delta K_{R/L}(x,t;x',t-\tau)=0
\eea
so that the two time kernel in the NESS is simply given by
\bea \label{final_kernel}
\underset{\substack{\ell\to\infty\\t\to\infty}}{\lim}K_{R/L}(x,t;x',t-\tau)= \int_{R/L,\tau} \frac{dk}{2\pi} e^{i\frac{k^2}{2}\tau}\chi_k^{R/L}(x)^*\chi_k^{R/L}(x')
\eea
where $\chi_k^{R/L}(x)$ are given in \eqref{chifinal}. Note that the phase shifts $\delta_k^{R/L}$ of the initial condition cancel 
in the product. 
\\

Note that the above derivation of the asymptotics of the kernel, although simple, would in principle require more care. In particular, the fact that $\delta \chi_{k}^{R/L}$ decays in the large time limit may not be sufficient to conclude that $\delta K_{R/L}$ also decays to zero. Indeed, inserting the overlap decomposition into
the above formula for the kernel leads to multiple integrals (up to a triple integral). Because of the singularity $\propto \frac{1}{k-k'}$ contained in each $\delta \chi_{k}^{R/L}$ [see Eq. (\ref{deltapsi})] much care must be taken to analyze the asymptotics. This was done in detail in \cite{GLS}, by a slightly
different route (the contour integral trick was performed on the variable $k$ rather than $k'$) in the case of a delta impurity. The present
final result agrees perfectly with the conclusions of Ref. \cite{GLS}, which shows that our simpler method works.
\\

{\bf Remark:}  Following the remark above we can surmise that the two-time kernel in the NESS takes the general 
form for any $x,x',\tau=O(1)$, in terms of the eigenfunctions of the evolution Hamiltonian $\hat H$ considered here (which does not
possess any bound state)
\bea \label{generalK} 
&\underset{\substack{\ell\to\infty\\t\to\infty}}{\lim}K_{R/L}(x,t;x',t-\tau) = \int_{R/L,\tau} \frac{dk}{2\pi}e^{i\frac{k^2}{2}\tau}\big( \phi_{-,k}(x)^*\phi_{-,k}(x')+ \phi_{+,k}(x)^*\phi_{+,k}(x') \\
& \pm ( \phi_{+,k}(x)^*\phi_{-,k}(x')e^{ik(\delta_k^+-\delta_k^-)}+ \phi_{-,k}(x)^*\phi_{+,k}(x')e^{-ik(\delta_k^+-\delta_k^-)})\big) \nonumber
\eea

Comparing with our previous work in the case of a delta impurity and $\tau=0$, we see that the "diagonal terms" called $A$ and $B$ there
(which are real and do not carry current) correspond to the first two terms respectively in \eqref{generalK}, while the terms noted $C$ and $D$ (which carry the
current in the NESS) correspond to the last two terms in \eqref{generalK}.

\section{Dynamics of initial eigenstates: ray regime $x=O(t)$}
\subsection{Large time limit}
In this section, we study the evolution of a single eigenstate $\psi_k^{R/L}(x,t)$ under the Hamiltonian $\hat H$, in the
regime of rays 
\be \label{xray} 
x = \xi t + y \quad , \quad y=O(1) 
\ee 
Again we first take the large system limit $\ell \to +\infty$ to avoid boundary effects.
Our starting point is thus the formula \eqref{largeellpsi} for $\psi_k^{R/L}(x,t)$. The difference with the previous regime (NESS) occurs 
when we take the double limit $t \to +\infty$ and $x \to +\infty$ according to \eqref{xray}. 
In that regime the integral part $\delta \chi_{k}^{R/L}(x,t)$ in \eqref{largeellpsi} no longer goes to zero.
Let us recall \eqref{deltapsi}
\bea \label{deltapsinew} 
&\delta \chi_k^{R/L}(x,t)=-\int_{\Gamma_-}\frac{dk'}{\pi}2e^{-i\frac{k'^2}{2}t}\frac{F^{R/L}_+(k,k',x)\pm {\rm sgn}(x)F^{R/L}_-(k,k',x)}{k-k'}\\
&+\int_0^{\infty}\frac{dk'}{2\pi}\sqrt{2}e^{-i\frac{k'^2}{2}t}\left(\cos(k'(|x|-\delta_{k'}^{+}))M^{R/L}(\phi_k^{R/L},\phi_{+,k'},a)+s(x)\cos(k'(|x|-\delta_{k'}^{-}))M^{R/L}(\phi_k^{R/L},\phi_{-,k'},a)\right) \nonumber \\
&=\psi_{k,1,+}^{R/L}(x,t)\pm {\rm sgn}(x)\psi_{k,1,-}^{R/L}(x,t)+\psi_{k,2}^{R/L}(x,t)
\eea
where the last term of the third line equals the second line. 

{\bf Large time limit of $\psi_{k,2}^{R/L}$:} in the limit $t \to \infty$, $\psi_{k,2}^{R/L}(\xi t+y,t)$ decays to zero. The large time behaviour can again be estimated using the stationary phase approximation, however we expect it to be dominated by the vicinity of $k'=|\xi|$.

{\bf Large time limit of $\psi_{k,1,\pm}^{R/L}$:}
For fixed $x= O(1)$, we have seen before that the term $e^{-i\frac{k'^2}{2}t} \cos(k'(|x|-\delta_{k'}^{\pm}))$ in $F^{R/L}_\pm (k,k',x)$ in the first 
integral in \eqref{deltapsinew} decays on $\Gamma_-$ for large time since $k$ has a negative imaginary part. In the present ray regime 
\eqref{xray} this term, $\cos(k'(|x|-\delta_{k'}^{\pm}))=\cos(k'(|\frac{x}{t} |t-\delta_{k'}^{\pm}))$, depends on time and $\psi_{k,1,+}^{R/L}(x,t)\pm {\rm sgn}(x)\psi_{k,1,-}^{R/L}(x,t)$ does not decay anymore on $\Gamma_-$
(and can even diverge). As shown in Fig. \ref{ContourXi}, we will deform the integration contour $\Gamma_-$ into a contour $\Gamma_-'$ such that the integral over  $\Gamma_-'$ decays in the large time limit. While deforming the integration contour over $k'$ we will eventually cross a pole at $k'=k$ producing an additional contribution to the infinite time limit of $\psi_{k,1,\pm}^{R/L}(x,t)$.
\begin{figure}
    \centering
    \includegraphics[width=\linewidth]{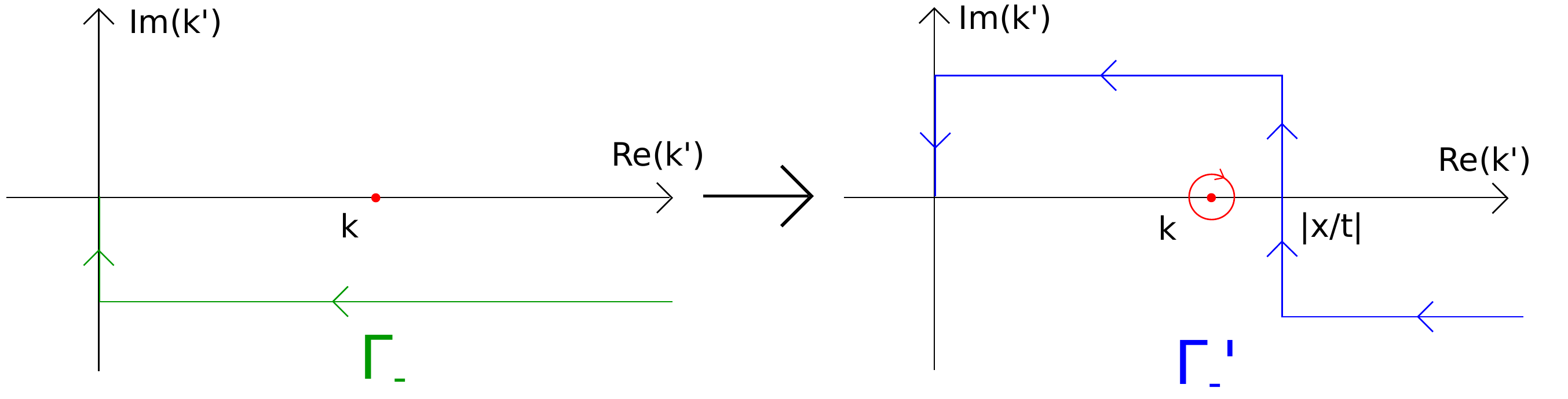}
    \caption{Illustration of the contours $\Gamma_-$ and $\Gamma_-'$ in the complex $k'$ plane. The contour $\Gamma_-$
    can be deformed into the union of the contour $\Gamma_-'$ and of the small clockwise red circle around $k'=k$. This circle is only present under the condition $k<|\frac{x}{t}|$.}
    \label{ContourXi}
\end{figure}
Let us work on $\psi_{k,1,+}^{R/L}$ (similar reasoning can be done for $\psi_{k,1,-}^{R/L}$). Writing the above cosine term as a sum of exponentials gives
\bea\label{cos}
e^{-i\frac{k'^2}{2}t}\cos(k'(|\frac{x}{t}|t-\delta_{k'}^{+}))=\frac{1}{2}e^{-i\frac{k'^2}{2}t}(e^{ik'(|\frac{x}{t}|t-\delta_{k'}^{+})}+e^{-ik'(|\frac{x}{t}|t-\delta_{k'}^{+})})
\eea

Then with $k'=k_1'+i k_2'$ (on $\Gamma_-$, $k_2'$ is negative but on $\Gamma_-'$ it will be either negative or positive). The term linear in $t$ in the two exponentials in \eqref{cos} have real part given by, respectively
\bea
Re[-i\frac{k'^2}{2}\pm ik'|\frac{x}{t}|]=k_2'(k_1' \mp |\frac{x}{t}|)
\eea
where $\pm$ corresponds to the first or second term of \eqref{cos}.
Because $k_2'(k_1' + |\frac{x}{t}|)<0$ for $k' \in \Gamma_-$, the second term of \eqref{cos},  $\frac{1}{2}e^{-i\frac{k'^2}{2}t}e^{-ik'(|\frac{x}{t}|t-\delta_{k'}^{+})}$ will always decay at infinite time. Therefore it has no large time contribution. On the contrary, $k_2'(k_1' - |\frac{x}{t}|)>0$ for $k' \in \Gamma_-$ and $k_1' < |\frac{x}{t}|$, hence the first term in \eqref{cos} $\frac{1}{2}e^{-i\frac{k'^2}{2}t}e^{ik'(|\frac{x}{t}|t-\delta_{k'}^{+})}$, diverges at large time. Hence $\Gamma_-$ must be deformed into $\Gamma_-'$ (see Fig. \ref{ContourXi}) constructed in such a way that 
that the product $k_2'(k_1' - |\frac{x}{t}|)$ remains negative. If during the deformation the contour crosses the pole 
of $\delta \chi_{k,1,+}^{R/L}(x,t)$ at $k'=k$, a residue part remains in the large time limit. On the other hand the contour integral over $\Gamma_-'$
vanishes at large time by similar arguments (i.e., the integrand decays exponentially except on the vertical axis with $k_1'=|\frac{x}{t}|$ and
near $k'=0$, but thanks to oscillating term the resulting decay is algebraic in time). 
Using \eqref{def_Fpm2}, the residue part gives the large time limit of \eqref{deltapsinew} in the ray regime as 
\bea
&\delta \chi_k^{R/L}(x= \xi t + y,t)%\underset{\substack{t\to\infty\\x\to\pm \infty \\ \frac{x}{t}\to\xi}}{\simeq}
\underset{t \to \infty}{\simeq} 2\pi i\frac{1}{\pi}2e^{-i\frac{k^2}{2}t}(\frac{e^{ik(|x|-\delta_{k}^{+})}}{2}\frac{\sin(k(\delta_k^+-\delta_k^{R/L}))}{2}\pm {\rm sgn}(x) \frac{e^{ik(|x|-\delta_{k}^{-})}}{2}\frac{\sin(k(\delta_k^--\delta_k^{R/L}))}{2})\theta(k<|\xi|)\\
&=e^{-i\frac{k^2}{2}t}e^{ik|x|}(e^{-ik\delta_k^{R/L}}\frac{1\pm {\rm sgn}(x)}{2}-e^{ik\delta_k^{R/L}}\frac{e^{-2ik\delta_k^+}\pm {\rm sgn}(x) e^{-2ik\delta_k^-}}{2}) \theta(k<|\xi|)\\
&=e^{-i\frac{k^2}{2}t}e^{ik|x|}(\frac{1\pm {\rm sgn}(x)}{2}(e^{-ik\delta_k^{R/L}}-\sr(k) e^{ik\delta_k^{R/L}})-\frac{1\mp {\rm sgn}(x)}{2}\st(k)e^{ik\delta_k^{R/L}}))\theta(k<|\xi|)
\eea
Here the factor $\theta(k<|\xi|)$ ensures a contribution only if we crossed the pole while deforming the contour.
In the last line we used \eqref{rtodelta}. In the end, since ${\rm sign}(x) = {\rm sign}(\xi)$ 
\bea
&\delta \chi_k^{R}(x=\xi t + y,t)\underset{t \to \infty}{\simeq} e^{-i\frac{k^2}{2}t}(\theta(x)\theta(k<\xi)e^{ikx}(e^{-ik\delta_k^{R}}-\sr(k) e^{ik\delta_k^{R}})-\theta(-x)\theta(k<-\xi)e^{-ikx}\st(k) e^{ik\delta_k^{R}}) \\
&\delta \chi_k^{L}(x=\xi t + y,t)\underset{t \to \infty}{\simeq}e^{-i\frac{k^2}{2}t}(\theta(-x)\theta(k<-\xi)e^{-ikx}(e^{-ik\delta_k^{L}}-\sr(k) e^{ik\delta_k^{L}})-\theta(x)\theta(k<\xi)e^{ikx}\st(k) e^{ik\delta_k^{L}}) 
\eea
These are the additional contributions to the large time limit of the wavefunctions which exist in the ray regime. It must be added
to the NESS contribution $e^{-i\frac{k^2}{2}t}\chi_k^{R/L}(x)$ obtained previously in \eqref{largeellpsi} and \eqref{chifinal}, leading to our final result in the ray regime (as given in the main text)
\bea
&\psi_k^{R/L}(x,t)\underset{\ell \to \infty}{\simeq} \frac{1}{\sqrt{\ell}} (e^{-i\frac{k^2}{2}t}\chi_{\xi,k}^{R/L}(x) + \delta \chi_{\xi,k}^{R/L}(x,t)) \\
&\underset{ t\to\infty}{\lim}\delta \chi_{\xi,k}^{R/L}(x=\xi t + y,t) = 0
\eea
To write the leading term at large time we have defined 
% \bea
% &\chi_{\xi,k}^{R}(x)=\theta(x)(e^{-ikx}e^{ik\delta_k^{R}}+\theta(k<\xi)e^{ikx}e^{-ik\delta_k^{R}})\\
% &+(\theta(x)\theta(k>\xi)\sr(k) e^{ikx}+\theta(-x)\theta(k>-\xi)\st(k) e^{-ikx}) e^{ik\delta_k^{R}} \\
% &\chi_{\xi,k}^{L}(x)=\theta(-x)(e^{ikx}e^{ik\delta_k^{L}}+\theta(k<-\xi)e^{-ikx}e^{-ik\delta_k^{L}})\\
% &+(\theta(-x)\theta(k>-\xi)\sr(k) e^{-ikx}+\theta(x)\theta(k>\xi)\st(k) e^{ikx}) e^{ik\delta_k^{L}} 
% \eea
% \bea
% &\chi_{\xi,k}^{R}(x)=\theta(-x)\theta(k>-\xi)\st(k) e^{-ikx} e^{ik\delta_k^{R}}\\
% &+\theta(x)(e^{-ikx}e^{ik\delta_k^{R}}+\theta(k<\xi)e^{ikx}e^{-ik\delta_k^{R}}+\theta(k>\xi)\sr(k) e^{ikx}e^{ik\delta_k^{R}})\\
% &\chi_{\xi,k}^{L}(x)=\theta(x)\theta(k>\xi)\st(k) e^{ikx} e^{ik\delta_k^{L}} \\
% &+\theta(-x)(\theta(k>-\xi)\sr(k) e^{-ikx}e^{ik\delta_k^{L}}+  e^{ikx}e^{ik\delta_k^{L}}+\theta(k<-\xi)e^{-ikx}e^{-ik\delta_k^{L}})
% \eea
\bea \label{chixi} 
&\chi_{\xi,k}^{R}(x)=\theta(-x)\theta(k>-\xi)\st(k) e^{ik\delta_k^{R}} e^{-ikx}+\theta(x)\left(e^{ik\delta_k^{R}}e^{-ikx}+(\theta(k<\xi)e^{-ik\delta_k^{R}}+\theta(k>\xi)\sr(k) e^{ik\delta_k^{R}})e^{ikx} \right)  \\
&\chi_{\xi,k}^{L}(x)=\theta(x)\theta(k>\xi)\st(k)  e^{ik\delta_k^{L}} e^{ikx} + \theta(-x)\left(e^{ik\delta_k^{L}} e^{ikx}+ (\theta(k<-\xi)e^{-ik\delta_k^{L}}+ \theta(k>-\xi)\sr(k) e^{ik\delta_k^{L}})e^{-ikx}\right)  
\eea
Note that the second line can be obtained from the first by the change $(x, \xi, R) \to (-x,-\xi,L)$, which is indeed a symmetry of the problem.
In the equations \eqref{chixi} the argument should be understood as $x = \xi t + y$. Hence there are fast oscillations in time
with factors $e^{\pm i k \xi t}$.
For completeness we also give the full expression of the part which decays in time (where here $\xi=x/t$) 
\bea
&\delta \chi_{\xi,k}^{R/L}(x,t)=-\int_{\Gamma_-} \frac{dk'}{\pi} \frac{2^{ -3/2}}{k-k'} e^{-i\frac{k'^2}{2}t} \left( e^{-ik'(x-\delta_{k'}^+)}\gamma_{k',k}^+\pm {\rm sgn}(x)  e^{ -ik'(x-\delta_{k'}^-)} \gamma_{k',k}^- \right)\\
&-\int_{\Gamma_-'} \frac{dk'}{\pi} \frac{2^{ -3/2}}{k-k'} e^{-i\frac{k'^2}{2}t} \left(  e^{ik'(x-\delta_{k'}^+)} \gamma_{k',k}^+ \pm {\rm sgn}(x) e^{ik'(x-\delta_{k'}^-)} \gamma_{k',k}^- \right)\nn\\
&+\int_0^{\infty}\frac{dk'}{2\pi}\sqrt{2}e^{-i\frac{k'^2}{2}t}\left(\cos(k'(|x|-\delta_{k'}^{+}))M^{R/L}(\phi_k^{R/L},\phi_{+,k'},a)\right.\nn\\
&\left.+{\rm sgn}(x)\cos(k'(|x|-\delta_{k'}^{-}))M^{R/L}(\phi_k^{R/L},\phi_{-,k'},a)\right)\nn
\eea
Where $\gamma_{k',k}^\pm=2^{3/2}\frac{F^{R/L}_{\pm}(k,k',x)}{\cos(k'(x-\delta_{k'}^\pm))}=2^{3/2}\frac{k'\sin(k'(\frac{a}{2}-\delta_{k'}^{\pm}))\cos(k(\frac{a}{2}-\delta_{k}^{R/L}))-k\sin(k(\frac{a}{2}-\delta_k^{R/L}))\cos(k'(\frac{a}{2}-\delta_{k'}^\pm))}{k+k'}$

\subsection{Kernel computation in the ray regime}
First we rewrite \eqref{largelK} in a slightly different manner adapted to the ray regime
\bea
\underset{\ell\to\infty}{\lim}K_{R/L}(x,t;x',t')&=\int_0^{\infty}\frac{dk}{2\pi}\left( f_{R/L}(k)-\theta(t'-t)\right)(e^{-i\frac{k^2}{2}t}\chi_{\xi,k}^{R/L}(x)+\delta \chi_{\xi,k}^{R/L}(x,t))^*\\
&\times(e^{-i\frac{k^2}{2}t'}\chi_{\xi',k}^{R/L}(x')+\delta \chi_{\xi',k}^{R/L}(x',t'))\nn\\
&=K_{{\rm ray},R/L}(x,t;x',t') + \delta K_{{\rm ray},R/L}(x,t;x',t')
\eea
where we have defined
\bea \label{Kraydef} 
K_{{\rm ray},R/L}(x,t;x',t')=\int_{0}^{\infty} \frac{dk}{2\pi}\left( f_{R/L}(k)-\theta(t'-t) \right) e^{i\frac{k^2}{2}(t-t')}\chi_{\xi,k}^{R/L}(x)^*\chi_{\xi',k}^{R/L}(x')
\eea
and $\delta K_{{\rm ray},R/L}$ is what remains in expanding the sums. 
% \bea
% \delta K_{{\rm ray},R/L}(x,t;x',t')=\int_{0}^{\infty} \frac{dk}{2\pi}\left( f_{R/L}(k)-\theta(t'-t) \right)\left((\delta \chi_{\xi,k}^{R/L}(x,t))^*\delta \chi_{\xi',k}^{R/L}(x',t')\right.\\
% &\left. + \; (e^{-i\frac{k^2}{2}t}\chi_{\xi,k}^{R/L}(x))^*\delta \chi_{\xi',k}^{R/L}(x',t')+(\delta \chi_{\xi,k}^{R/L}(x,t))^*e^{-i\frac{k^2}{2}(t+\tau)}\chi_{\xi',k}^{R/L}(x')\right)
% \eea
%Where $K_{{\rm ray},R/L}$ and $\delta K_{{\rm ray},R/L}$ are just $K_{\infty,R/L}$ and $\delta K_{R/L}$ where we replaced $\chi_{k}^{R/L}(x)$ by $\chi_{\xi,k}^{R/L}(x)$ and $\delta \chi_{k}^{R/L}(x,t)$ by $\delta \chi_{\xi,k}^{R/L}(x,t)$.

Let us now consider the large time limit $t \to +\infty$ with 
\be 
t' = t - \tau \quad , \quad x= \xi t + y \quad , \quad x'= \xi t' + y'  \quad, \quad \tau, y,y'=O(1)
\ee 
The first property is that 
\bea
\underset{\substack{t\to\infty\\x/t\to \xi\\x'/t \to \xi'}}{\lim}\delta K_{{\rm ray},R/L}(x,t;x',t-\tau)=0 \;.
\eea
This is shown by proceeding similarly as in Section \ref{sectionKernelNess}. 
%in order to go to the limit $\left\{\substack{t\to\infty\\x/t\to \xi\\x'/t \to \xi'}\right.$ 
Special care has to be taken of each time dependent exponential that $\delta K_{{\rm ray},R/L}$ contains. Additionally we have to give special attention to the vertical part in the contour $\Gamma_-'$ at $Re[k']=|\frac{x}{t}|$. This kind of integral was already studied in \cite{GLS} and found to decay algebraically in time.

We now inject the expressions for $\chi_{\xi,k}^{R/L}(x)$ given in \eqref{chixi} in the definition of $K_{{\rm ray},R/L}$ in
\eqref{Kraydef}. Upon expanding the product $\chi_{\xi,k}^{R/L}(x)^*\chi_{\xi,k}^{R/L}(x')$, all terms contain a factor $e^{ik(\pm x \pm x')}=e^{ik(\pm \xi \pm \xi')t}e^{ik(\pm y \pm y')}$. If $\pm \xi \pm \xi'\neq 0$, in the large time limit the corresponding term will give a zero contribution because of oscillatory behaviour of the integral over $k$. However if $\xi'= \pm \xi$, the contribution of this term is non zero, and the factor simplifies as $e^{ik(\pm x \pm x')}=e^{ik(\pm y \pm y')}$. This is how we obtain the final result \eqref{Kxi1} given in the main text for the total kernel along identical rays and opposite rays, respectively
\bea
&K^+_\xi(y,y',\tau) = \lim_{t \to +\infty}  K_{{\rm ray},R}(\xi t + y ,\xi t + y',t-\tau)
+  K_{{\rm ray},L}(\xi t + y ,\xi t + y',t-\tau) \\
&K^-_\xi(y,y',\tau) = \lim_{t \to +\infty}  K_{{\rm ray},R}(\xi t + y ,-\xi t + y',t-\tau)
+  K_{{\rm ray},L}(\xi t + y ,-\xi t + y',t-\tau) 
\eea
In the calculation one uses that ${\sf t}(k) {\sf r}(k)^* + {\sf t}(k)^* {\sf r}(k)=0$ and $|{\sf r}(k)|^2 + |{\sf t}(k)|^2=1$. 
Note that the symmetry of the system  $K(x,t,x',t')|_{L,R}=K(-x,t,-x',t')|_{R,L}$ implies the following relations 
\be\label{symray}
K^+_{-\xi}(y,y',\tau)|_{L,R} = K^+_\xi(-y,-y',\tau)|_{R,L} \quad , \quad K^-_{-\xi}(y,y',\tau)|_{L,R} = K^-_\xi(-y,-y',\tau)|_{R,L}
\ee

\section{Large time wave function in the presence of bound states}
\label{sec:bound} 

In this section we allow for possible bound states for both potentials $V_0(x)$ and $V(x)$.
We assume that $V_0(x)$ may have a discrete set of bound states $\Lambda_b^{R/L}$ with eigenenergies $E_0=- \frac{\kappa^2}{2}$ 
(with $\kappa>0$). We denote their corresponding eigenfunctions by $\phi^{R/L}_\kappa(x)$ 
which decay exponentially far from the defect as $\phi^R_\kappa(x) \propto e^{- \kappa x}$ as ${x \to +\infty}$
and $\phi^L_\kappa(x) \propto e^{- \kappa |x|}$ as ${x \to -\infty}$. Similarly we assume 
that $V(x)$ may have a discrete set of bound states $\Lambda_b$ with eigenenergies $E=- \frac{(\kappa')^2}{2}$ 
(with $\kappa'>0$). We denote their corresponding eigenfunctions by $\phi_{\kappa'}(x)$ 
which decay exponentially far away from the defect as $\phi_{\kappa'}(x) \propto e^{- \kappa' x}$ as $|x| \to +\infty$.
We use the same notations as before for the states with positive energies. 
By extension we also denote $\psi^{R/L}_\kappa(x,t)$ the evolution under $H$ of $\psi^{R/L}_\kappa(x,t=0)=\phi_\kappa^{R/L}(x)$.

{\bf Overlaps.} To study the large $\ell$ and large time limit of 
$\psi^{R/L}_{k}(x,t)$ and $\psi^{R/L}_{\kappa}(x,t)$ we need to make the following natural assumptions 
about the overlaps which take the form
\bea \label{overlaps} 
&\braket{\phi_{k'}|\phi_k^{R/L}}=\frac{\gamma^{R/L}_{k',k}}{\ell(k-k')} \quad , \quad \braket{\phi_{\kappa'}|\phi_\kappa^{R/L}}=\gamma^{R/L}_{\kappa',\kappa}\\
&\braket{\phi_{\kappa'}|\phi_k^{R/L}}=\frac{\gamma^{R/L}_{\kappa',k}}{\sqrt{\ell}} \quad , \quad \braket{\phi_{k'}|\phi_\kappa^{R/L}}=\frac{\gamma^{R/L}_{k',\kappa}}{\sqrt{\ell}} \nonumber 
\eea
where the $\gamma$ factors (with a slight abuse of notation in their arguments) remain finite in the large $\ell$ limit.
Furthermore we assume that they are smooth functions of $k' \in \mathbb{R}^+$. 
Note that for finite $\ell$ these overlaps are all smaller than unity in absolute value (since the eigenstates 
are normalized). In order to understand the $\ell$ factors in \eqref{overlaps} let us recall that for a scattering
state the normaliazation constant is $O(1/\sqrt{\ell})$, whereas for a bound state it is $O(1)$. The first overlap $\braket{\phi_{k'}|\phi_k^{R/L}}$ between two scattering states was already discussed in detail in the previous sections. The second one $\braket{\phi_{\kappa'}|\phi_\kappa^{R/L}}$ is the overlap between two bound states and therefore is of order zero in $\ell$. The two last ones are overlaps between a bound state and a scattering state. 
They were considered in \cite{GLS} in the case of delta impurity (i.e., with $a=0$). Similarly for a square well potential of size $a$ it is easy to check
that $\braket{\phi_{\kappa}|\phi_k}
%\int_0^{a/2}dy\frac{1}{\sqrt{a}}\cos(\kappa y)\frac{1}{\sqrt{\ell}}\cos(k y)
\sim\sqrt{\frac{a}{\ell}}$. 

\subsection{Evolution of the initial eigenstates} 

We now proceed with the computation of $\psi^{R/L}_\kappa(x,t)$ by extending the decomposition \eqref{exact_psik} to including bound states.
We will separate the sum over eigenstates of $H$ into sums over bound and scattering states. We denote $\Lambda$ the lattice of all eigenstates of $\hat H$ for a finite size system. We split it between bound states $\Lambda_b$ and scattering states $\Lambda_s$ (i.e., $\Lambda_s = \Lambda^+ \cup \Lambda^- $) such that $\Lambda = \Lambda_b \cup \Lambda_s$. In the large system size limit $\Lambda_s$ becomes continuous (and equal to $\mathbb{R}^+$) whereas $\Lambda_b$ remains discrete
and finite. We use the same separation for the lattice of eigenstates of $\hat H_0$, $\Lambda^{R/L} = \Lambda_b^{R/L} \cup \Lambda_s^{R/L}$.\\
The initial state can be either a scattering state, in which case
\bea \label{scattering1} 
&\psi^{R/L}_k(x,t)=\sum_{\kappa'\in \Lambda_b }\phi_{\kappa'}(x)e^{i\frac{\kappa'^2}{2}t}\braket{\phi_{\kappa'}|\phi_k^{R/L}} + \sum_{k'\in\Lambda_s}\phi_{k'}(x)e^{-i\frac{k'^2}{2}t}\braket{\phi_{k'}|\phi_k^{R/L}}\\
&=\sum_{\kappa'\in\Lambda_b}\phi_{\kappa'}(x)e^{i\frac{\kappa'^2}{2}t}\frac{\gamma_{\kappa',k}^{R/L}}{\sqrt{\ell}} + \sum_{k'\in\Lambda_s}\phi_{k'}(x)e^{-i\frac{k'^2}{2}t}\frac{\gamma_{k',k}^{R/L}}{\ell(k-k')} \nonumber 
\eea
or a bound state, and one has
\bea \label{bound1} 
&\psi_\kappa^{R/L}(x,t)=\sum_{\kappa'\in\Lambda_b}\phi_{\kappa'}(x)e^{i\frac{\kappa'^2}{2}t}\braket{\phi_{\kappa'}|\phi_\kappa^{R/L}} + \sum_{k'\in\Lambda_s}\phi_{k'}(x)e^{-i\frac{k'^2}{2}t}\braket{\phi_{k'}|\phi_\kappa^{R/L}}\\
&=\sum_{\kappa'\in\Lambda_b}\phi_{\kappa'}(x)e^{i\frac{\kappa'^2}{2}t}\gamma_{\kappa',\kappa}^{R/L} + \sum_{k'\in\Lambda_s}\phi_{k'}(x)e^{-i\frac{k'^2}{2}t}\frac{\gamma_{k',\kappa}^{R/L}}{\sqrt{\ell}} \nonumber 
\eea
%{\bf Remark:} For $|x|>a/2$, the bound states $\phi_\kappa$ and $\psi_\kappa$ exhibit an exponential decay $\sim e^{-\kappa |x|}$. 

We should note on the equations \eqref{scattering1} and \eqref{bound1} that all terms contribute in the large $\ell$ limit. Indeed in that limit one replaces
$\frac{1}{\ell} \sum_{k' \in \Lambda_s} \to 2 \int_{\mathbb{R}^+} \frac{dk'}{2 \pi}$,
and takes into account that $\phi_k$ contains a factor $1/\sqrt{\ell}$ while $\phi_\kappa=O(1)$ and that
$\psi_k = O(\frac{1}{\sqrt{\ell}})$ while $\psi_\kappa = O(1)$.

Furthermore, in that large $\ell$ limit, one can decompose the evolution of the initial scattering states as 
\bea
&\psi^{R/L}_k(x,t)\underset{\ell\to\infty}{\simeq}\frac{1}{\sqrt{\ell}}\left(\tilde{\chi}_{k}^{R/L}(x,t)+\delta \chi_k^{R/L}(x,t)\right)\\
&\tilde{\chi}_{k}^{R/L}(x,t)=e^{-i\frac{k^2}{2}t}\chi_k^{R/L}(x)+\sum_{\kappa'\in\Lambda_b}\phi_{\kappa'}(x)e^{i\frac{\kappa'^2}{2}t}\gamma_{\kappa',k}^{R/L}
\eea
where $\chi_k^{R/L}(x)$ and $\delta \chi_k^{R/L}(x,t)$ are given by the same formula as in the absence of bound states. In addition, 
we recall that $\underset{t\to\infty}{\lim}\delta \chi_k^{R/L}(x,t)=0$. Hence, in the large time limit one has
\be 
\underset{t\to\infty}{\lim}\psi^{R/L}_k(x,t)=\tilde{\chi}_{k}^{R/L}(x,t)
\ee 
As compared to the results of the previous section, it contains additional terms due to the bound states of $V(x)$, which leads to oscillations
in time. In addition the amplitudes depend on the details of the initial potential $V_0(x)$, hence there is some memory of the initial state.

On the other hand, in the evolution of an initial bound state, the second term in \eqref{bound1} decays to zero
at large time (from an oscillating integral) hence it admits the large time limit
\be
\psi^{R/L}_\kappa(x,t) \underset{ \substack{\ell\to\infty\\t\to\infty} }{\simeq} \sum_{\kappa'\in\Lambda_b} \phi_{\kappa'}(x) e^{i\frac{\kappa'^2}{2}t} \gamma_{\kappa',\kappa}^{R/L} 
\ee
which does not depend on the scattering states at all, and also carries some memory of the initial state (and potential $V_0(x)$).

\subsection{Large time limit of the kernel} 

Let us now study the two time kernel. From its definition \eqref{timedepkernel} it can be separated in two parts, i.e., the contribution of the
initial scattering states and the contribution of the initial bound states, as
\bea
&&K_{R/L}(x,t;x',t')=K_{R/L}^{(1)}(x,t;x',t')+K_{R/L}^{(2)}(x,t;x',t') \\
&& K_{R/L}^{(1)}(x,t;x',t') = \sum_{\kappa\in\Lambda_b^{R/L}} \left( f_{R/L}(i\kappa) - \theta(t'-t)\right)\psi_\kappa(x,t)^* \psi_\kappa(x',t') \\
&& K_{R/L}^{(2)}(x,t;x',t') = \sum_{k\in\Lambda_s^{R/L}} \left( f_{R/L}(k) - \theta(t'-t)\right) \psi_k(x,t)^* \psi_k(x',t)
\eea
where we use the notation $f_{R/L}(i\kappa)= 1/(1+ e^{- \beta_{R/L} (\frac{\kappa^2}{2} + \mu_{R/L})})$ for the Fermi factor of the bound states. 
\\

Let us now perform the double limit $\ell \to +\infty$ followed by $t,t' \to +\infty$ on each part.
\\

{\bf Double limit of $K^{(1)}$ }. Let us insert the expression \eqref{bound1} into the formula for $K^{(1)}$. One finds

\bea
&&K_{R/L}^{(1)}(x,t;x',t')=\sum_{\kappa\in\Lambda_b^{R/L}} \left( f_{R/L}(i\kappa) - \theta(t'-t)\right) \left( \sum_{\kappa' \in \Lambda_b }\phi_{\kappa'}(x)e^{i\frac{\kappa'^2}{2}t}\gamma_{\kappa',\kappa}^{R/L} + \sum_{k' \in \Lambda_s } \phi_{k'}(x) e^{-i\frac{k'^2}{2}t} \frac{\gamma_{k',\kappa}^{R/L}}{\sqrt{\ell}} \right)^*\\
&&\times \left( \sum_{\kappa'' \in \Lambda_b } \phi_{\kappa''}(x') e^{i\frac{\kappa''^2}{2}t'} \gamma_{\kappa'',\kappa}^{R/L} + \sum_{k'' \in \Lambda_s } \phi_{k''}(x') e^{-i\frac{k''^2}{2}t'} \frac{\gamma_{k'',\kappa}^{R/L}}{\sqrt{\ell}} \right) \nonumber 
\eea
In the large $\ell$ limit the sums $\sum_{k'\in \Lambda_s}$ and $\sum_{k'' \in \Lambda_s}$ turn into integrals and upon expanding
all four terms are of the same order. The expression simplifies however in the large time limit. Indeed the integrals over scattering
states contain oscillating terms, such as $e^{-i\frac{k'^2}{2}t}$ and $e^{-i\frac{k''^2}{2}t'}$, and thus decay to zero 
in the large time limit. The only remaining part in the double limit is therefore the part containing only discrete sums, which reads
\bea
 K_{R/L}^{(1)}(x,t;x',t') \underset{\substack{\ell\to\infty\\t,t'\to\infty}}{\simeq} \sum_{ \kappa',\kappa'' \in \Lambda_b } \phi_{\kappa'}(x)  \phi_{\kappa''}(x') e^{-i(\frac{\kappa'^2}{2}t-\frac{\kappa''^2}{2}t')} \sum_{\kappa \in \Lambda_b }\left( f_{R/L}(i\kappa) - \theta(t'-t)\right) {\gamma_{\kappa',\kappa}^{R/L}}^* \gamma_{\kappa'',\kappa}^{R/L} 
\eea
Note that for this expression to hold we only need $t,t'$ both large, but $\tau=t'-t$ does not have to be $O(1)$, it can also be large.

{\bf Double limit of $K^{(2)}$} Let us insert the expression \eqref{scattering1} into the formula for $K^{(2)}$. One finds
\bea
&K_{R/L}^{(2)}(x,t;x',t')=\sum_{k \in \Lambda_s^{R/L}} \left( f_{R/L}(k) - \theta(t'-t)\right)  \left( \sum_{\kappa' \in \Lambda_b }\phi_{\kappa'}(x) e^{i\frac{\kappa'^2}{2}t} \frac{\gamma_{\kappa',k}^{R/L}}{\sqrt{\ell}} + \sum_{k'  \in \Lambda_s } \phi_{k'}(x) e^{-i\frac{k'^2}{2}t} \frac{\gamma_{k',k}^{R/L}}{\ell(k-k')} \right)^* \nn \\
&\times\left(\sum_{\kappa'' \in \Lambda_b }\phi_{\kappa''}(x') e^{i\frac{\kappa''^2}{2}t'}\frac{\gamma_{\kappa'',k}^{R/L}}{\sqrt{\ell}} + \sum_{k'' \in \Lambda_s }\phi_{k''}(x') e^{-i\frac{k''^2}{2}t'} \frac{\gamma_{k'',k}^{R/L}}{\ell(k-k'')} \right) 
\eea
In the Large $\ell$ limit as shown before the sums over scattering states $\sum_{k' \in \Lambda_s } \phi_{k'}(x) e^{-i\frac{k'^2}{2}t} \frac{c_{k',k}}{\ell(k-k')}$ converge to $\frac{1}{\sqrt{\ell}}(e^{-i\frac{k^2}{2}t}\chi_{k}^{R/L}(x)+\delta \chi_{k}^{R/L}(x,t))$ leading to % $\phi_{k}(x)c_{k,k}e^{-i\frac{k^2}{2}t}$
\bea
&\underset{\ell\to\infty}{\lim} K_{R/L}^{(2)}(x,t;x',t') =\int_{R/L,t'-t} \frac{dk}{2\pi}   \left(
\sum_{\kappa' \in \Lambda_b }\phi_{\kappa'}(x) e^{i\frac{\kappa'^2}{2}t} \gamma_{\kappa',k}^{R/L} 
+ e^{-i\frac{k^2}{2}t}\chi_{k}^{R/L}(x)
+\delta \chi_{k}^{R/L}(x,t) \right)^*\\
&\times\left(\sum_{\kappa'' \in \Lambda_b }\phi_{\kappa''}(x') e^{i\frac{\kappa''^2}{2}t'}\gamma_{\kappa'',k}^{R/L} 
+ e^{-i\frac{k^2}{2}t'}\chi_{k}^{R/L}(x')
+\delta \chi_{k}^{R/L}(x',t')\right) \nonumber 
\eea
Expanding the terms in parenthesis we find that only two terms survive in the large time limit. Indeed all the other cross terms
contain integrals with factors oscillating in time which decay to zero, or $\delta \chi_{k}^{R/L}$ terms which also decay to zero. Hence
in the large time limit $t, t' \to +\infty$ we are left with
% \bea
% &\underset{\ell\to\infty}{\lim} K_{R/L}^2(x,t;x',t') =\int_{R/L,t'-t} \frac{dk}{2\pi} \left((\sum_{\kappa' \in \Lambda_b }\phi_{\kappa'}(x) e^{i\frac{\kappa'^2}{2}t} \gamma^{R/L}_{\kappa',k})^* \sum_{\kappa'' \in \Lambda_b }\phi_{\kappa''}(x') e^{i\frac{\kappa''^2}{2}t'}\gamma^{R/L}_{\kappa'',k}\right. \\
% &+(\sum_{\kappa' \in \Lambda_b }\phi_{\kappa'}(x) e^{i\frac{\kappa'^2}{2}t} \gamma^{R/L}_{\kappa',k})^* e^{-i\frac{k^2}{2}t'}\chi_{k}^{R/L}(x') +
% (e^{-i\frac{k^2}{2}t}\chi_{k}^{R/L}(x))^* \sum_{\kappa'' \in \Lambda_b }\phi_{\kappa''}(x') e^{i\frac{\kappa''^2}{2}t'}\gamma^{R/L}_{\kappa'',k} \\
% &+ (\sum_{\kappa' \in \Lambda_b }\phi_{\kappa'}(x) e^{i\frac{\kappa'^2}{2}t} \gamma^{R/L}_{\kappa',k} )^* \delta \chi_{k}^{R/L}(x',t') +
% \delta \chi_{k}^{R/L}(x,t)^* \sum_{\kappa'' \in \Lambda_b }\phi_{\kappa''}(x') e^{i\frac{\kappa''^2}{2}t'}\gamma^{R/L}_{\kappa'',k} \\
% &  +\underbrace{\left. ( e^{-i\frac{k^2}{2}t}\chi_{k}^{R/L}(x)+\delta \chi_{k}^{R/L}(x,t))^*(e^{-i\frac{k^2}{2}t'}\chi_{k}^{R/L}(x')+\delta \chi_{k}^{R/L}(x',t')) \right)}_{K_{\infty,R/L}(x,t;x',t) + \delta K_{R/L}(x,t;x',t')} 
% \eea
\bea
&K_{R/L}^{(2)}(x,t;x',t')\underset{\substack{\ell\to\infty\\t,t'\to\infty}}{\simeq}K_{R/L}^s(x,t;x',t')\\
&+\sum_{\kappa',\kappa'' \in \Lambda_b }\phi_{\kappa'}(x)\phi_{\kappa''}(x')e^{-i(\frac{\kappa'^2}{2}t-\frac{\kappa''^2}{2}t')}\int_{0}^\infty \frac{dk}{2\pi}\left(f_{R/L}(k)-\theta(t'-t)\right){\gamma^{R/L}_{\kappa',k}}^*\gamma^{R/L}_{\kappa'',k} \nonumber 
\eea
where we denote $K_{R/L}^s(x,t;x',t')$ the kernel where all contributions of bound states are excluded, which has been computed in Section 
\ref{sectionKernelNess}. Note that $K_{R/L}^s(x,t;x',t')$ is non zero only if $t-t'=O(1)$.

Putting together the limits of $K^{(1)}$ and $K^{(2)}$ we finally obtain the large time limit of the kernel as $K=K_R + K_L$ with
\bea \label{largetimeKb} 
& K_{R/L}(x,t;x',t') \underset{\substack{\ell\to\infty\\t,t'\to\infty}}{\simeq}  \int_0^{\infty} \frac{dk}{2\pi}\left(f_{R/L}(k)-\theta(t'-t) \right)e^{-i\frac{k^2}{2}(t'-t)}{\chi_k^{R/L}}^*(x) \chi_k^{R/L}(x') \\
&+ \sum_{ \kappa',\kappa'' \in \Lambda_b } \phi_{\kappa'}(x)  \phi_{\kappa''}(x') e^{-i(\frac{\kappa'^2}{2}t-\frac{\kappa''^2}{2}t')} C_{\kappa',\kappa''}^{R/L}
\nonumber
\eea
where
\be \label{C} 
C_{\kappa',\kappa''}^{R/L}=\sum_{\kappa \in \Lambda_b^{R/L}}\left( f_{R/L}(i\kappa) - \theta(t'-t)\right)  {\gamma^{R/L}_{\kappa',\kappa}}^* \gamma^{R/L}_{\kappa'',\kappa}+ \int_0^{+\infty} \frac{dk}{2\pi} \left( f_{R/L}(k) - \theta(t'-t)\right)
\gamma^{R/L *}_{\kappa',k} \gamma^{R/L}_{\kappa'',k}
\ee
The kernel \eqref{largetimeKb} thus has an additional contribution from the bound states of $\hat{H}$, which is spatially localized in the vicinity of the impurity since 
the wave functions $\phi_\kappa(x) \sim e^{-\kappa|x|}$ decay exponentially away from the impurity. This contribution has oscillations in time independently in $t'$ and in $t$. If one sets $t'=t-\tau$ the time dependent exponential factor in the second part of \eqref{largetimeKb} 
can be rewritten as $e^{-i(\frac{\kappa'^2-\kappa''^2}{2}t+\frac{\kappa''^2}{2}\tau)}$, thus if $\hat H$ has more than one bound state 
the kernel at fixed $\tau$ exhibits oscillations in $t$ at large time. For example one can compute the large time density and current which differ from the case without bound state as follows (note that they exhibits oscillations if there is more than one bound state)
\bea
&&\rho(x,t) \underset{t \to \infty}{\simeq} \rho_\infty (x) +\delta \rho (x,t)\\
&&\delta \rho (x,t)=\sum_{ \kappa',\kappa'' \in \Lambda_b } \phi_{\kappa'}(x)  \phi_{\kappa''}(x) e^{-\frac{it}{2}(\kappa'^2-\kappa''^2)}C_{\kappa',\kappa''}^{R/L}\\
&&J(x,t) \underset{t \to \infty}{\simeq} J_\infty +\delta J (x,t)\\
&&\delta J (x,t)=\frac{1}{2i}\sum_{ \kappa',\kappa'' \in \Lambda_b } (\phi_{\kappa'}(x) \partial_x \phi_{\kappa''}(x) -\phi_{\kappa''}(x) \partial_x \phi_{\kappa'}(x)   )e^{-\frac{it}{2}(\kappa'^2-\kappa''^2)}C_{\kappa',\kappa''}^{R/L} \\
&& = \frac{{\rm sgn}(x)}{2i}\sum_{ \kappa',\kappa'' \in \Lambda_b }(\kappa'-\kappa'') \phi_{\kappa'}(x)  \phi_{\kappa''}(x) e^{-\frac{it}{2}(\kappa'^2-\kappa''^2)}C_{\kappa',\kappa''}^{R/L} \text{    if } |x|>\frac{a}{2}
\eea
Finally note that the coefficient $C_{\kappa',\kappa''}^{R/L}$ 
contains a memory of the details of the initial condition, in particular it depends on the overlap of the initial states of $\hat H_0$  (bound or scattering state), with the bound states of the propagating Hamiltonian $\hat H$. If $\hat H$ does not have bound states, $C_{\kappa',\kappa''}^{R/L}=0$.
%{\red this was written but i think it's false it depends on the overlap of the initial bound states of $V_0$ with the scattering states of $\hat H$. If $V_0$ does not have bound states, this second term is absent. }

Interestingly, one sees that in \eqref{C} the sum over the intermediate initial states $|\phi^{R/L}_{\kappa} \rangle$ and $|\phi^{R/L}_{k} \rangle$ 
in the product of overlap can be 
performed and one can rewrite the amplitude \eqref{C} as the following matrix element of the initial single-particule density matrix
\be 
C_{\kappa',\kappa''}^{R/L} = \langle \phi_{\kappa'} | \frac{1}{1 + e^{ \beta_{R/L} (\hat H_0^{R/L} - \mu_{R/L}) } } | \phi_{\kappa''} \rangle
\ee 
where $\hat H_0^{R/L}$ are the initial single particle Hamiltonian of the right and left part of the system, respectively.
The last term in \eqref{largetimeKb} is thus the exact two time kernel one would obtain if the single particle time evolution was described
by the projection of $\hat H$ on the linear subspace spanned by all bound states only. In general it oscillates with a set of frequencies 
$\frac{\kappa'^2-\kappa''^2}{2}$. However, if one considers its zero frequency component (e.g. upon time averaging) we see that
only the terms $\kappa''=\kappa'$ remain. In that case it equals the so called diagonal approximation and the 
$C_{\kappa',\kappa'}^{R/L}$ are interpreted as occupation numbers, which can be predicted from a restriction 
of the GGE to this subspace spanned by bound states (see e.g. 
\cite{DLSM2019} Eq. (40) and second column page 6). This is automatically the case when a single bound state is present,
e.g. for the delta impurity.
\\

{\bf Special examples}:  
In the particular case $V(x)=g\delta(x)$ ($g<0$ in order to have a bound state) and $V_0(x)=\underset{g\to\infty}{\lim}g\delta(x)$, there is one bound state for $\kappa'=-g$ and the overlap is
\be
\braket{\phi_{-g}|\phi_k^{R/L}}\underset{\ell\to\infty}{\simeq}\sqrt{-g}\sqrt{\frac{4}{\ell}}\frac{k}{k^2+g^2}
\ee
which leads to
\be
C_{-g,-g}^{R/L}=-2g\int_{0}^{\infty}\frac{dk}{\pi}\left( f_{R/L}(k)-\theta(t'-t) \right)\frac{k^2}{(k^2+g^2)^2}
\ee
and we recover the result for the kernel 
\be
K_b^{R/L}(x,t;x',t')=2g^2e^{g(|x|+|x'|)}e^{-i\frac{g^2}{2}(t-t')} \int_{0}^{\infty} \left( f_{R/L}(k)-\theta(t'-t) \right) \frac{dk}{\pi}\frac{k^2}{(k^2+g^2)^2} \;,
\ee
which coincides with the result obtained in \cite{GLS}. 
\\

\section{Wigner function in the NESS}

In this section we compute the Wigner function $W(x,p)$ associated to the NESS. At the end we compare it 
with the so-called semi-classical form for the Wigner function which was put forward in
\cite{Prosen2018} for a discrete version of the model and in \cite{GLS} for the present model.
It was argued there that the semi-classical form should be exact in the NESS far from the defect, i.e for $x \to +\infty$
(a semi-classical form also exists for the ray regime). This is correct as we confirm below, 
and the semi-classical Wigner function contains information about quantum correlations between two points $x,x'$ far
from the defect and with $x-x'=O(1)$. However there are other correlations at symmetric points $x \approx - x'$ far from 
the defect but with $x+x'=O(1)$. Here we show that they are encoded in an additional, singular
part of the Wigner function, which contains a $\delta(p)$ term which cannot be anticipated
from semi-classical arguments.

Let us recall that the Wigner function and the kernel are related by a Fourier transform formula which is
valid at all times. Here we consider it in the infinite time limit in the NESS and it reads
\be\label{Wdef}
W(x,p)=\int_{-\infty}^{+\infty} \frac{dy}{2\pi} e^{i p y} K_\infty(x + \frac{y}{2}, x - \frac{y}{2})
\ee
where we recall that 
\be
K_{R/L}(x,x')=\int_0^{+\infty} \frac{dk}{2\pi} f_{R/L}(k) \chi_k^{R/L,*}(x) \chi_k^{R/L}(x')
\ee
Using the expressions for $\chi_k^{R/L}(x)$ in \eqref{chiR}, \eqref{chiL}, after a tedious
calculation one finds 

\bea
W(x,p) &&= W_1(x,p) +W_2(x,p)\\ \label{exactW} 
W_1(x,p) &&=\int_0^{+\infty} \frac{dk}{2\pi}  f_R(k)  \left[ \theta(x) \left( \frac{\sin 2(p+k)x}{\pi(p+k)}+|\sr(k)|^2 \frac{\sin 2(p-k)x}{\pi(p-k)} + Re[\sr(k)e^{2ikx}]\frac{2\sin 2px}{\pi p} \right)  \right. \\
&& \left.  - \theta(-x) |\st (k)|^2 \frac{\sin 2(p+k)x}{\pi(p+k)}  +Re[\st(k)\frac{e^{2i|x|(p+k)}}{\pi} \frac{i}{(k+p)+i\epsilon}]  \right] \nn\\
&& + \int_0^{+\infty} \frac{dk}{2\pi}  f_L(k)  \left[ -\theta(-x) \left( \frac{\sin 2(p-k)x}{\pi(p-k)}+|\sr(k)|^2 \frac{\sin 2(p+k)x}{\pi(p+k)} + Re[\sr(k)e^{-2ikx}]\frac{2\sin 2px}{\pi p} \right)     \right. \nn \\
&&\left. +\theta(x) |\st (k)|^2 \frac{\sin 2(p-k)x}{\pi(p-k)} + Re[\st(k)\frac{e^{2i|x|(k-p)}}{\pi} \frac{i}{(k-p)+i\epsilon}] \right] \nn \\
 W_2(x,p) && = \int_0^{+\infty} \frac{dk}{2\pi} Re \left[ f_R(k) \sr^*(k) \st(k) \frac{e^{2i(|x|p-kx)}}{\pi} \frac{i}{p+i\epsilon}  - f_L(k) \sr(k) \st^*(k) \frac{e^{-2i(|x|p-kx)}}{\pi} \frac{i}{-p+i\epsilon} \right]
\eea
 This is the final and complete result for the Wigner in the NESS. We have assumed that the defect width $a\ll 1$ so we can neglect the contributions from the 
 region inside the defect (it is thus exact for the delta function impurity). However the previous formula remains correct for finite $a$ and $|x|>\frac{a}{2}$. 
 One can note that $W$ has the following symmetry property  $W(x,p)|_{R,L}=W(-x,-p)|_{L,R}$. This symmetry can be understood from the kernel symmetry previously highlighted $K_\infty (x,x')|_{R,L}=K_\infty (-x,-x')|_{L,R}$ and the Wigner function definition \eqref{Wdef}.
 
We can now analyze the main interesting features of this result. First, let us consider the limit $|x| \to +\infty$, i.e.,
far from the defect. Using the identity
\be  \label{identitysine} 
\lim_{|x| \to +\infty} \frac{\sin q x}{\pi q} = {\rm sgn}(x) \delta(q) 
\ee 
we find that $W_1(x,p)$ for $x>0$ converges to 
\be 
W_1(x,p) \to W_{\rm sc}(x,p) = \frac{1}{2 \pi} ((T(p) f_L(p) + R(p) f_R(p) ) \Theta(p) +  f_R(p) \Theta(-p)  ) 
\ee 
and for $x<0$ converges to 
\be 
W_1(x,p) \to W_{\rm sc}(x,p) =  \frac{1}{2 \pi} ((T(p) f_R(p) + R(p) f_L(p)  ) \Theta(-p) +  f_L(p) \Theta(p)  ) 
\ee 
which are exactly the semi-classical expressions $W_{\rm sc}(x,p)$ for the Wigner function \cite{GLS}. 
Note that the limit arises using the identity \eqref{identitysine} in the first, second, fourth, sixth, seventh, ninth
terms in \eqref{exactW}. The third and eighth term vanish in the limit. 
The fifth and tenth terms also go to zero in the $|x| \to \infty$ limit. Indeed the factor $e^{2i|x|k}$ allows to move the integration contour over $k$ 
in the upper half of the complex plane, while the poles $k= \pm p-i\epsilon $ are in the complex lower half plane.

Next, we note that in $W_2(x,p)$ there is an unusual $\delta(p)$ component, indeed, using the identity for real $q$
\be 
\frac{1}{q \mp i \epsilon} = P.V. \frac{1}{q} \pm i \pi \delta(q) 
\ee 
we obtain
\bea
W_2(x,p) && =\delta(p) \int_0^{+\infty} \frac{dk}{2\pi}   Re\left[f_R(k) \sr^*(k) \st(k) e^{2i(|x|p-kx)} - f_L(k) \sr(k) \st^*(k) e^{-2i(|x|p-kx)} \right]\\
&&- P.V. \frac{1}{\pi p} \int_0^{+\infty} \frac{dk}{2\pi} Im\left[f_R(k)\sr^*(k) \st(k) e^{2i(|x|p-kx)}
+f_L(k) \sr(k) \st^*(k) e^{-2i(|x|p-kx)} \right] 
\eea

This non semi-classical characteristic of the Wigner function is related to the fact that the
correlations between symmetric points $\pm z$ do not vanish, even when those are very far apart, i.e., one has
\be  \label{limits} 
\lim_{z \to \pm \infty} K_\infty(z + y , - z + y') = K_{0^\pm}^-(y,y') 
\ee 
where the $K_{0^\pm}^-(y,y')$ are also equal to the $\xi=0^\pm$ limits of the kernel $K_{\xi}^-(y,y')$ in the ray regime
given in \eqref{Kxi1} and \eqref{symray}, which are
\bea
&K_{0^+}^-(y,y')=\int_{L,R} \frac{dk}{2\pi} \sr(k) \st(k)^* e^{-ik(y+y')} \\
&K_{0^-}^-(y,y')=\int_{L,R} \frac{dk}{2\pi} \sr^*(k) \st(k) e^{ik(y+y')} 
\eea
recalling that $\sr(k) \st(k)^* = - \sr(k)^* \st(k)$. 

Let us now explain why the existence of the non zero limits \eqref{limits} imply quite generally that the Wigner function
$W(x,p)$ must have a $\delta(p)$ piece. To this aim let us come back to the integral in \eqref{Wdef}. Since the
integrand goes to constant values at $y = \pm \infty$ we see that the Fourier integral leads to
\be \label{WK} 
W(x,p) = \tilde W(x,p) + \delta(p) \frac{1}{2} ( K_{0^+}^-(x,x) + K_{0^-}^-(x,x) ) + P.V.\frac{i}{2 \pi p} ( K_{0^+}^-(x,x) e^{2i p |x|}  - K_{0^-}^-(x,x) e^{-2i p |x|})
\ee 
where $\tilde W(x,p)$ is a regular part. We have used that $\int dy e^{i p y} (K_+ \theta(y- 2 |x|) + K_- \theta(- 2|x| -y)) = \frac{K_+ e^{2i p |x|}}{\epsilon - i p} 
+ \frac{K_- e^{-2i p |x|}}{\epsilon + i p}$. One can now check that indeed the predicted coefficients of $\delta(p)$ and $1/p$ in \eqref{WK} agree with
the result of the calculation of $W(x,p)$ above. While it is easy to match the $\delta(p)$ part in the two formulas, matching the oscillating prefactors in 
the $1/p$ part require some care. 

In summary, although $W_2(x,p)$ tends to zero at large $|x|$ (since the coefficient of $\delta(p)$ is an oscillating integral), at finite $|x|$ it has a form which is intrinsically non semi-classical, which encodes the correlations between distant opposite points. 
This calculation is a bit reminiscent of the (much simpler) one for the Wigner function of a single particle in a half-space which
also exhibits a delta function part far from the wall (see Appendix A in \cite{WignerBenjamin2021}).

\section{Observables}

\subsection{Density-density correlations}

Here we recall the relation between the quantum correlation function of the density operator and the space-time kernel. We start with the case of 
equal-time correlations, and discuss later the multi-time case. %We focus on the two point correlation $m=2$. 

\subsubsection{Equal time correlations} 

For equal time it is easier to work in the Schr\"odinger representation and denote $\ket{\Psi(t)}$ the 
$N$ body wavefunction at time $t$ (respectively the density matrix $\hat D(t)$). Here $\ket{\Psi(t)}$
is a Slater determinant (respectively $\hat D(t)$ is Gaussian) since it is the case at initial time
and this property is preserved by the dynamics. Here we denote $\langle \dots \rangle = \langle \Psi(t) | \dots | \Psi(t) \rangle$ 
the averages in that $N$ body state, and we omit for simplicity the time dependence. 

The fermion positions $x_i$, $i=1,\dots,N$ are known to form a determinantal point process (DPP).
One defines its two point correlation function \cite{us_review_JPA} 
\be \label{R2} 
R_2(x,x') = \langle \sum_{i \neq j} \delta(x-x_i) \delta(x'-x_j) \rangle =
\langle \hat \rho(x) \hat \rho(x') \rangle - \rho(x) \delta(x-x')  
\ee 
with $\rho(x)=\langle \hat \rho(x) \rangle$ and $\hat \rho(x)$ is the density operator. The main property
of a DPP is that 
\be \label{R2det} 
\rho(x)= K(x,x) \quad , \quad 
R_2(x,y)= K(x,x) K(y,y)- K(x,y)^2 
\ee 
where $K(x,x')\equiv K(x,t,x',t)$ is the equal time kernel. Hence the connected 
density-density correlation function is related to the kernel by 
\be 
\langle \hat \rho(x) \hat \rho(x') \rangle^c  = - K(x,x')^2 +  \rho(x) \delta(x-x')   
\ee 
One can check this formula using the standard second quantization anti-commuting operators $c_x^\dagger$ and 
$c_x$, with $\{c_x,c_{x'}\}=\{c_x^\dagger,c_{x'}^\dagger\} =0$ and 
$\{c^\dagger_x,c_{x'}\}=\delta(x-x')$. One has
\be  \label{rhorho1} 
\langle \hat \rho(x) \hat \rho(x') \rangle = \langle c_x^\dagger c_x  c_{x'}^\dagger c_{x'} \rangle 
= \langle c_x^\dagger c_x \rangle \delta(x-x') + \langle c_{x'}^\dagger c_x^\dagger c_x c_{x'} \rangle 
\ee 
Let us examine for simplicity the case of two fermions $N=2$, with energies $\epsilon_1$, $\epsilon_2$, and the state $|\Psi(t) \rangle = |\epsilon_1 \epsilon_2 \rangle = c^\dagger_{\epsilon_1} c^\dagger_{\epsilon_2} |0 \rangle$. Its bra is noted $\langle \epsilon_2 \epsilon_1 | = \langle 0 | c_{\epsilon_2} c_{\epsilon_1}$. 
One has 
\be
\langle c_{x'}^\dagger c_x^\dagger c_x c_{x'} \rangle = | c_x c_{x'}  c^\dagger_{\epsilon_1} c^\dagger_{\epsilon_2} |0 \rangle|^2 
\ee 
which is the square of the Slater determinant hence also equal to the $2 \times 2$ determinant of the
kernel. Thus \eqref{rhorho1} is indeed consistent with \eqref{R2} and \eqref{R2det}. 

\subsubsection{Multi-time correlations}

% This could be consistent with saying that 
% \be 
% R_2(x,y) = K(x,x) K(y,y)- K(x,y)^2 = - \langle c_x^\dagger c_y^\dagger c_x c_y \rangle 
% =  \langle c_x^\dagger c_x  c_y^\dagger c_y \rangle - \langle c_x^\dagger c_x \rangle \delta(x-y) 
% \ee 
% Indeed one has, in the particular case of a 2 body state 
% \be 
% - \langle c_x^\dagger c_y^\dagger c_x c_y \rangle =  \langle c_y^\dagger c_x^\dagger c_x c_y \rangle
% = | c_x c_y  c^\dagger_{k_1} c^\dagger_{k_2} |0 \rangle|^2 
% \ee 
% which is the square of the Slater determinant hence the $2 \times 2$ determinant of the
% kernel. One uses that $c^\dagger_k = \sum_{x'} \phi_k(x') c^\dagger_{x'}$ which leads to
% $\{ c^\dagger_k , c_x \}= \phi_k(x)$.

% One has $K(x,t;x',t') = {\rm Tr} \hat D T c^\dagger_{x,t}  c_{x',t'} $
% where $T c^\dagger_{x,t}  c_{x',t'} = c^\dagger_{x,t}  c_{x',t'} \theta(t-t') - c_{x',t'}  c^\dagger_{x,t}  \theta(t'-t) $. 

Similarly there is a formula which gives the multi-time/multi-space $m$ point quantum correlations of the density operator. 
We are interested in the (time ordered) quantum averages 
\be  \label{RmtoDens} 
{\cal C}_m(x_1,t_1;...;x_m,t_m) = \langle \hat \rho(x_m,t_m) \ldots \hat \rho(x_1,t_1) \rangle \quad , \quad t_1 \leq t_2 \dots \leq t_m
\ee 
Here $\langle \dots \rangle = {\rm Tr}( \hat D  \dots )$ denotes the quantum average, 
where $\hat D$ is the density matrix which characterizes the initial state, assumed to be Gaussian. 
The $\hat \rho(x,t)$ are the density operators and we use now (as in the main text) the Heisenberg representation
$\hat \rho(x,t) = e^{i {\cal H} t} \hat \rho(x) e^{- i {\cal H} t}$ where ${\cal H}$ is the many body Hamiltonian.

For non-interacting fermions, it can be related to the multi space-time point
correlation function of an extended determinantal point process, and when all space-time points $(x_i,t_i)$, $i=1,\dots,m$ are distinct, 
the formula reads, see e.g \cite{us_eq_dyn} (Section VIII C) and \cite{DLSM2019arXiv} (see Section VI) 
\be \label{Rdet} 
{\cal C}_m(x_1,t_1;...;x_m,t_m)= \det_{1 \leq i,j\leq m} K(x_i, t_i; x_j, t_j) \;,
\ee 
where the explicit expression for the extended kernel is given in the text, see \eqref{initial_kernel}, \eqref{timedepkernel}. 
It is also defined in second quantization notation as 
\be 
K(x,t ; x', t') = \langle T c^\dagger_{x,t} c_{x',t'} \rangle 
\ee 
where we recall the definition of the time ordering
\be 
T c^\dagger_{x,t}  c_{x',t'} = c^\dagger_{x,t}  c_{x',t'} \theta(t \geq t') - c_{x',t'}  c^\dagger_{x,t}  \theta(t<t')
= c^\dagger_{x,t}  c_{x',t'} - \{ c_{x',t'} ,  c^\dagger_{x,t} \} \theta(t<t')
\ee
\\

{\bf Remark about ordering}. Although the r.h.s. in \eqref{Rdet} is symmetric under exchanges of the space time points,
it is not the case for \eqref{RmtoDens}. The order of the time matters. Indeed one has
\be 
\{ c^\dagger_{x,t} , c_{x',t'} \} = \langle x' | e^{- i H (t'-t) } | x \rangle 
\ee 
which leads to
\be \label{commut} 
[ c^\dagger_{x,t} c_{x,t}, c^\dagger_{x',t'} c_{x',t'} ] =  \langle x | e^{- i H (t-t') } | x' \rangle \langle c^\dagger_{xt} c_{x't'} \rangle 
- \langle x' | e^{- i H (t'-t) } | x \rangle \langle c^\dagger_{x't'} c_{x t} \rangle  \;.
\ee

\subsubsection{Response function}

The commutator of the density $\hat \rho(x,t)= c^\dagger_{xt} c_{xt}$ in \eqref{commut} is related to the response function of the system, $\chi(x,t;x',t')$,
to a perturbation of the chemical potential (equivalently of the external potential). 
Consider an infinitesimal perturbation of the many body evolution Hamiltonian $\hat {\cal H}$ of the form
$\delta \hat {\cal H}(t) =\int dx \hat \rho(x,t) f(x,t)$. Then, to linear order in~$f$
\be  \label{lin} 
\delta \langle \hat \rho(x,t) \rangle= \int dx' dt' f(x',t') \chi(x,t;x',t') \quad , \quad 
\chi(x,t;x',t') = - i \theta(t-t') \langle [\hat \rho(x,t) , \hat \rho(x',t')] \rangle
\ee 
Let us recall the main idea of this standard result of linear response theory. Writing
the new Hamiltonian as $\hat {\cal H} + \delta \hat {\cal H}(t)$ one easily checks that for any many body state $|\Psi(t)\rangle = e^{-i \hat {\cal H} t} U(t) |\Psi(0)\rangle $
where $U(t) = T e^{- i \int_{-\infty}^t \delta {\cal H}_{\cal H}(t')dt'}$, where $\delta {\cal H}_{\cal H}(t')$ is the perturbation Hamiltonian
in the Heisenberg representation of ${\cal H}$, i.e., $\delta {\cal H}_{\cal H}(t)= e^{i \hat {\cal H} t}  \delta {\cal H}(t) e^{-i \hat {\cal H} t}$.
Then one obtains to linear order in the perturbation (indicating explicitly each Heisenberg representation) 
$\delta \langle \hat \rho(x,t) \rangle := \langle \hat \rho_{{\cal H} + \delta {\cal H}} (x,t) \rangle 
- \langle \hat \rho_{{\cal H}}(x,t) \rangle =
\langle U(t)^{-1} \hat \rho_{{\cal H}}(x,t) U(t) \rangle - \langle \hat \rho_{{\cal H}}(x,t) \rangle=
i \int_{-\infty}^t dt' \langle [\delta {\cal H}_{\cal H}(t) , \hat \rho_{\cal H}(x,t)]\rangle $, which leads
to \eqref{lin}. Here using \eqref{commut} we obtain 
\be \label{response}
\chi(x,t;x',t') = - i \theta(t-t')\left[
{\rm sgn}(t'-t)\Big |\langle x | e^{- i H(t-t')}|x' \rangle \Big |^2 + 2 i \, {\rm Im}\left(\langle x | e^{-i H(t-t')} |x'\rangle  K(x,t;x',t')\right) \right] \;.
\ee 
which is valid for any non-interacting fermion system. 

In the large time limit $t,t'\to\infty$, $\tau=t'-t\simeq O(1)$, and in the absence of bound states, the response function becomes only a function of $\tau$ and
\be
\lim_{t \to +\infty} \chi(x,t;x',t-\tau) = \chi_\infty (x,x',\tau)=  - \theta(\tau)\left[
- i {\rm sgn}(\tau)\Big |\langle x | e^{- i H \tau}|x' \rangle \Big |^2 + 2  \, {\rm Im}\left(\langle x | e^{-i H \tau} |x'\rangle  K_\infty(x,x',\tau)\right) \right] \;.
\ee 
where $K_\infty(x,x',\tau)$ is the kernel in the NESS from \eqref{Ktau}, given explicitly in \eqref{kernel_extended_NESS}, and %the propagator is given by
\be
\langle x | e^{- i H \tau}|x' \rangle = \int_0^\infty \frac{dk}{\pi}   \Phi_{+,k}^*(x)  \Phi_{+,k}(x') e^{- i \frac{k^2}{2} \tau} +  \int_0^\infty \frac{dk}{\pi}   \Phi_{-,k}^*(x)  \Phi_{-,k}(x') e^{- i \frac{k^2}{2} \tau} \;,
\ee
with $\Phi_{\pm,k}^*(x) = \lim_{\ell \to \infty} \frac{\phi_{\pm,k}(x)}{c_{\ell,\pm,k}}$ is the continuous basis, which is the limit of the discrete basis of the eigenfunctions \eqref{phi} of $\hat H$.

\subsection{Current correlation formula}

Finally, for completeness let us present a formula which allows to compute the correlation functions of the current. First we recall the formula for the current as a function of the current operator
\bea
J_{x,t}=\left< \hat J_{x,t} \right> \quad, \quad \hat J_{x,t}=\frac{1}{2i}(\partial_{x'} - \partial_x)\left[ c_{x,t}^\dagger c_{x',t}\right]_{x'=x}
\eea

 It reads for $t_1 \leq t_2 \dots \leq t_m$ and all space-time points distinct
\bea
\left<\hat J_{x_m,t_m} ... \hat J_{x_1,t_1}\right>&=\left(\frac{1}{2i}\right)^m\prod_{i=1}^m(\partial_{y_i}-\partial_{x_i})\left<c_{x_m,t_m}^\dagger c_{y_m,t_m}...c_{x_1,t_1}^\dagger c_{y_1,t_1}\right> |_{x_i=y_i} \nn \\
&=\left(\frac{1}{2i}\right)^m\prod_{i=1}^m(\partial_{y_i}-\partial_{x_i})\det_{1\leq i,j \leq m}K(x_i,t_i;y_j,t_j)|_{x_i=y_i} \nn \\
&=\sum_{\sigma\in\{0,1\}^m}\det_{1 \leq i,j \leq m}\frac{1}{2i}(-\partial_{x_i})^{\sigma_i}\partial_{y_j}^{1-\sigma_j}K(x_i,t_i;y_j,t_j)|_{x_i=y_i}
\eea
\\

\subsection{Commutation of products of kernels and large time limit} 

As shown above, the calculation of correlation functions requires to study products of kernels.
Let us restrict here to equal time correlations. It was shown above that the kernel $K(x,t;x',t)$ has a finite limit $K_\infty(x,x',0)$ for large $\ell$ and for large time in the NESS regime.
It is thus tempting to insert this finite limit $K_\infty(x,x',0)$ into the above determinantal formula (\ref{Rdet}) to obtain the
large time limit of the correlation functions. It turns out that this is correct here, but there is a subtle point to be discussed.

Indeed in \cite{DLSM2019, DLSM2019arXiv} the case of non interacting fermions in a trap was studied. It was found that the kernel
has a finite large time limit, equal in that case to the diagonal ensemble prediction, which coincides with the GGE prediction
for the kernel. 
However it was shown that the large time limit of the equal time correlation function {\it does not} coincide with the
determinantal formula where the diagonal kernel is inserted. This means that in this context the GGE does not describe 
the multi-point correlation functions. 

Here the situation is different since (i) there is no trap since we take
the limit $\ell \to +\infty$ before the large time limit (ii) the kernel converges to the NESS (different from the GGE). 
One can show that the additional terms which lead to the discrepancy described above 
decrease to zero at large $\ell$. Let us give now a very schematic explanation of this fact.

Using the notations of \cite{GLS} we write very schematically the kernel as a triple discrete sums over momenta
\be 
K \sim \frac{1}{\ell^3}\sum_{k,k_a,k_b} \frac{1}{(k-k_a)(k-k_b)} e^{- \frac{i}{2} (k_a^2-k_b^2) t} 
\ee
where for simplicity in the summand we kept only the poles and the time oscillations. As shown in \cite{GLS}
the diagonal term $k_a=k_b$ is one of the two leading terms in the large $\ell$ and large time limit (it does not depend 
on time). To see that its limit is $O(1)$ one can argue that in the large $\ell$ limit it is dominated
by $k_a=k_b$ close to $k$ to $O(1/\ell)$, and one is left with a single sum $\frac{1}{\ell} \sum_k \to \int dk = O(1)$.

Consider now the product of two of these kernels (as occurs in the correlation functions). It looks schematically like
\bea
K K \sim\frac{1}{\ell^6}\sum_{k,k_a,k_b} \sum_{k',k_a',k_b'}\frac{1}{(k-k_a)(k-k_b)} e^{- \frac{i}{2} (k_a^2-k_b^2)t}  \frac{1}{(k'-k_a')(k'-k_b')} e^{- \frac{i}{2} (k_a'^2-k_b'^2)t} 
\eea
There are two types of terms which are independent of time:

(i) $k_a = k_b$ and $k_a'=k_b'$. This gives simply the square of the diagonal part of the kernel
described above, as expected. 

(ii) $k_a = k_b'$ and $k_a'=k_b$. This is an exchange term which is problematic in the case of the fermions in a trap.
Here however we see that because of the poles the leading contribution comes from the case where
al six momenta are within $O(1/\ell)$. This then leads to a single sum but now with 
a factor $\frac{1}{\ell^2} \sum_k \to 0$, hence it becomes $O(1/\ell)$ at large $\ell$.

Similar "phase space volume" arguments can be made for the other terms (which are current-carrying, see \cite{GLS}),
as well as for higher order correlations, 
and allow to argue that the asymptotic quantum state is a determinantal point process based on the
NESS kernel.

\section{Two point density correlation: explicit calculation} 

\subsection{General formula} 

Let us specify to two space time points $(x_1,t_1)$, $(x_2,t_2)$. For {\it distinct} space time points one has
\be 
\langle \hat \rho(x_2,t_2) \hat \rho(x_1,t_1) \rangle = K(x_1,t_1;x_1,t_1) K(x_2,t_2;x_2,t_2) - 
K(x_1,t_1;x_2,t_2) K(x_2,t_2;x_1,t_1) 
\ee 
Next one takes the large time limit. One sets $t_1=t$, $t_2=t+\tau$, $x_1=x$, $x_2=x'$ and one 
recalls that $\lim_{t \to +\infty} K(x,t;x',t-\tau)=K_\infty (x,x',\tau)$. For $\tau >0$
this gives the equation \eqref{density_correl} in the text, which we recall here
\bea 
 C(x,x',\tau) &=& \lim_{t \to +\infty} \langle \hat \rho(x',t + \tau) \hat \rho(x,t) \rangle^c \\
&& =
\lim_{t \to +\infty} \left( 
\langle \hat \rho(x',t + \tau) \hat \rho(x,t) \rangle- 
\langle \hat \rho(x',t + \tau) \rangle \langle \hat \rho(x,t) \rangle \right) \\
&& =
 - K_\infty(x,x',-\tau) K_\infty(x',x,\tau) 
\eea

% This
% leads to $\tau>0$
% \be
% \lim_{t \to +\infty} \langle \hat \rho(x_2,t + \tau) \hat \rho(x_1,t) \rangle =
% \rho_\infty(x_1) \rho_\infty(x_2) - K_\infty(x_1,x_2;-\tau) K_\infty(x_2,x_1,\tau)
% \ee
% \be
% \lim_{t \to +\infty} \langle \hat \rho(x_2,t + \tau) \hat \rho(x_1,t) \rangle^c =
%  - K_\infty(x_1,x_2;-\tau) K_\infty(x_2,x_1,\tau)
% \ee

Let us now recall the formula for the kernel in the NESS
% \bea
% &K_\infty(x>\frac{a}{2},x'>\frac{a}{2},\tau)=\int_{L,R}\frac{dk}{2\pi}e^{i(\frac{k^2}{2}\tau-  k (x-x'))}|\st(k)|^2\nn\\
% %&\int_{k_R,\tau}\frac{dk}{2\pi}e^{-i\frac{k^2}{2}\tau}\left((\sr(k)e^{ik(x+x')}+e^{ik(x-x')})+cc\right)\\
% &+\int_{R,\tau}\frac{dk}{\pi}e^{i\frac{k^2}{2}\tau}\left(\cos(k(x-x'))+{\rm Re}[\sr(k)e^{ik(x+x')}]\right)\nn\\
% &K_\infty(x>\frac{a}{2},x'<\frac{-a}{2},\tau)=\int_{R,\tau}\frac{dk}{\pi}e^{i\frac{k^2}{2}\tau}{\rm Re}[\st(k)e^{ik(x-x')}]\nn\\
% &+\int_{L,R}\frac{dk}{2\pi}e^{i\frac{k^2}{2}\tau}\frac{\sr(k)\st(k)^*-\sr(k)^*\st(k)}{2}e^{-ik(x+x')}
% \eea
\bea\label{kernel_extended_NESS2}
&K_\infty(x>\frac{a}{2},x'>\frac{a}{2},\tau)=\int_{L,R}\frac{dk}{2\pi}e^{i(\frac{k^2}{2}\tau-  k (x-x'))}|\st(k)|^2 + \int_{-\infty}^\infty \frac{dk}{2\pi}\left(f_R(k)-\theta(-\tau)\right)e^{i\frac{k^2}{2}\tau}\left(e^{ik(x-x')}+\sr(k)e^{ik(x+x')}\right)\nn\\
&K_\infty(x>\frac{a}{2},x'<\frac{-a}{2},\tau)=\int_{0}^\infty\frac{dk}{2\pi}\left(f_R(k)-\theta(-\tau)\right)e^{i\frac{k^2}{2}\tau}\st(k)e^{ik(x-x')} + \int_{0}^\infty\frac{dk}{2\pi}\left(f_L(k)-\theta(-\tau)\right)e^{i\frac{k^2}{2}\tau}\st^*(k)e^{-ik(x-x')}\nn\\
&+\int_{L,R}\frac{dk}{2\pi}e^{i\frac{k^2}{2}\tau}\frac{\sr(k)\st(k)^*-\sr(k)^*\st(k)}{2}e^{-ik(x+x')}
\eea
together with the other regions obtained using the symmetry
$K_\infty(x,x',\tau)|_{L,R}=K_\infty(-x,-x',\tau)|_{R,L}$. Note that this symmetry implies
\be  \label{symC} 
C(x,x',\tau)  = C(-x,-x',\tau)|_{L \leftrightarrow R}
\ee 

We will now compute the density correlation focusing on $T=0$. Because of the symmetry \eqref{symC} we
only need to study the two sectors (i) $x>0,x'>0$ and (ii) $x>0,x'<0$. In each case we will
study separately the large $\tau$ behavior in the two sub-cases $x,x'=O(1)$ and $x,x'=O(\tau)$. This will lead to
the phase diagram at the end. The special case without a defect will be discussed at the end.

\subsection{First sector $x>0$, $x'>0$}

% \bea
% && \lim_{t \to +\infty} \langle \hat \rho(x',t + \tau) \hat \rho(x,t) \rangle^c =
%   - 
% \left( 
%  \int_{k_R}^{k_L} \frac{dk}{2\pi} e^{- i k (x-x') -i\frac{k^2}{2}\tau}|\st(k)|^2
%  - \int_{k_R}^{+\infty} \frac{dk}{\pi} e^{-i\frac{k^2}{2}\tau}\left( \cos k(x-x') +{\rm Re}[\sr(k)e^{ ik (x+x')}]\right) \right) \nn 
% \\ 
% &&  \times \left( 
%  \int_{k_R}^{k_L} \frac{dk}{2\pi} e^{i k (x-x') + i\frac{k^2}{2}\tau}|\st(k)|^2
%  +\int_0^{k_R} \frac{dk}{\pi}   e^{i\frac{k^2}{2}\tau}\left( \cos k(x-x')  +{\rm Re}[\sr(k)e^{ ik (x+x')}]\right) \right) 
% \eea 

In the sector $x>0$, $x'>0$ using the explicit form of the kernel $K_\infty$ one obtains
% \bea
% && \lim_{t \to +\infty} \langle \hat \rho(x',t + \tau) \hat \rho(x,t) \rangle^c =
%   - 
% \left( 
%  \int_{k_R}^{k_L} \frac{dk}{2\pi} e^{- i k (x-x') -i\frac{k^2}{2}\tau}|\st(k)|^2
%  - \int_{|k|>k_R} \frac{dk}{2\pi}  e^{-i\frac{k^2}{2}\tau}\left( e^{i k(x-x')} +\sr(k)e^{ ik (x+x')}\right) \right) \nn 
% \\ 
% &&  \times \left( 
%  \int_{k_R}^{k_L} \frac{dk}{2\pi} e^{i k (x-x') + i\frac{k^2}{2}\tau}|\st(k)|^2
%  +\int_{|k|<k_R} \frac{dk}{2\pi} 
%  e^{i\frac{k^2}{2}\tau}\left( e^{i k(x-x')}  +\sr(k)e^{ ik (x+x')}\right) \right) 
% \eea 

\bea \label{ABC} 
&&C(x,x',\tau) = -(\mathbb{A}-\mathbb{B})(\mathbb{A}^*+\mathbb{C})\\
&& \mathbb{A} = \int_{k_R}^{k_L} \frac{dk}{2\pi} e^{- i k (x-x') -i\frac{k^2}{2}\tau}|\st(k)|^2 \\
&& \mathbb{B} =  \mathbb{B}_1 +  \mathbb{B}_2, \quad  \mathbb{B}_1=\int_{|k|>k_R} \frac{dk}{2\pi}  e^{-i\frac{k^2}{2}\tau} e^{-i k(x-x')} ,\quad \mathbb{B}_2=\int_{|k|>k_R} \frac{dk}{2\pi}  e^{-i\frac{k^2}{2}\tau}\sr(k)e^{ ik (x+x')}\\
&& \mathbb{C} = \mathbb{C}_1 + \mathbb{C}_2,\quad \mathbb{C}_1 = \int_{|k|<k_R} \frac{dk}{2\pi}
 e^{i\frac{k^2}{2}\tau} e^{i k(x-x')}  \quad \mathbb{C}_2 = \int_{|k|<k_R} \frac{dk}{2\pi}
 e^{i\frac{k^2}{2}\tau}\sr^*(k)e^{ - ik (x+x')} \label{ABC} 
\eea
\\

{\bf Case $\tau \to \infty$ with $x,x'=O(1)$} Let us now consider the limit of large $\tau$ with fixed $x,x'$. 
In that limit the integrands in Eqs. \eqref{ABC} exhibit a saddle point at $k=0$ which may or may not be in the
interval of integration. In the first case we can use the saddle point method and in the second case
we use the following formula for any smooth function $f(k,x,x')$ 
%\be
%\int_a^b dk f(k,x,x') e^{-i\frac{k^2}{2}\tau}=-i\left[\frac{f(k,x,x')}{k\tau} e^{-i\frac{k^2}{2}\tau}\right]_{a}^b + O(\tau^{-2})
%\ee
\be \label{ipp}
\int_a^b dk f(k) e^{i\tau(\alpha k-\frac{k^2}{2})}=i\left[\frac{f(k)}{(k-\alpha)\tau} e^{i\tau(\alpha k-\frac{k^2}{2})}\right]_{a}^b + O(\tau^{-2})
\ee
which is shown by successive integration by parts, assuming that $[a,b]$ does not contain $k=\alpha$. 
This leads to the following asymptotic behaviors for large $\tau$ (we assume $k_R<k_L$). Here we will use $\alpha=0$ resulting in
\bea
&& \mathbb{A} \simeq - \sum_{R/L} \pm  \frac{i}{2\pi k_{R/L}\tau}e^{- i k_{R/L} (x-x') -i\frac{k_{R/L}^2}{2}\tau}|\st(k_{R/L})|^2 \\
&& \mathbb{B} \simeq - \frac{i}{\pi k_{R}\tau}e^{ -i\frac{k_{R}^2}{2}\tau} \left( \cos k_{R} (x-x') + Re[\sr(k_R)e^{ i k_{R} (x+x')} ] \right)     \\
&& \mathbb{C} \simeq  (1+\sr(0)) \int \frac{dk}{2\pi} e^{i\frac{k^2}{2}\tau} = \sqrt{\frac{i}{2\pi \tau}} (1+\sr(0))
\eea
where $\pm$ stands for  $R/L$. Note that the leading estimate for $\mathbb{C}$ is proportional to
$1 + \sr(0)$. However, it turns out that $\sr(0)=-1$ for most potentials, see Ref. \cite{chadan,aktosun,egorova}, where this is
shown under the condition $\int dx (1+|x|)|V(x)| < \infty $. There are however exceptional examples where this is not true: these are explored in Ref. \cite{ahmed} and correspond to potentials with bound states, in the limit where one of the bound state energy approaches zero. 
In the generic case $\sr(0)=-1$ one needs to push the asymptotics to the next order, which turns out to be $O(\tau^{-1})$.
Indeed one sees that the combinations $\mathbb{B}^* + \mathbb{C}$ is an integral for $k$ over the real axis, leading to
\be 
\mathbb{B}^* + \mathbb{C} = \int \frac{dk}{2\pi} e^{i\frac{k^2}{2}\tau}\left( e^{ik(x-x')} + \sr(k) e^{ik(x+x')}\right)
\sim \frac{c}{\tau^{3/2}} (2 x x' + i (x+x')r'(0) + \frac{r''(0)}{2})) = O(\tau^{-3/2}) 
\ee 
where $c=\int \frac{dk}{2\pi} k^2 e^{i\frac{k^2}{2}}= (i-1)/(2 \sqrt{\pi})$. Hence 
\be 
\mathbb{C} = - \mathbb{B}^* + O(\tau^{-3/2}) \label{cbstar}
\ee

% \be
% \mathbb{C} \sim \frac{1}{\tau} (\sr'(0)-2ix')\int \frac{dk}{2\pi} k e^{i\frac{k^2}{2}}
% \ee

Performing the products and keeping the leading terms at large time this leads to, schematically
\be
C(x,x',\tau) \sim \begin{cases}
\tau^{-\frac{3}{2}} \times \text{oscillations} (1+\sr(0)) \quad \sr(0)\neq -1 \\
\tau^{-2} \times \text{oscillations plus constant} \quad \sr(0) = -1
\end{cases}
\ee

More precisely, there are two cases to consider. In the generic case $\sr(0)=-1$ one finds
\bea
&&C(x,x',\tau) \sim - |\mathbb{A}-\mathbb{B}|^2 + O(\tau^{-5/2})  
\sim -\tau^{-2}|g_L(x,x') e^{-i\frac{k_L^2}{2}\tau}+g_R(x,x') e^{-i\frac{k_R^2}{2}\tau}|^2 + O(\tau^{-5/2}) \\
&&g_L(x,x')= \frac{1}{2\pi k_{L}}e^{- i k_{L} (x-x') }|\st(k_{L})|^2\\
&&g_R(x,x')=\frac{1}{2\pi k_{R}}\left(-
e^{- i k_{R} (x-x')}|\st(k_{R})|^2+   2\cos k_{R} (x-x') + 2 Re[\sr(k_R)e^{ i k_{R} (x+x')} ] \right) 
%&&C(x,x',\tau) \sim \tau^{-2}(c_1 +c_2 \cos((x-x')(k_R-k_L) + \frac{k_R^2-k_L^2}{2}\tau) + c_3 e^{-i\frac{k_R^2-k_L^2}{2}\tau})\\
%&& c_1= -\sum_{R/L} \frac{|\st(k_{R/L})|^4}{(2\pi k_{R/L}  )^2} + \frac{|\st(k_R)|^2 e^{ik_R(x-x')}}{2\pi^2 k_R^2 }(\cos k_R(x-x') + Re[\sr(k_R)e^{ik_R(x+x')}])\\
%&&c_2=\frac{|\st(k_R)\st(k_L)|^2}{2\pi^2 k_R k_L}\\
%&& c_3= - \frac{|\st(k_L)|^2 e^{ik_L(x-x')}}{2\pi^2 k_R k_L }(\cos k_R(x-x') + Re[\sr(k_R)e^{ik_R(x+x')}])
\eea
 %Note that there is a non-oscillating part given by the first term in $c_1$. 
This leads to oscillating terms in space and time, together with a non-oscillating component
\be 
C(x,x',\tau) \sim \frac{1}{(2 \pi)^2 \tau^2} \left[ \left( \frac{1}{k_L^2} |\st(k_L)|^4 + \frac{1}{k_R^2} ( 1+ |\sr(k_R)|^2)^2 \right) 
+ \text{oscillating} \right] \label{oscillating_same}
\ee

In the second case where $1+r(0) \neq 0$ one finds, to leading order for large $\tau$,
% \begin{eqnarray} \label{largetaureflected}
% &&\lim_{t \to +\infty} \langle \hat \rho(x',t + \tau) \hat \rho(x,t) \rangle^c \sim \frac{e^{- i \frac{\pi}{4}}}{(2 \pi \tau)^{3/2}} (1+r(0))  \\
% &&\times \Bigg(e^{-i \frac{k_R^2 \tau}{2}}\left( |\sr(k_R)|^2 e^{-i k_R(x-x')} + e^{ik_R(x-x')} + 2 Re(\sr(k_R)e^{i k_R(x+x')}) \right) + e^{-i\frac{k_L^2}{2}\tau - i k_L(x-x')} |\st(k_L)|^2\Bigg) \nonumber 
% %2 \sr'(k_R)\cos{(k_R(x+x'))} - 2\sr''(k_R)\sin{(k_R)}(x+x')
% \end{eqnarray}
\begin{eqnarray} \label{largetaureflected}
&&C(x,x',\tau) \sim \frac{e^{- i \frac{\pi}{4}}}{(2 \pi \tau)^{3/2}} (1+r(0))  \\
&&\times \Bigg(\frac{1}{k_R} e^{-i \frac{k_R^2 \tau}{2}}\left( |\sr(k_R)|^2 e^{-i k_R(x-x')} + e^{ik_R(x-x')} + 2 Re(\sr(k_R)e^{i k_R(x+x')}) \right) + \frac{1}{k_L} e^{-i\frac{k_L^2}{2}\tau - i k_L(x-x')} |\st(k_L)|^2\Bigg) \nonumber 
%2 \sr'(k_R)\cos{(k_R(x+x'))} - 2\sr''(k_R)\sin{(k_R)}(x+x')
\end{eqnarray}
This formula matches Eq. \eqref{nodefect} given below for the decay in the absence of a defect setting $\sr(k)=0$ and $\st(k)=1$.
\\

{\bf Case $\tau \to \infty$ with $x,x'=O(\tau)$}. We now consider the
decay of the density correlation function when $x, x'$ are scaled with $\tau$, as  $x=\zeta \tau, x'=\zeta' \tau$
with $\zeta,\zeta'=O(1)$.

%\bea
%&& \int_{k_F}^\infty \frac{dk}{2\pi}f(k,x,x')e^{i\tau(ak-\frac{k^2}{2})}\sim\int_0^\infty \frac{-i dp}{2\pi} f(k_F,x,x') e^{i\tau a k_F}e^{-i \tau \frac{k_R^2}{2}}e^{-\tau p (k_F-a)} \quad a<k_F\\
%&&\sim \frac{-i  f(k_F,x,x') e^{i\tau a k_F}e^{-i \tau \frac{k_F^2}{2}}}{2\pi \tau (k_F-a)}\\
%&& \int_{-\infty}^{k_F} \frac{dk}{2\pi}f(k,x,x')e^{i\tau(ak-\frac{k^2}{2})} \simeq \int_0^\infty \frac{i dp}{2\pi} f(k_F,x,x') e^{i\tau a k_F}e^{-i \tau \frac{k_R^2}{2}}e^{-\tau p (a-k_F)} \quad a>k_F\\
%&&\sim \frac{i  f(k_F,x,x') e^{i\tau a k_F}e^{-i \tau \frac{k_F^2}{2}}}{2\pi \tau (a-k_F)}
%\eea

%\bea
%&& \int_{k_F}^\infty \frac{dk}{2\pi}g(k)e^{i\tau(ak-\frac{k^2}{2}})\sim\frac{g(a)}{\sqrt{2\pi\tau}} e^{i\tau \frac{a^2}{2}}e^{-\frac{i\pi}{4}}   \quad a>k_F\\
%&& \int_{-\infty}^{k_F} \frac{dk}{2\pi}g(k)e^{i\tau(ak-\frac{k^2}{2}})\sim\frac{g(a)}{\sqrt{2\pi\tau}} e^{i\tau \frac{a^2}{2}}e^{-\frac{i\pi}{4}}   \quad a<k_F
%\eea
Again we use the saddle point method. The behavior of each term depends on whether the saddle point is inside or outside the integration interval. If it is outside  we will use \eqref{ipp}. Otherwise we use the usual saddle point formula.
%\bea
%&& \mathbb{A} = \int_{k_R}^{k_L} \frac{dk}{2\pi} e^{if_\mathbb{A}(k)\tau}|t(k)|^2 \\
%&& f_{\mathbb{A}}(k)=-k(\zeta-\zeta')-\frac{k^2}{2}
%\eea
 The term $\mathbb{A}$ has a saddle point at $k=k_1=\zeta'-\zeta$ such that %$, \quad f_{\mathbb{A}}(k^*)=\frac{(\zeta-\zeta')^2}{2}$
 \bea
 && \mathbb{A} \sim \begin{cases}
 \frac{|\st(\zeta'-\zeta)|^2}{\sqrt{2\pi\tau}}e^{i\tau \frac{(\zeta-\zeta')^2}{2}-\frac{i \pi}{4}}\mbox{ if } k_1 \in [k_R,k_L]\\
 %\frac{1}{2}\frac{|\st(k_s)|^2}{\sqrt{2\pi\tau}}e^{-i\tau (k_s(\zeta-\zeta')+\frac{k_s^2}{2})-\frac{i \pi}{4}} 
\frac{1}{2\pi i \tau}\left( \frac{|\st(k_R)|^2 e^{i\tau (\zeta'-\zeta)) k_R}e^{-i \tau \frac{k_R^2}{2}}}{\zeta-\zeta'+k_R}- \frac{|\st(k_L)|^2e^{i\tau (\zeta'-\zeta)) k_L}e^{-i \tau \frac{k_L^2}{2}}}{\zeta-\zeta'+k_L}\right) \mbox{ if } k_1 \notin [k_R,k_L]
 \end{cases} \label{asympt_A}
 \eea 
% Where $k_s=\begin{cases} k_R \mbox{ if } k^*<k_R \\ k_L \mbox{ if } k^* > k_L \end{cases}$
 
% \bea
% && \mathbb{B} = \mathbb{B}_1+ \mathbb{B}_2 \\
% &&\mathbb{B}_{1}=\int_{|k|>k_R} \frac{dk}{2\pi}  e^{if_{\mathbb{B},1}(k)\tau}\\
%  &&\mathbb{B}_{2}=\int_{|k|>k_R} \frac{dk}{2\pi} \sr(k) e^{if_{\mathbb{B},2}(k)\tau} \\
% && f_{\mathbb{B},1}(k) = -k(\zeta-\zeta')-\frac{k^2}{2}\\
%  && f_{\mathbb{B},2}(k) = k(\zeta+\zeta')-\frac{k^2}{2}
%\eea
 The term $\mathbb{B}_1$ has a saddle point at $k=k_1$ and  $\mathbb{B}_2$ has a saddle point at  $k=k_2= \zeta' + \zeta$  such that
% $k_1^*=\zeta'-\zeta,k_2^*=\zeta'+\zeta, \quad f_{\mathbb{B},1}(k_1^*)=\frac{(\zeta-\zeta')^2}{2},f_{\mathbb{B},2}(k_2^*)=\frac{(\zeta+\zeta')^2}{2}$
\bea
&& \mathbb{B}_1 \sim \begin{cases}
\frac{e^{i\tau \frac{(\zeta-\zeta')^2}{2}-\frac{i \pi}{4}}}{\sqrt{2\pi\tau}}\mbox{ if } k_R < |k_1| \\
%\frac{\cos(\tau k_R(\zeta-\zeta')) e^{-i\tau \frac{k_R^2}{2}-\frac{i \pi}{4}}}{\sqrt{2\pi\tau}}
\frac{-ie^{-i\tau \frac{k_R^2}{2}}}{2\pi  \tau}\left( \frac{e^{-i\tau k_R(\zeta-\zeta')}}{k_R+(\zeta-\zeta')}+ \frac{e^{i\tau k_R(\zeta-\zeta')}}{k_R-(\zeta-\zeta')}\right)\mbox{ if } k_R > |k_1|
\end{cases}\\
&& \mathbb{B}_2 \sim \begin{cases}
\frac{\sr(\zeta+\zeta')e^{i\tau \frac{(\zeta+\zeta')^2}{2}-\frac{i \pi}{4}}}{\sqrt{2\pi\tau}}\mbox{ if } k_R < |k_2| \\
%\frac{Re[\sr(k_R)e^{i\tau k_R(\zeta+\zeta')}] e^{-i\tau \frac{k_R^2}{2}-\frac{i \pi}{4}}}{\sqrt{2\pi\tau}}
\frac{-ie^{-i\tau \frac{k_R^2}{2}}}{2\pi  \tau}\left( \frac{\sr(k_R) e^{i\tau k_R(\zeta+\zeta')}}{k_R-(\zeta+\zeta')} + \frac{ \sr^*(k_R) e^{-i\tau k_R(\zeta+\zeta')}}{k_R+(\zeta+\zeta')}\right)\mbox{ if } k_R > |k_2|
\end{cases}
\eea

%\bea
% && \mathbb{C} = \mathbb{C}_1+ \mathbb{C}_2 \\
% &&\mathbb{C}_{1}=\int_{|k|<k_R} \frac{dk}{2\pi}  e^{if_{\mathbb{C},1}(k)\tau}\\
%  &&\mathbb{C}_{2}=\int_{|k|<k_R} \frac{dk}{2\pi} \sr(k) e^{if_{\mathbb{C},2}(k)\tau} \\
% && f_{\mathbb{C},1}(k) = k(\zeta-\zeta')+\frac{k^2}{2}\\
%  && f_{\mathbb{C},2}(k) = k(\zeta+\zeta')+\frac{k^2}{2}
%\eea
 The term $\mathbb{C}_1$ has a saddle point at $k=k_1$ and  $\mathbb{C}_2$ has a saddle point at $k=k_2$  such that
%$k_1^*=\zeta'-\zeta,k_2^*=-(\zeta'+\zeta), \quad f_{\mathbb{C},1}(k_1^*)=-\frac{(\zeta-\zeta')^2}{2},f_{\mathbb{C},2}(k_2^*)=-\frac{(\zeta+\zeta')^2}{2}$

% \bea
% && \mathbb{C}_1 \sim \begin{cases}
% \frac{e^{-i\tau \frac{(\zeta-\zeta')^2}{2}+\frac{i \pi}{4}}}{\sqrt{2\pi\tau}}\mbox{ if } k_R > |k_1| \\
% %\frac{e^{\pm i\tau k_R(\zeta-\zeta')} e^{-i\tau \frac{k_R^2}{2}+\frac{i \pi}{4}}}{2\sqrt{2\pi\tau}}
% \frac{-ie^{i\tau\frac{k_R^2}{2}}}{2\pi \tau} \left( \frac{e^{-ik_R(\zeta'-\zeta)\tau}}{\zeta'-\zeta-k_R}-\frac{e^{ik_R(\zeta'-\zeta)\tau}}{\zeta'-\zeta+k_R}\right)  \mbox{ if } k_R < |k_1|
% \end{cases}\\
% && \mathbb{C}_2 \sim \begin{cases}
% \frac{\sr^*(\zeta+\zeta')e^{-i\tau \frac{(\zeta+\zeta')^2}{2}+\frac{i \pi}{4}}}{\sqrt{2\pi\tau}}\mbox{ if } k_R > |k_2| \\
% %&&\frac{\sr(\pm k_R)e^{\pm i\tau k_R(\zeta+\zeta')} e^{i\tau \frac{k_R^2}{2}+\frac{i \pi}{4}}}{2\sqrt{2\pi\tau}}\\
% \frac{ie^{i\tau\frac{k_R^2}{2}}}{2\pi \tau} \left( \frac{\sr(k_R)e^{ik_R(\zeta'+\zeta)\tau}}{\zeta'+\zeta+k_R}-\frac{\sr^*(k_R)e^{-ik_R(\zeta'+\zeta)\tau}}{\zeta'+\zeta-k_R}\right)  \mbox{ if } k_R < |k_2|
% \end{cases}
% \eea

\bea
&& \mathbb{C}_1 \sim \begin{cases}
\frac{e^{-i\tau \frac{(\zeta-\zeta')^2}{2}+\frac{i \pi}{4}}}{\sqrt{2\pi\tau}}\mbox{ if } k_R > |k_1| \\
%\frac{e^{\pm i\tau k_R(\zeta-\zeta')} e^{-i\tau \frac{k_R^2}{2}+\frac{i \pi}{4}}}{2\sqrt{2\pi\tau}}
\frac{-ie^{i\tau\frac{k_R^2}{2}}}{2\pi \tau} \left( \frac{e^{ik_R(\zeta-\zeta')\tau}}{k_R + \zeta-\zeta'}+\frac{e^{-ik_R(\zeta-\zeta')\tau}}{k_R-(\zeta-\zeta')}\right)  \mbox{ if } k_R < |k_1|
\end{cases}\\
&& \mathbb{C}_2 \sim \begin{cases}
\frac{\sr^*(\zeta+\zeta')e^{-i\tau \frac{(\zeta+\zeta')^2}{2}+\frac{i \pi}{4}}}{\sqrt{2\pi\tau}}\mbox{ if } k_R > |k_2| \\
%&&\frac{\sr(\pm k_R)e^{\pm i\tau k_R(\zeta+\zeta')} e^{i\tau \frac{k_R^2}{2}+\frac{i \pi}{4}}}{2\sqrt{2\pi\tau}}\\
\frac{-ie^{i\tau\frac{k_R^2}{2}}}{2\pi \tau} \left( \frac{\sr^*(k_R)e^{-ik_R(\zeta'+\zeta)\tau}}{k_R-(\zeta'+\zeta)} + \frac{\sr(k_R)e^{ik_R(\zeta'+\zeta)\tau}}{k_R+(\zeta'+\zeta)}\right)  \mbox{ if } k_R < |k_2|
\end{cases}
\eea

%where $\pm$ correspond to $\begin{cases} + \to k_1^*,k_2^*> k_R \\ - \to k_1^*,k_2^*< -k_R \end{cases}$

Let us now build a phase diagram for the decay of correlations in the plane $\zeta,\zeta'$, represented in Fig. \ref{suppphasediag}
using the formula \eqref{ABC}
\be 
C(x,x',\tau) = -(\mathbb{A}-\mathbb{B}_1-\mathbb{B}_2)(\mathbb{A}^*+\mathbb{C}_1+\mathbb{C}_2)
\ee
Here we restrict to $\zeta>0, \zeta'>0$. First we note that the sub-cases for the decay of 
$\mathbb{B}_1$ and $\mathbb{C}_1$ are mutually exclusive, and similarly for $\mathbb{B}_2$ and $\mathbb{C}_2$. Next we note that a total correlation decay $C(x,x',\tau) \sim \tau^{-1}$
can only be generated by either:

{\it Region 1} the term $- \mathbb{A} \mathbb{A}^*$ with $\mathbb{A}, \mathbb{A}^* \sim \tau^{-1/2}$, which
is realized for $\zeta'-\zeta \in [k_R,k_L]$, corresponding to a first region in the Fig. \ref{suppphasediag}

{\it Region 2} the term $\mathbb{B}_2 \mathbb{C}_1 $
with $\mathbb{B}_2 , \mathbb{C}_1 \sim \tau^{-1/2}$ which
is realized for $|\zeta'-\zeta| < k_R < \zeta'+\zeta$, corresponding to a second region in the Fig. \ref{suppphasediag}.

These two regions 1 and 2 do not overlap since $k_R<k_L$, and their union form the region in blue in the figure where the decay of the correlation is $C(x,x',\tau) \sim \tau^{-1}$.
Note the other combination, $\mathbb{B}_1 \mathbb{C}_2 $
with $\mathbb{B}_2 , \mathbb{C}_1 \sim \tau^{-1/2}$ leads to an empty region. 

Next, because of the complementary behavior of $\mathbb{B}_1$ and $\mathbb{C}_1$ the only other
possible decay for the total correlation decay is $C(x,x',\tau) \sim \tau^{-3/2}$ which holds in the complementary of
the blue region in Fig. \ref{suppphasediag}. 
\\

%We will have a $\tau^{-1}$ behaviour in the following section of $\zeta>0,\quad \zeta'>0$ (and a $\tau^{-3/2}$ decay otherwise)
%\bea
%&& A \sim \tau^{-1/2} \quad \mbox{ if } k_R<\zeta' - \zeta < k_L \quad \mbox{band of slop $\zeta$ between $k_R$ and $k_L$} \\
%&& B_1 \sim \tau^{-1/2} \cap C_2\sim \tau^{-1/2} \quad \mbox{ if } k_R<|\zeta' - \zeta| \cap k_R > \zeta' + \zeta  \quad \mbox{never realised}\\
%&& B_2 \sim \tau^{-1/2} \cap C_1 \sim \tau^{-1/2} \quad \mbox{ if } k_R>|\zeta' - \zeta| \cap k_R < \zeta' + \zeta 
%\eea

One can now use the symmetry \eqref{symC} to obtain the corresponding phase diagram in the region $x<0,x'<0$ and $\zeta<0,\zeta'<0$. It implies that the phase diagram is symmetric with respect to the transformation $\begin{cases}
\zeta \to -\zeta\\
\zeta' \to -\zeta'\\
k_R \leftrightarrow k_L
\end{cases}$ Therefore the two previous regions 1 and 2 are mapped to the following regions 1' and 2' with the same decay:\\

{\it Region 1'} for $\zeta-\zeta' \in [k_R,k_L]$ (%notice that there seems to be no exchange of $k_R,k_L$. This is because 
since the condition comes from a saddle being inside the integration interval)
%such that the exchange only changes $C(x,x',\tau)$ by a minus sign.

{\it Region 2'} for $|\zeta'-\zeta|<k_L<-(\zeta+\zeta')$

These regions are indicated in Fig. \ref{suppphasediag}. Note that in that case they overlap. 

%{\red P: can we understand the physics of the case where everything (holes and particles) are controlled by diffusion, i.e. saddle point}

\begin{figure}[h]
\centering
\includegraphics[scale=0.7]{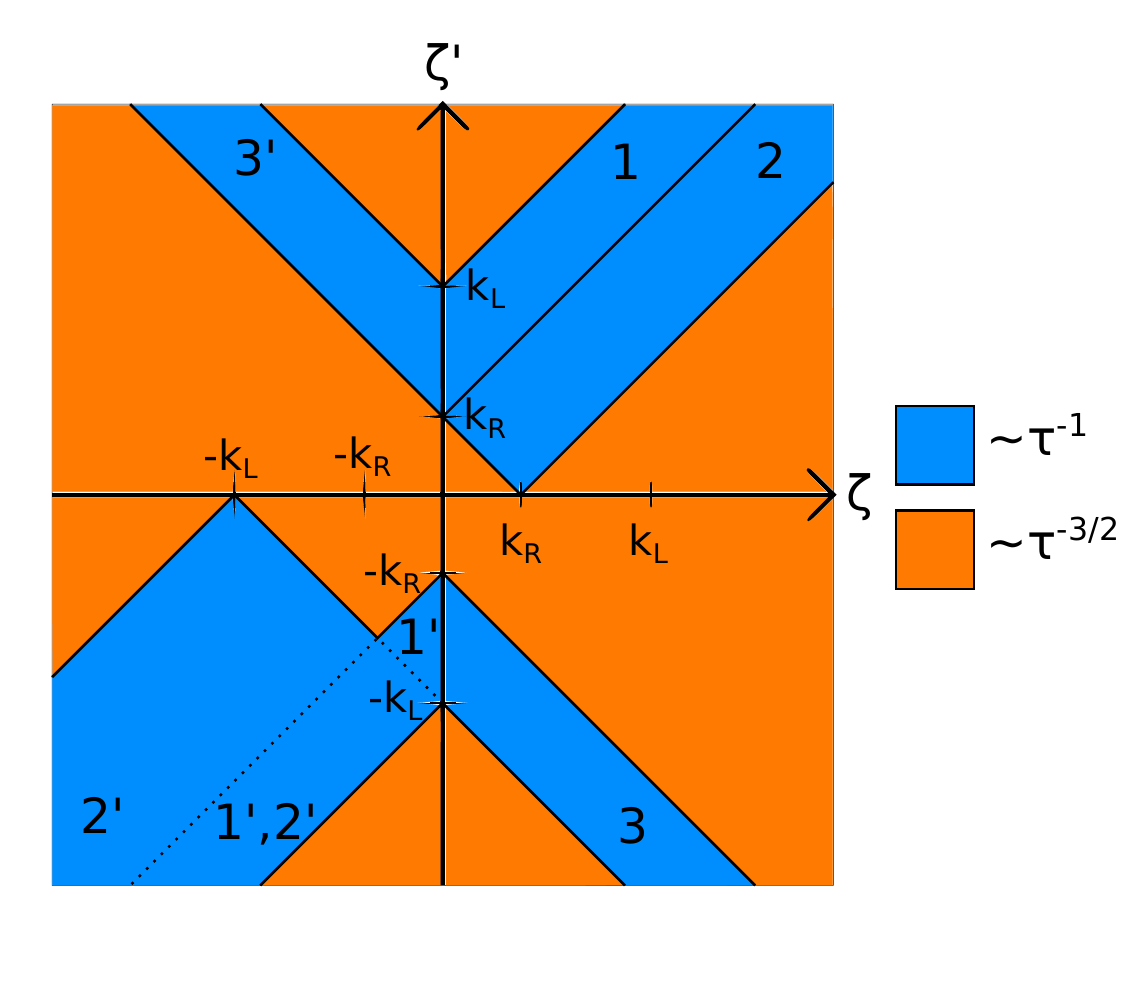}
\caption{Phase diagram of the large $\tau$ behavior of the density correlation function $C(x,x;\tau)$ in the regime 
$x=\zeta \tau,\quad x'=\zeta' \tau$ with $\zeta,\zeta'=O(1)$. The different regions where the decay are $\sim \tau^{-1}$ (blue)
or $\sim \tau^{-3/2}$ (orange) are represented in the plane $(\zeta,\zeta')$ following the analysis given in the text (here $k_R<k_L$).
The meaning of the labels of the regions $1$, $1'$, $2$, $2'$, $3$, $3'$ are explained in the text. Note that {\it Region 2} disappear if there is no impurity, {\it Region 1} in the equilibrium case ($k_L=k_R$), and {\it Region 3} if one of the previous condition is met.
 }\label{suppphasediag}
\end{figure} 

\subsection{Second sector  $x>0, \quad x'<0$}
In the sector $x>0, \quad x'<0, \quad \tau >0$, using the explicit expression of $K_\infty$ one obtains

% \bea
% && \lim_{t \to +\infty} \langle \hat \rho(x',t + \tau) \hat \rho(x,t) \rangle^c =
%   - K_{R,L}(x,x',-\tau) K_{R,L}(x',x,\tau)\\
%   && =- K_{R,L}(x,x',-\tau) K_{L,R}(-x',-x,\tau)\\
%   &&= -\left( \int_{k_R}^{k_L}\frac{dk}{2\pi} tr(k) e^{-i\frac{k^2}{2}\tau}e^{-ik(x+x')} -\int_{|k|>k_R} \frac{dk}{2\pi} \st(k)e^{-i\frac{k^2}{2}\tau}e^{ik(x-x')} \right)\\
%   && \times\left( \int_{k_L}^{k_R}\frac{dk}{2\pi} tr(k) e^{i\frac{k^2}{2}\tau}e^{ik(x+x')} +\int_{|k|<k_L} \frac{dk}{2\pi} \st(k)e^{i\frac{k^2}{2}\tau}e^{ik(x-x')} \right)
% \eea
% with $tr(k) = \frac{\sr(k)\st^*(k)-\sr^*(k)\st(k)}{2}$.\\
% Good formula is
\bea
&& C(x,x',\tau) =
  - K_\infty (x,x',-\tau) K_\infty (x',x,\tau)\\
  && =- K_\infty(x,x',-\tau) K_\infty(-x',-x,\tau)|_{R \leftrightarrow L}\\
  &&= -\left( \int_{k_R}^{k_L}\frac{dk}{2\pi} \sr(k) \st^*(k) e^{-i\frac{k^2}{2}\tau}e^{-ik(x+x')} -\int_{k_R}^{+\infty} \frac{dk}{2\pi} \st(k)e^{-i\frac{k^2}{2}\tau}e^{ik(x-x')} -\int_{k_L}^{+\infty} \frac{dk}{2\pi} \st^*(k)e^{-i\frac{k^2}{2}\tau}e^{-ik(x-x')}  \right)\\
  && \times\left( \int_{k_L}^{k_R}\frac{dk}{2\pi}  \sr(k) \st^*(k) e^{i\frac{k^2}{2}\tau}e^{ik(x+x')} +\int_{0}^{k_L} \frac{dk}{2\pi} \st(k)e^{i\frac{k^2}{2}\tau}e^{ik(x-x')} +\int_{0}^{k_R} \frac{dk}{2\pi} \st^*(k)e^{i\frac{k^2}{2}\tau}e^{-ik(x-x')} \right) \nonumber 
\eea 
This expression can be written in a compact way
\bea
&& C(x,x',\tau) = - \left( \mathbb{A}' - \mathbb{B}'\right) \left( \mathbb{A}^{\prime *} + \mathbb{C}'\right)\\
&& \mathbb{A}'= \int_{k_R}^{k_L}\frac{dk}{2\pi}  \sr(k) \st^*(k) e^{-i\frac{k^2}{2}\tau}e^{-ik(x+x')}\\
&& \mathbb{B}'= \left( \int_{k_R}^{+\infty} +  \int_{-\infty}^{-k_L} \right) \frac{dk}{2\pi} \st(k)e^{-i\frac{k^2}{2}\tau}e^{ik(x-x')}\\
&& \mathbb{C}'= \int_{-k_R}^{k_L} \frac{dk}{2\pi} \st(k)e^{i\frac{k^2}{2}\tau}e^{ik(x-x')}
\eea

{\bf Case $\tau \to \infty$ with $x,x'=O(1)$.} First we focus on the regime $x,x'$ fixed and $\tau \to \infty$. Following the method used in the same regime to determine $\mathbb{A}, \mathbb{B}, \mathbb{C}$ one obtains
\bea
&& \mathbb{A}' \simeq \sum_{R/L} \pm  \frac{\sr(k_{R/L})\st^*(k_{R/L})}{2\pi i k_{R/L}\tau}e^{- i k_{R/L} (x+x') -i\frac{k_{R/L}^2}{2}\tau} \\
&& \mathbb{B}' \simeq  \frac{1}{2 \pi i \tau}\left( \frac{\st(k_R)}{k_R} e^{ik_R(x-x')}e^{ -i\frac{k_{R}^2}{2}\tau} +  \frac{\st^*(k_L)}{k_L} e^{-ik_L(x-x')}e^{ -i\frac{k_{L}^2}{2}\tau} \right)     \\
&& \mathbb{C}' \sim \begin{cases}
\sqrt{\frac{i}{2\pi\tau}}\st(0) \quad \st(0) \neq 0\\
- \mathbb{B}^{\prime *} + O(\tau^{-3/2}) 
%\frac{c}{\tau^{3/2}} ( i (x-x')\st'(0) + \frac{\st''(0)}{2}))\quad \st(0)= 0
\end{cases}
\eea
where the last estimate is obtained by the same reasoning as the one leading to \eqref{cbstar}. 
%With $c$ defined in \eqref{cbstar}. 
Note that since $\sr(0)=-1$ implies $\st(0)=0$ we must again consider the same two cases as above,
leading to the same power law.
%  Under regularity conditions from \cite{aktosun} the two following conditions verify $\st(0)=0 \Leftrightarrow \sr(0)=-1$ such that the decays follow the same power law for $xx'>0$ and $xx'<0$.

\be
C(x,x',\tau) \sim \begin{cases}
\tau^{-\frac{3}{2}} \times \text{oscillations} \times \st(0) \quad \st(0)\neq 0 \\
\tau^{-2} \times \text{oscillations plus constant} \quad \st(0)= 0
\end{cases}
\ee
More precisely if $\st(0) = 0$ then
\bea
&&C(x,x',\tau) \sim - |\mathbb{A}'-\mathbb{B}'|^2 + O(\tau^{-5/2})  
\sim -\tau^{-2}|g_L'(x,x') e^{-i\frac{k_L^2}{2}\tau}+g_R'(x,x') e^{-i\frac{k_R^2}{2}\tau}|^2 + O(\tau^{-5/2}) \\
&&g_L'(x,x')= -\frac{1}{2\pi k_L} \left( \sr(k_{L})\st^*(k_{L}) e^{- i k_{L} (x+x') } + \st^*(k_L) e^{-ik_L(x-x')} \right)\\
&&g_R'(x,x')=  \frac{1}{2\pi k_R}\left(  \sr(k_{R})\st^*(k_{R})e^{- i k_{R} (x+x') } - \st(k_R) e^{ik_R(x-x')}  \right) \;.
\eea
This leads to oscillating terms in space and time, together with a non-oscillating component
\be 
C(x,x',\tau) \sim \frac{1}{(2 \pi)^2 \tau^2} \left( \frac{1}{k_R^2} (1- |\sr(k_R)|^4) +  \frac{1}{k_L^2} (1- |\sr(k_L)|^4) 
+ \text{oscillating} \right) \quad, \quad x>0 \quad, \quad x'<0 \;.
\ee
Note that the non-oscillating part is symmetric under the permutation of left (L) and right (R), which is not the case of the non-oscillating part for $x,x'>0$ -- see Eq. (\ref{suppphasediag}).

% \bea
% &&C(x,x',\tau) \simeq \tau^{-2} ( c_1' + c_2' e^{i\frac{k_L^2-k_R^2}{2}\tau} + c_3' e^{-i\frac{k_L^2-k_R^2}{2}\tau}) \\
% && c_1' = \frac{1}{(2\pi)^2}\left( \frac{\sr^*(k_R)\st(k_R)}{k_R^2}(\st(k_R)e^{2ik_R x} - \sr(k_R) \st^*(k_R)) - \frac{\sr^*(k_L)\st(k_L)}{k_L^2}(\st^*(k_L)e^{2ik_L x'} + \sr(k_L) \st^*(k_L)) \right)\\
% && c_2'= -\frac{\sr^*(k_L) \st(k_L)e^{ik_L(x+x')}}{(2\pi)^2 k_R k_L}\left( \st(k_R) e^{ik_R(x-x')} - \sr(k_R)\st^*(k_R) e^{-ik_R(x+x')} \right)\\
% && c_3'= \frac{\sr^*(k_R) \st(k_R)e^{ik_R(x+x')}}{(2\pi)^2 k_R k_L}\left( \st^*(k_L) e^{-ik_L(x-x')} + \sr(k_L)\st^*(k_L) e^{-ik_L(x+x')} \right)
% \eea
%Note the non-oscillating part given by $c_1'$. 

The second case when $\st(0)\neq 0$ gives
\bea
C(x,x',\tau)\simeq && \frac{e^{-i \pi/4}\st(0)}{(2\pi \tau)^{3/2}}\left( \frac{e^{-i\frac{k_R^2}{2} \tau}}{k_R}(\st(k_R) e^{ik_R(x-x')} - \sr(k_R) \st^*(k_R)e^{-ik_R(x+x')}) \right. \\
&& \left. \frac{e^{-i\frac{k_L^2}{2} \tau}}{k_L}(\st^*(k_L) e^{-ik_L(x-x')} + \sr(k_L) \st^*(k_L)e^{-ik_L(x+x')}) \right) \nonumber 
\eea
This formula matches the case \eqref{nodefect} without barrier when one takes $\begin{cases}
\st(k)=1\\
\sr(k)=0
\end{cases}$.\\
Finally, the symmetry \eqref{symC} allows to obtain the decay of the density correlation for $x<0, x'>0$.\\

{\bf Case $\tau \to \infty$ with $x,x'=O(\tau)$, $x>0,x'<0$}.
We now investigate the decay of the correlation function in the regime $x=\zeta \tau ,\quad x'=\zeta' \tau $
with $\zeta>0$ and $\zeta'<0$.  

The term $\mathbb{A}'$ has a saddle point at $k=-k_2=-(\zeta + \zeta')$
\be
\mathbb{A}'=\begin{cases}
\frac{\sr^*(\zeta+\zeta')\st(\zeta+\zeta')}{\sqrt{2\pi \tau}}e^{i\tau \frac{(\zeta+\zeta')^2}{2}}e^{-\frac{i\pi}{4}} \quad -k_2 \in [k_R,k_L] \Leftrightarrow -\zeta -k_L <\zeta'<-\zeta -k_R\\
\frac{-i}{2\pi \tau}\left( \frac{\sr(k_R)\st^*(k_R)e^{-ik_R(\zeta+\zeta')\tau}e^{-i\tau \frac{k_R^2}{2}}}{k_R+\zeta+\zeta'}-\frac{\sr(k_L)\st^*(k_L)e^{-ik_L(\zeta+\zeta')\tau}e^{-i\tau \frac{k_L^2}{2}}}{k_L+\zeta+\zeta'}\right) \quad -k_2 \notin [k_R,k_L]
\end{cases}
\ee

The term $\mathbb{B}'$ has a saddle point at $k=-k_1=\zeta-\zeta'$
%\be
%\mathbb{B}=\begin{cases}
%\frac{\st(\zeta-\zeta')}{\sqrt{2\pi \tau}}e^{i\tau \frac{(\zeta-\zeta')^2}{2}} e^{i(\zeta-\zeta')(y-y')}e^{-\frac{i\pi}{4}} \quad |k^*|> k_R\\
%\frac{ie^{-i\tau \frac{k_R^2}{2}}}{2\pi \tau}\left( \frac{\st(k_R)e^{i(x-x')k_R}}{\zeta-\zeta'-k_R}+\frac{\st^*(k_R)e^{-i(x-x')k_R}}{\zeta-\zeta'+k_R}\right) \quad |k^*| <k_R
%\end{cases}
%\ee

\be
\mathbb{B}'=\begin{cases}
\frac{\st(\zeta-\zeta')}{\sqrt{2\pi \tau}}e^{i\tau \frac{(\zeta-\zeta')^2}{2}} e^{-\frac{i\pi}{4}} \quad -k_1> k_R \cup -k_1<-k_L \Leftrightarrow \zeta'<\zeta-k_R \cup \zeta'>\zeta + k_L\\
\frac{i}{2\pi \tau}\left( \frac{\st(k_R)e^{ik_R(\zeta-\zeta')\tau -i\tau \frac{k_R^2}{2}}}{\zeta-\zeta'-k_R}-\frac{\st^*(k_L)e^{-ik_L(\zeta-\zeta')\tau -i\tau \frac{k_L^2}{2}}}{\zeta-\zeta'+k_L}\right) \quad -k_L<-k_1< k_R \Leftrightarrow \zeta - k_R < \zeta' < \zeta + k_L
\end{cases}
\ee

The term $\mathbb{C}'$ has a saddle point at $k=k_1$
%\be
%\mathbb{C}=\begin{cases}
%\frac{\st(\zeta'-\zeta)}{\sqrt{2\pi \tau}}e^{-i\tau \frac{(\zeta-\zeta')^2}{2}} e^{i(\zeta'-\zeta)(y-y')}e^{\frac{i\pi}{4}} \quad |k^*|< k_L\\
%\frac{-ie^{i\tau \frac{k_R^2}{2}}}{2\pi \tau}\left( \frac{\st(k_L)e^{i(x-x')k_L}}{\zeta-\zeta'+k_R}-\frac{\st^*(k_L)e^{-i(x-x')k_L}}{\zeta-\zeta'-k_R}\right) \quad |k^*| >k_L
%\end{cases}
%\ee

\be
\mathbb{C}'=\begin{cases}
\frac{\st^*(\zeta-\zeta')}{\sqrt{2\pi \tau}}e^{-i\tau \frac{(\zeta-\zeta')^2}{2}} e^{\frac{i\pi}{4}} \quad -k_R <k_1< k_L \Leftrightarrow \zeta - k_R < \zeta' < \zeta + k_L\\
\frac{-i}{2\pi \tau}\left( \frac{\st(k_L)e^{ik_L(\zeta-\zeta')\tau +i\tau \frac{k_L^2}{2} }}{\zeta-\zeta'+k_L}-\frac{\st^*(k_R)e^{-ik_R(\zeta-\zeta')\tau+ i\tau \frac{k_R^2}{2} }}{\zeta-\zeta'-k_R}\right)  k_1 >k_L \cup k_1 <-k_R \Leftrightarrow\zeta + k_L < \zeta' \cup \zeta' < \zeta - k_R
\end{cases}
\ee

We can now complete the phase diagram of Fig. \ref{suppphasediag}. We can have two 
different types of decay in the following regions of the quadrant $\zeta>0,\quad \zeta'<0$ :

{\it Region 3}: the term $-\mathbb{A}^{\prime *}\mathbb{A}'$ generates a decay $C(x,x',\tau) \sim \tau^{-1}$ for $-(\zeta+\zeta') \in  [k_R, k_L] $.
This region is represented in Fig. \ref{suppphasediag}.\\

The decay in the remainder of the quadrant $\zeta>0,\quad \zeta'<0$ is $C(x,x',\tau) \sim \tau^{-3/2}$. This is because $\mathbb{B}'$ and $\mathbb{C}'$ are complementary, hence if one of them decays as $\sim \tau^{-1/2}$ the other one decays as $\sim \tau^{-1}$.

Once again the symmetry \eqref{symC} permits to recover the opposite region of the diagram, namely $\zeta <0, \zeta'>0$. 
Indeed the symmetry maps the {region 3} to the {region 3'}, $\zeta+\zeta' \in  [k_R, k_L]$.
\\

%\bea
%&& \tau^{-1} \quad AA \quad \mbox{ if } -\zeta -k_L <\zeta'<-\zeta -k_R \\
%&& \tau^{-1} \quad BC \quad \mbox{ impossible }\\
%&& \tau^{-3/2} \mbox{this is the rest of the plan because B and C are complementary}
%\eea

{\bf Remark}.
Note that there are exceptional cases, which we will not study in detail here, 
where some of the decay amplitudes obtained here vanish, in particular:\\
%the previous computation turns out to be false either\\
(a) some regions of the diagram if $k_L$ or $k_R$ is fine tuned to coincide exactly with one of the zeroes of $\sr(k)$ and $\st(k)$ (see for instance the second line of Eq. (\ref{asympt_A}))\\
(b) on specific point of the diagram where the saddle point $k_1$ and $k_2$ coincides with the zeroes of $\sr(k)$ and $\st(k)$.
Finally, in the phase diagram of \eqref{suppphasediag} there are also transition regions between the different sectors
that would deserve a further study.

\subsection{Case without defect} 

In the case without a defect, one sets $\st(k)=1$ and $\sr(k)=0$ and one obtains, for any $x,x'$
\bea\label{kernel_extended_NESS3}
&K_\infty(x,x',\tau) = \int_{0}^{+\infty} \frac{dk}{2\pi} (f_L(k)-\theta(-\tau))  e^{i(\frac{k^2}{2}\tau-  k (x-x'))} 
+ \int_{-\infty}^{0} \frac{dk}{2\pi} (f_R(k) -\theta(-\tau)) e^{i(\frac{k^2}{2}\tau-  k (x-x'))}
\eea
% which comes from 
% \bea\label{kernel_extended_NESS4}
% &K_\infty(x>\frac{a}{2},x'>\frac{a}{2},\tau)=\int_{L,R}\frac{dk}{2\pi}e^{i(\frac{k^2}{2}\tau-  k (x-x'))} +\int_{-\infty}^\infty \frac{dk}{2\pi}\left(f_R(k)-\theta(-\tau)\right)e^{i\frac{k^2}{2}\tau} e^{ik(x-x')} \nn\\
% &K_\infty(x>\frac{a}{2},x'<\frac{-a}{2},\tau)=\int_{0}^\infty\frac{dk}{2\pi}\left(f_R(k)-\theta(-\tau)\right)e^{i\frac{k^2}{2}\tau}e^{ik(x-x')} + \int_{0}^\infty\frac{dk}{2\pi}\left(f_L(k)-\theta(-\tau)\right)e^{i\frac{k^2}{2}\tau}e^{-ik(x-x')}
% \eea
It is easy to obtain the large $\tau$ asymptotic at zero temperature in that case either by a direct calculation or from
matching with the above calculations. It reads in the regime $x,x' = O(1)$
\bea \label{nodefect} 
&&C(x,x',\tau) \underset{\tau \to \infty}{ \simeq } \frac{e^{- i \frac{\pi}{4}}}{(2 \pi \tau)^{3/2}} \Bigg(\frac{e^{-i \frac{k_R^2 \tau}{2}+ik_R(x-x')}}{k_R} + \frac{e^{-i\frac{k_L^2}{2}\tau - i k_L(x-x')}}{k_L} \Bigg) 
\eea

% At coinciding points, with $x>a$ (we can choose $\tau>0$ without loss of generality)
% \bea
% && \lim_{t \to +\infty} \langle \hat \rho(x,t + \tau) \hat \rho(x,t) \rangle^c =
%  - K_\infty(x,x;-\tau) K_\infty(x,x,\tau) \\
%  && = - 
% \left( 
%  \int \frac{dk}{2\pi} (f_L(k)-f_R(k)) e^{-i\frac{k^2}{2}\tau}|\st(k)|^2
%  +\int \frac{dk}{\pi} (f_R(k)-1) e^{-i\frac{k^2}{2}\tau}\left(1 +{\rm Re}[\sr(k)e^{2 ik x}]\right) \right) 
% \\ 
% &&  \times \left( 
%  \int \frac{dk}{2\pi} (f_L(k)-f_R(k)) e^{i\frac{k^2}{2}\tau}|\st(k)|^2
%  +\int \frac{dk}{\pi} f_R(k)  e^{i\frac{k^2}{2}\tau}\left(1+{\rm Re}[\sr(k)e^{2 ik x}]\right) \right) 
% \eea

% Consider now the regime $x=\zeta \tau$, $x'= \zeta' \tau$, with $\tau \to +\infty$ and $\zeta,\zeta'=O(1)$. One finds that 
% for $\zeta \zeta' >0$, $\mathbb{B}_2=0,\quad \mathbb{C}_2=0$ which implies that the region 2 does not exist anymore. If $\zeta, \zeta'<0$, $\mathbb{A}=0$ 
% and the correlation always decays as $C(x,x',\tau) \sim \tau^{-3/2}$. This is represented in the Fig. \ref{fig_without}. 
 
 \begin{figure}[t]
\centering
\includegraphics[scale=0.7]{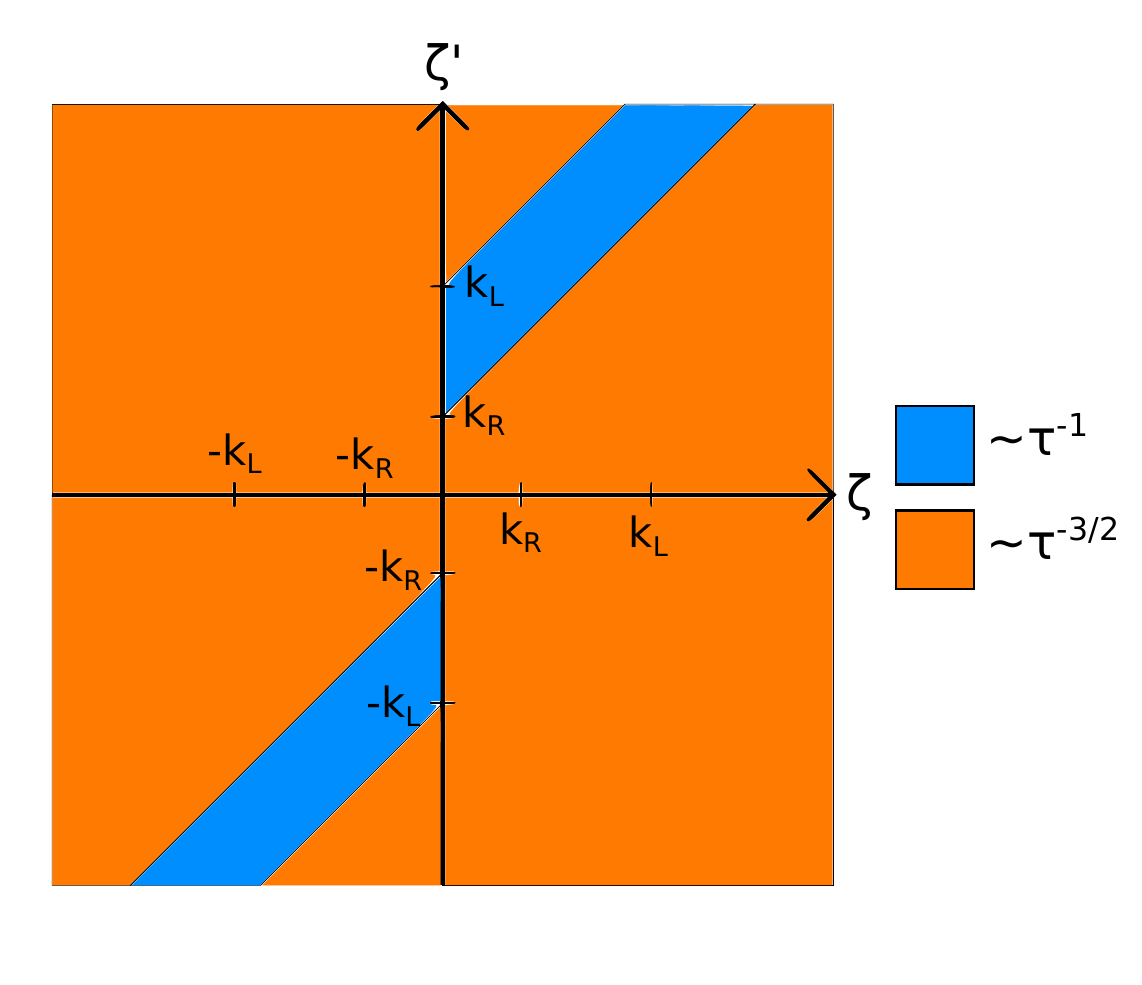}
\caption{Phase diagram of the decay of the density correlation $C(x,x',\tau)$ -- similar to Fig. \ref{suppphasediag} -- in the absence 
of a defect.}\label{fig_without}
\end{figure}
 
\bigskip

\bigskip

\section{General potential $V(x)$} 

 In this section we generalise our results to the case of a potential $V(x)$ with no a priori symmetry with respect to $x \to -x$, but which again vanishes outside $x \in [-a/2,a/2]$.
This case is treated separately as it is technically slightly more involved. For simplicity we also assume no bound states, although these
can be incorporated as in Section \ref{sec:bound}. In a first stage we obtain the eigenfunctions of $\hat H$, and in a second
stage we study the dynamics of the initial eigenstates.

\subsection{Eigenstates of $\hat H$}

For a general potential $V(x)$, with no a priori symmetry, it turns out that the eigenfunctions can be chosen in the form
\be \label{formH} 
\phi_{\sigma,k}(x)=\tilde{c}_{\ell,\sigma,k} \cos(k(|x|-\delta_{k}^\sigma))\left( P_k^\sigma \theta(x<-\frac{a}{2})+ \sigma P_k^{-\sigma} \theta(x>\frac{a}{2}) \right)
\ee
where $\sigma$ takes the value $+1$ and $-1$. One has defined the phase shifts as
\be \label{defphaseshiftsgeneral} 
e^{-2ik\delta_k^\sigma}=\frac{1}{2}(\sr_R(k)+\sr_L(k) + \sigma \sqrt{(\sr_R(k)-\sr_L(k))^2+4\st(k)^2})
\ee
and the amplitudes $P_k$ and $p_k$ as
\be\label{defp}
P_k^\sigma=\frac{1}{\sqrt{1+|p_k^\sigma|^2}} \quad , \quad p_k^\sigma=\frac{e^{-2ik \delta_k^\sigma}-\sr_L(k)}{\st(k)}=\frac{\st(k)}{e^{-2ik \delta_k^\sigma}-\sr_R(k)}
= \frac{\sr_R(k)-\sr_L(k)+\sigma \sqrt{(\sr_R(k)-\sr_L(k))^2+ 4 \st^2(k)}}{2 \st(k)} \ee
and at this stage $k$ is arbitrary. Before proceeding note the useful identities
\be \label{identity_pk}
p_k^+ p_k^{-}=-1 \quad , \quad P_k^\sigma p_k^\sigma = \sigma P_k^{-\sigma} \quad , \quad {P_k^+}^2 + {P_k^-}^2 = 1 \;.
\ee 
The first comes from the definition of $p_k^\sigma$ (last equation in \eqref{defp}). Next one notes that $p_k^\sigma$ is real, and, because $p_k^+>p_k^-$ we have $p_k^+>0$ and $p_k^-<0$. 
One then checks from the definition of $P_k^\sigma$ (first equation in \eqref{defp}), 
together with the previous identity, that $P_k^+ p_k^+ = P_k^-$ and $P_k^- p_k^- = -P_k^+$,
as well as the last relation in \eqref{identity_pk}.  Next, for the definition 
\eqref{defphaseshiftsgeneral} of the phase shifts to be meaningful (and give $k \delta_k^\sigma \in [0,2 \pi])$
we need to check that the r.h.s is a complex number of modulus unity. This is done using 
the parameterization \eqref{param}, and \eqref{defphaseshiftsgeneral} becomes
\be 
e^{-2ik\delta_k^\sigma}=  e^{ i \frac{\varphi_R(k) + \varphi_L(k)}{2}} \left( \cos \theta(k) \cos (\frac{\varphi_R(k) - \varphi_L(k)}{2}) + \sigma i \sqrt{\cos^2 \theta(k) \sin^2 \frac{\varphi_R(k) - \varphi_L(k)}{2} + \sin^2 \theta(k)} \right)  
\ee 

which is clearly of unit modulus.

In the case of an even potential $V(x)=V(-x)$, one has $\sr_R(k)=\sr_L(k)= \sr(k)$ and 
\eqref{defphaseshiftsgeneral} becomes 
\be 
 e^{-2ik\delta_k^{\pm}} = \sr(k) \pm \st(k) 
\ee 
with the convention that $\sqrt{\st(k)^2}=\st(k)$ which we use here and below (the other choice amounts to redefine $\sigma \to -\sigma$). 
This is identical to \eqref{rtodelta}, which indicates that $\sigma=+1$ becomes the even eigenfunction, and $\sigma=-1$ the 
odd one. Indeed one checks that in this limit $p_k^+=1$, $p_k^-=-1$, hence $P_k^+=P_k^-=\frac{1}{\sqrt{2}}$. Thus the eigenfunctions
\eqref{formH} recover the ones given in \eqref{phi} with $\tilde c_{\ell,\sigma,k}  = \sqrt{2} c_{\ell,\sigma,k}$ (up to an irrelevant global minus
sign in the definitions of $\phi_{-,k}(x)$).

Let us now justify the form \eqref{formH} for the eigenfunctions of $\hat H$. Similarly to the symmetric case, outside of the impurity (i.e., for $|x|>a/2$) the particles display free propagation and the eigenfunctions thus take the form of Eq. \eqref{def_phi} where the four amplitudes are related as in \eqref{Srelation} via
the S matrix \eqref{def_S}. Eliminating $B_k$ and $C_k$ they take the form for $|x|>a/2$
\be \label{formAD} 
\phi_{k}(x) = ( A_k e^{i k x} + (\sr_L(k) A_k + \st(k) D_k)  e^{-i k x} ) \theta(x< - \frac{a}{2}) +
 ( (\st(k) A_k + \sr_R(k) D_k)  e^{i k x} + D_k e^{- i k x} ) \theta(x> \frac{a}{2})
\ee 
which depends on two complex parameters $A_k$ and $D_k$ and thus forms a two-dimensional subspace.
One can check that for each value of $\sigma=\pm 1$ the eigenfunction $\phi_{\sigma,k}(x)$ in Eq. (\ref{formH}) is indeed of the above 
form (\ref{formAD}) with a specific choice for the ratio
\be 
\frac{D_k}{A_k}= p_k^\sigma \;.
\ee 
To show that one uses the definitions \eqref{defp} and the identities \eqref{identity_pk} mentioned above.  

Once we have parameterized the eigenfunctions in that way, it is immediate to see that the 
boundary conditions $\phi_{\sigma,k}(\pm \frac{\ell}{2})=0$ become equivalent to the quantification condition 
\be \label{quantificationgeneral} 
k \in \Lambda^\sigma \quad \Leftrightarrow \quad e^{-ik \ell}= - e^{-2ik\delta_k^\sigma} 
\ee
which defines the lattices $\Lambda^\sigma$ in momentum space.

Note that on the full line $x \in \mathbb{R}$ the eigenfunctions $\phi_{+,k}(x)$ and $\phi_{-,k}(x)$ are degenerate in energy 
and form a basis of the two dimensional eigenspace. For $x \in [-\ell/2,\ell/2]$ with the vanishing boundary conditions,
these functions remain eigenfunctions, with now different energies. The above parametrization and choice of 
ratio $\frac{D_k}{A_k}$ is the convenient one to study these "even" boundary conditions. 

Finally, it is easy to see that the squared norm of the wave function in \eqref{formH} 
behaves as $\simeq \frac{\ell}{4} {\tilde c}_{\ell,\sigma,k}^2 ({P_k^+}^2 + {P_k^-}^2)$. Using the
last identity in \eqref{identity_pk} we see that, as before, the normalization constant behaves at large $\ell$ as ${\tilde{c}}_{\ell,\sigma,k}\underset{\ell\to\infty}{\simeq} \sqrt{\frac{4}{\ell}}$.
\\

{\bf Remark:} The same method can be applied to the eigenstates of the initial Hamiltonian $\hat H_0$. 
In that case the system is cut in two halves and $\st(k)=0$. Eq. \eqref{defphaseshiftsgeneral} recovers the
two phase shifts denoted $\delta_k^{R/L}$ (and already defined above in \eqref{phaseshiftH00}) 
\be
e^{-2ik\delta_k^{R/L}} = \sr_{R/L}^0(k) 
\ee 
where we chose by convention $\sqrt{(\sr_R(k)-\sr_L(k))^2}=\sr_R(k)-\sr_L(k)$ and $\sigma_R=+1$, $\sigma_L=-1$. 
One has $p_k^R=+\infty$, $p_k^L=0$, $P_k^L=1$, $P_k^R=0$ (equivalently $A_{k}^{R}=0$, $D_{k}^{L}=0$) and
one finds for $|x|>a/2$ 
\bea\label{formH0}
&\phi_k^{R}(x)=\tilde{c}_{\ell,\sigma,k}\cos(k(|x|-\delta_k^R))\theta(x>\frac{a}{2})\\
&\phi_k^{L}(x)=\tilde{c}_{\ell,\sigma,k}\cos(k(|x|-\delta_k^L))\theta(x<-\frac{a}{2})
\eea
and we recover \eqref{phiRL}. The quantification condition is again
\be
e^{-ik\ell}= -e^{-2ik\delta_k^{R/L}} \quad , \quad  k \in \Lambda^{R/L}
\ee
which defines the two initial lattices $\Lambda^{R/L}$ in momentum space.

\subsection{Dynamics of initial eigenstates}

Now we want to find the large time behavior of a single eigenstate with initial condition $\psi_k^{R/L}(x,t=0) = \phi_k^{R/L}(x)$. 
We start from the decomposition \eqref{exact_psik} at finite time and $\ell$, which we write in a slightly different way as a sum of two components 
\bea\label{psisigma}
&\psi_k^{R/L}(x,t)=\psi_{+,k}^{R/L}(x,t)+ \psi_{-,k}^{R/L}(x,t)\\
&\psi_{\sigma,k}^{R/L}(x,t)=\underset{k'\in \Lambda^\sigma}{\sum}\phi_{\sigma,k'}(x)e^{-i\frac{k'^2}{2}t}\braket{\phi_{\sigma,k'}|\psi_k^{R/L}(t=0)}%\frac{\gamma_{k',k}^{R/L}}{\ell(k-k')}
\eea

In the overlap, we leave aside again the contribution of the region of the impurity (i.e., $x\in [-\frac{a}{2}, \frac{a}{2}]$) which leads to a decay at large time.
In the other region, $x \in [-\frac{\ell}{2},-\frac{a}{2}] \cup [\frac{a}{2},\frac{\ell}{2}]$, using that all terms are plane waves as in the argument leading to \eqref{overlap0}
we can write
\bea\label{overlaps}
 &\braket{\phi_{\sigma,k'}|\psi_k^{R}(t=0)} \simeq \int_{a/2}^{\ell/2} \phi_{\sigma,k'}^*(x) \phi^R_{k}(x) \, dx  = \frac{1}{(k')^2-k^2} ( \phi_{\sigma, k'}^{* \prime}(\frac{a}{2}) \phi^R_{k}(\frac{a}{2}) - \phi_{\sigma, k'}^*(\frac{a}{2}) \phi^{R,\prime}_{k}(\frac{a}{2}))\\
 &\braket{\phi_{\sigma,k'}|\psi_k^{L}(t=0)} \simeq \int_{-\ell/2}^{-a/2} \phi_{\sigma,k'}^*(x) \phi^L_{k}(x) \, dx  = -\frac{1}{(k')^2-k^2} ( \phi_{\sigma, k'}^{* \prime}(-\frac{a}{2}) \phi^L_{k}(-\frac{a}{2}) - \phi_{\sigma, k'}^*(-\frac{a}{2}) \phi^{L,\prime}_{k}(-\frac{a}{2})) \nonumber
\eea
This is valid for $k' \in \Lambda^\sigma$ and $k \in \Lambda^{R/L}$ since we have used explicitly the vanishing of the
wave-functions at $x=\pm \ell/2$. Note that these expressions contain a pole when $k' \to k$ in the large $\ell$ limit.

In order to compute the sum over $k' \in \Lambda^\sigma$  in \eqref{psisigma} in the double large $\ell$ and large $t$ limit, we use the same method as before.
We first extend the formula for $\phi_{\sigma,k'}(x)$ \eqref{formH} and the formula \eqref{overlaps} for the overlaps, in the complex plane for $k'$.
This is done via an analytic continuation of the phase shifts $k' \delta^\sigma_{k'}$ and the amplitudes $P_{k'}^\sigma$. We will
not give details since this method was described above. We use the formula \eqref{sumtoint} and obtain
\be
\psi_{\sigma,k}^{R/L}(x,t)=\int_{\Gamma_0}\frac{dk'}{2\pi} g_{\delta^\sigma,\ell}(k') \phi_{\sigma,k'}(x)e^{-i\frac{k'^2}{2}t}(\ell \braket{\phi_{\sigma,k'}|\psi_k^{R/L}(t=0)})
\ee
In the double limit only the residue of the pole at $k'=k$ contributes
\be\label{doublelim}
\psi_{\sigma,k}^{R/L}(x,t)\underset{\substack{\ell\to \infty \\ t \to \infty}}{\simeq}  i g_{\delta^\sigma,\ell}(k) \phi_{\sigma,k}(x)e^{-i\frac{k^2}{2}t} {\rm Res}_{k'=k} (\ell \braket{\phi_{\sigma,k'}|\psi_k^{R/L}(t=0)})
\ee
(for $k \in \Lambda^{R/L}$, $g_{\delta^\sigma,\ell}(k)$ is independent of $\ell$ as shown in \eqref{prefactor1}).

The last step is to compute the residue ${\rm Res}_{k'=k}  \braket{\phi_{\sigma,k'}|\psi_k^{R}(t=0)} \simeq {\rm Res}_{k'=k}\int_{a/2}^{\ell/2} \phi_{\sigma,k'}^*(x) \phi^R_{k}(x) \, dx $, i.e., from \eqref{overlaps} 
\bea
Res_{k'=k}\int_{a/2}^{\ell/2} \phi_{\sigma,k'}^*(x) \phi^R_{k}(x) \, dx  = \frac{1}{2k} ( \phi_{\sigma,k}^{* \prime}(\frac{a}{2}) \phi^R_{k}(\frac{a}{2}) - \phi_{\sigma,k}^*(\frac{a}{2}) \phi^{R,\prime}_{k}(\frac{a}{2}))
\eea
where here $\phi^*_{\sigma,k}(\frac{a}{2})$ is the analytic continuation of $\phi^*_{\sigma,k'}( \frac{a}{2})$ to 
$k'=k$ (in a different lattice). This analytic continuation $\phi^*_{\sigma,k}(x)$ satisfies ${\phi^*_{\sigma,k}}''(x)= - k^2 \phi^*_{\sigma,k}(x)$ outside
of the interval $[-a/2,a/2]$,
hence the r.h.s. above is independent of $a$ (from the Wronskian theorem) and one can as well replace $a/2 \to \ell/2$ leading to
\bea \label{overlap2} 
&{\rm Res}_{k'=k} \int_{a/2}^{\ell/2}   dx \phi^*_{\sigma,k'}(x) \phi^R_{k}(x) = -\frac{1}{2 k} 
 \phi^*_{\sigma,k}(\frac{\ell}{2}) \phi^{R \prime}_{k}(\frac{\ell}{2})\\
 &{\rm Res}_{k'=k} \int_{-\ell/2}^{-a/2}   dx \phi^*_{\sigma,k'}(x) \phi^L_{k}(x) = \frac{1}{2 k} 
 \phi^*_{\sigma,k}(-\frac{\ell}{2}) \phi^{L \prime}_{k}(-\frac{\ell}{2}) \nonumber 
\eea 
using that $\phi^{R/L}_{k}(\frac{\ell}{2})=0$. Now the point is that $\phi^*_{\sigma,k}(\frac{\ell}{2})$ does not vanish since $k$ does not belong to $\Lambda^\sigma$.
To evaluate $\phi^*_{\sigma,k}(\frac{\ell}{2})$ we use \eqref{formH} leading to 
\bea \label{phiedge} 
&\phi_{\sigma,k} (\frac{\ell}{2} ) =\tilde{c}_{\ell,\sigma,k} \cos(k(\frac{\ell}{2}-\delta_k^\sigma)) \sigma P_k^{-\sigma}\\
&\phi_{\sigma, k}(- \frac{\ell}{2} ) = \tilde{c}_{\ell,\sigma,k} \cos(k(\frac{\ell}{2}-\delta_k^\sigma)) P_k^\sigma \nonumber 
\eea
We need to evaluate these expressions for $k \in \Lambda^{R/L}$, i.e., to use the quantification condition
\be \label{q} 
e^{-ik\ell}= - e^{-2ik\delta_k^{R/L}}
\ee
Substituting \eqref{q} into \eqref{phiedge} we obtain 
\bea
&\phi_{\sigma,k}( \frac{\ell}{2})=i\tilde{c}_{\ell,\sigma,k}e^{ik(\frac{\ell}{2}-\delta_k^{R/L})}\sin(k(\delta_k^{R}-\delta_k^\sigma))\sigma P_k^{-\sigma} \quad k\in \Lambda^R\\
&\phi_{\sigma,k}(- \frac{\ell}{2})=i\tilde{c}_{\ell,\sigma,k}e^{ik(\frac{\ell}{2}-\delta_k^{R/L})}\sin(k(\delta_k^{L}-\delta_k^\sigma)) P_k^\sigma \quad k\in \Lambda^L
\eea
Next we evaluate using \eqref{phiRL}
\bea
&{\phi_k^{R}}'(\frac{\ell}{2})=-k c^0_{\ell,R,k}\sin(k(\frac{\ell}{2}-\delta_k^R))\\
&{\phi_k^{L}}'(-\frac{\ell}{2})=k c^0_{\ell,L,k}\sin(k(\frac{\ell}{2}-\delta_k^L))
\eea
and substitute again \eqref{q} which leads to
\bea
&{\phi_k^{R}}'(\frac{\ell}{2})=i k c^0_{\ell,R,k} e^{ik(\frac{\ell}{2}-\delta_k^R)}\\
&{\phi_k^{L}}'(-\frac{\ell}{2})=-ikc^0_{\ell,L,k}e^{ik(\frac{\ell}{2}-\delta_k^L)} 
\eea
Putting all terms together in \eqref{overlap2} one finds in the large $\ell$ limit using that $\tilde{c}_{\ell,\sigma,k} c^0_{\ell,R/L,k} \simeq \frac{4}{\ell}$
\bea \label{resfin2} 
&{\rm Res}_{k'=k} \int_{a/2}^{\ell/2}  dx \phi_{\sigma,k'}^*(x) \phi^R_{k}(x) = - \frac{2}{\ell} \sigma P_k^{-\sigma} \sin(k(\delta_k^{R}-\delta_k^\sigma))\\
&{\rm Res}_{k'=k} \int_{-\ell/2}^{-a/2}  dx \phi_{\sigma,k'}^*(x) \phi^L_{k}(x) = - \frac{2}{\ell}P_k^\sigma\sin(k(\delta_k^{L}-\delta_k^\sigma)) \nonumber 
\eea

Consider now again the formula \eqref{doublelim} for the large time limit of $\psi_{\sigma,k}^{R/L}(x,t)$. Inserting
into this formula the expression \eqref{prefactor1} for $- 2i g_{\delta^\pm,\ell}(k)$, the expression \eqref{formH} for $\phi_{\sigma,k}(x)$ and 
the expression for the residues \eqref{resfin2} we obtain
\bea \label{psipsi0} 
&\psi_{k,\pm}^{R}(x,t)\underset{\substack{\ell\to\infty\\t \to \infty}}{\simeq} \frac{1}{\sqrt{\ell}}e^{ik\delta_k^{R}}e^{-i\frac{k^2}{2}t}(e^{ik(|x|-2\delta_{k}^\pm)}+e^{-ik|x|})P_k^{\mp}\left(\pm P_k^\pm \theta(x<-\frac{a}{2})+ P_k^{\mp} \theta(x>\frac{a}{2}) \right)\\
&\psi_{k,\pm}^{L}(x,t)\underset{\substack{\ell\to\infty\\t \to \infty}}{\simeq}\frac{1}{\sqrt{\ell}}e^{ik\delta_k^{L}}e^{-i\frac{k^2}{2}t}(e^{ik(|x|-2\delta_{k}^\pm)}+e^{-ik|x|})P_k^{\pm}\left(P_k^\pm \theta(x<-\frac{a}{2}) \pm  P_k^{\mp} \theta(x>\frac{a}{2}) \right) \nonumber 
\eea

Finally we recombine $\psi_{+,k}^{R/L}$ and $\psi_{-,k}^{R/L}$ from \eqref{psisigma} to obtain the final wave-function $\psi_{k}^{R/L}$ in the double limit
\bea \label{psipsi} 
&\psi_{k}^{R}(x,t)\underset{\substack{\ell\to\infty\\t \to \infty}}{\simeq}\frac{1}{\sqrt{\ell}}e^{ik\delta_k^{R}}e^{-i\frac{k^2}{2}t} \left( (e^{-2ik\delta_{k}^+}-e^{-2ik\delta_{k}^-})e^{-ikx}P_k^+ P_k^-\theta(x<-\frac{a}{2}) \right. \\
& \left. + ((e^{ik(x-2\delta_{k}^+)}+e^{-ikx}){P_k^-}^2 + (e^{ik(x-2\delta_{k}^-)}+e^{-ikx}){P_k^+}^2 )\theta(x>\frac{a}{2})\right) \nonumber \\
&\psi_{k}^{L}(x,t)\underset{\substack{\ell\to\infty\\t \to \infty}}{\simeq}\frac{1}{\sqrt{\ell}}e^{ik\delta_k^{L}}e^{-i\frac{k^2}{2}t} \left( (e^{-2ik\delta_{k}^+}-e^{-2ik\delta_{k}^-})e^{ikx}P_k^+ P_k^-\theta(x>\frac{a}{2}) \right. \\
& \left. + ((e^{ik(-x-2\delta_{k}^+)}+e^{ikx}){P_k^+}^2 + (e^{ik(-x-2\delta_{k}^-)}+e^{ikx}){P_k^-}^2)\theta(x<-\frac{a}{2})\right) \nonumber 
\eea

These expressions can be simplified using a set of identities. 
Using \eqref{defp} we first show that 
\be 
\frac{(\sr_R(k)-\sr_L(k))^2}{ \st^2(k)} %= \sin(\frac{\phi_R-\phi_L}{2}) \cot \theta 
= {p_k^+}^2 + {p_k^-}^2 -2
\ee 
Hence one has using \eqref{defphaseshiftsgeneral} and again \eqref{defp}
\be \label{id1} 
e^{-2ik\delta_k^+} - e^{-2ik\delta_k^-} = \sqrt{\st(k)^2} \sqrt{{p_k^+}^2 + {p_k^-}^2 +2 }
 = \frac{\st(k)}{P_k^+ P_k^-} 
\ee 
Next, from \eqref{defp} one has 
\bea
&\sr_R(k) = e^{-2ik\delta_k^+} + p_k^- \st(k)\\
&\sr_L(k) = e^{-2ik\delta_k^-} - p_k^- \st(k) \nonumber 
\eea
where we used that $p^+_k=-1/p^-_k$. These equations imply
\bea
&\sr_R(k) + \sr_L(k) = e^{-2ik\delta_k^+} + e^{-2ik\delta_k^-}\\
&\sr_R(k) - \sr_L(k) = (e^{-2ik\delta_k^+} - e^{-2ik\delta_k^-})(1-2 {P_k^+}^2) \nonumber 
\eea
where in the second equation we used \eqref{id1} and $p_k^- P_k^-= - P_k^+$. 
The solution of this system for $\sr_R(k),\sr_L(k)$ is 
\bea\label{assymetricrelation}
&\sr_R(k)=e^{-2ik\delta_k^+} {P_k^-}^2 + e^{-2ik\delta_k^-} {P_k^+}^2\\
&\sr_L(k)=e^{-2ik\delta_k^+} {P_k^+}^2 + e^{-2ik\delta_k^-}{P_k^-}^2 \nonumber 
\eea
where we used ${P_k^+}^2 + {P_k^-}^2=1$.

Starting from \eqref{psipsi}, making use of the identities \eqref{id1} and \eqref{assymetricrelation}, and using
again ${P_k^+}^2 + {P_k^-}^2=1$, we obtain our final result, for $|x| >a/2$
\bea
&\psi_{k}^{R}(x,t)\underset{\substack{\ell\to\infty\\t \to \infty}}{\simeq}\frac{1}{\sqrt{\ell}}e^{ik\delta_k^{R}}e^{-i\frac{k^2}{2}t} \left( e^{-ikx} \st(k) \theta(x<-\frac{a}{2})  + (e^{-ikx} + e^{ikx} \sr_R(k)) \theta(x>\frac{a}{2})\right)\\
&\psi_{k}^{L}(x,t)\underset{\substack{\ell\to\infty\\t \to \infty}}{\simeq}\frac{1}{\sqrt{\ell}}e^{ik\delta_k^{L}}e^{-i\frac{k^2}{2}t} \left( e^{ikx} \st(k)\theta(x>\frac{a}{2})  + (e^{ikx} + e^{-ikx} \sr_L(k)) \theta(x<-\frac{a}{2})\right) \;.
\eea

{\bf Remark:} If in the manipulations leading to \eqref{psipsi0} (in the paragraph after
\eqref{resfin2}) we do not insert the explicit form for $\phi_{\sigma,k}(x)$, 
we obtain a formula for any $x=O(1)$ (including the region inside the impurity). 
A natural prediction for the asymmetric potential is thus that, for any $x \in \mathbb{R}$ one has
$\psi_{k}^{R/L}(x,t)\underset{\substack{\ell\to\infty\\t\to\infty}}{\simeq}\frac{1}{\sqrt{\ell}}e^{-i\frac{k^2}{2}t}\chi_k^{R/L}(x)$
where 
\bea
&& \frac{1}{\sqrt{\ell}} \chi_{k}^{R}(x)= \phi_{+,k}(x)e^{ik(\delta^{R}_k-\delta_k^{+})} P_k^- - 
\phi_{-,k}(x)e^{ik(\delta^{R}_k-\delta_k^{-})}P_k^+ \\
&& \frac{1}{\sqrt{\ell}} \chi_{k}^{L}(x)= \phi_{+,k}(x)e^{ik(\delta^{L}_k-\delta_k^{+})} P_k^+ +
\phi_{-,k}(x)e^{ik(\delta^{L}_k-\delta_k^{-})}P_k^-
\eea 
This generalizes the formula \eqref{remarksym} which holds for symmetric potentials. The latter is recovered since
in that case $P_k^\pm=1/\sqrt{2}$ (taking into account the minus sign difference in the definition of $\phi_{-,k}(x)$ in the two sections.)
\\

% {\red P: Gabriel, below is what you wrote, please check that it is identical to what I got
% \be
% \frac{1}{\sqrt{\ell}} \chi_{k}^{R/L}(x)=(\phi_{+,k}(x)e^{ik(\delta^{R/L}_k-\delta_k^{+})} P_k^\mp \mp \phi_{-,k}(x)e^{ik(\delta^{R/L}_k-\delta_k^{-})}P_k^\pm )
% \ee
% where $\pm$ stands for $\pm=\begin{cases}
% + \quad R\\
% - \quad L
% \end{cases}$ and $\mp$ stands for the reverse. 
% } 

% {\red P: Gabriel verifie que tu es d'accord que $\phi_{-}$ tend avec un signe $-$. On pourrait signaler ca deja plus haut, peux tu 
% le faire (en bleu ou rouge)? Je ne pense pas une bonne idee de redefinir $\phi_{-}$ car on va faire des erreurs
% partout, il vaut mieux juste signaler cette petite discrepancy dans les conventions. Fais le stp si tu es d'accord.}

{\bf Remark:} The above calculation can be extended to the ray regime with an asymmetric impurity, and the result is that the large time behavior is obtained from formula \eqref{chixi} by simply changing $\sr$ by $\sr_R$ (resp $\sr_L$) in $\chi_{\xi,k}^{R}$ (resp $\chi_{\xi,k}^{L}$).

\end{widetext}


\begin{thebibliography}{100}
\bibitem{footnotetau} 
Alternatively one could define the large $t$ limit of
$K(\xi t+y,t;\xi' (t -\tau) +y',t-\tau)$ which amounts to
a simple shift $y' \to y' - \xi' \tau$ in our formula.
\bibitem{Lippman} 
 Note that here we considered a finite box in the large $\ell$ limit. One can
 instead consider directly the system on the infinite axis and use the Lippman-Schwinger
 approach, see e.g. \cite{andrea_moving}. Although these are equivalent, the present method allows a control
 of finite size effects.
\end{thebibliography}
\end{document}